\documentclass[a4paper,twocolumn,11pt,accepted=2021-04-22]{quantumarticle}
\pdfoutput=1

\usepackage[utf8]{inputenc}
\usepackage[english]{babel}
\usepackage[T1]{fontenc}
\usepackage{amsmath}

\usepackage[numbers,sort&compress]{natbib}
\usepackage{hyperref}

\usepackage{dsfont}
\usepackage[normalem]{ulem}
\usepackage{mathtools}
\usepackage[makeroom]{cancel}
\usepackage{bbm}

\usepackage{amsfonts}
\usepackage{color}
\usepackage{amssymb}
\usepackage{graphicx}

\definecolor{henrik}{rgb}{1,.4,0}

\newcommand{\tr}{\mathrm{Tr}} %old
\newcommand{\id}{\mathbbm{1}}

\newcommand{\1}{\mathrm{id}}

\renewcommand{\1}{\id}

\newcommand{\ket}[1]{\left.\left|{#1}\right.\right\rangle}

\newcommand{\bra}[1]{\left.\left\langle{#1}\right.\right|}

\newcommand{\ketbra}[2]{\ket{#1} \!\! \bra{#2}}

\newcommand{\rom}[1]{\uppercase\expandafter{\romannumeral #1\relax}}

\usepackage{amssymb}             % AMS Math
\newcommand*\pmat[1]{\begin{pmatrix}#1\end{pmatrix}}

\begin{document}

\title{Weakly invasive metrology: quantum advantage and physical implementations}

\author{M. Perarnau-Llobet}
    \affiliation{Max-Planck-Institut f\"ur Quantenoptik, D-85748 Garching, Germany.}
    \affiliation{Munich Center for Quantum Science and Technology (MCQST), Schellingstrasse 4, D-80799 M\"unchen}
    \affiliation{D\' epartement de Physique Appliqu\' ee, Universit\' e de Gen\`eve, Gen\`eve, Switzerland} 
    
    \author{D. Malz}
 \affiliation{Max-Planck-Institut f\"ur Quantenoptik, D-85748 Garching, Germany.}
 \affiliation{Munich Center for Quantum Science and Technology (MCQST), Schellingstrasse 4, D-80799 M\"unchen}

\author{J. I. Cirac}
 \affiliation{Max-Planck-Institut f\"ur Quantenoptik, D-85748 Garching, Germany.}
 \affiliation{Munich Center for Quantum Science and Technology (MCQST), Schellingstrasse 4, D-80799 M\"unchen}

%\orcid{0000-0003-0290-4698}
%\thanks{You can use the \texttt{\textbackslash{}email}, \texttt{\textbackslash{}homepage}, and \texttt{\textbackslash{}thanks} commands to add additional information for the preceding \texttt{\textbackslash{}author}. If applicable, this can also be used to indicate that a work has previously been published in conference proceedings.}

\begin{abstract}
We consider the estimation of a Hamiltonian parameter of a  set of highly photosensitive samples, which are damaged after a few photons $N_{\rm abs}$ are absorbed, for a total time $T$. The samples are modelled as a two mode photonic system, where photons simultaneously acquire information on the unknown parameter and are absorbed at a fixed rate.  We show that arbitrarily intense coherent states can obtain information at a rate that scales at most linearly with $N_{\rm abs}$ and $T$, whereas quantum states with finite intensity  can overcome this bound. We characterise the  quantum advantage  as a function of $N_{\rm abs}$ and $T$, as well as its robustness to imperfections (non-ideal detectors, finite preparation and measurement rates for quantum photonic states).  
%The quantum advantage scales like $N/N_{\rm abs}$ (where $N$ is the number of photons in the quantum state),  so that few-photon entangled states can already surpass the classical bound when $N_{\mathrm{abs}}$ is small.
  We discuss an implementation in cavity QED, where Fock states are both prepared and measured by coupling atomic ensembles to the cavities.  We show that superradiance, arising due to a collective coupling between the cavities and the atoms, can be exploited for improving the speed and efficiency of the measurement.  
\end{abstract}

\maketitle

\section{Introduction}

One exciting prospect of quantum technologies are more precise photonic measurements through the use of quantum correlations~\cite{Giovannetti2011,Tth2014,demkowicz-dobrzanski15,Dowling2015,polino2020photonic}.
%Quantum advantage in photonic metrology is a well-established field of research~\cite{Giovannetti2011,Tth2014,demkowicz-dobrzanski15,Dowling2015,polino2020photonic}.
Given a fixed (average) photon number $N$, a variety of quantum states surpasses the shot-noise limit imposed by coherent states, notably squeezed, NOON or twin Fock states~\cite{Caves1981,holland93,Bollinger1996}. %States such as NOON or twin Fock states achieve the maximum advantage, reaching the so-called Heisenberg scaling~\cite{holland93,Bollinger1996}. 
%\dm{Somehow the folloing sentence doesn't sound particularly strong.} This advantage is lost if the number of photons is not restricted, since the variance in the measurements decreases with the intensity even for coherent states. 
One of the motivations to limit  $N$ is that some photosensitive materials 
%, such as biological samples~\cite{Ono2013,Taylor2013,Taylor2014,Taylor2016,Whittaker2017,Cole2014},
may be damaged when they absorb too many photons. However, in this case it may be more meaningful to constrain the number of photons absorbed by the sample $N_{\mathrm{abs}}$, instead of~$N$. %~\cite{Wolfgramm2012,Taylor2016}.
Examples of delicate samples include   biological~\cite{Neuman1999,Peterman2003,Cole2014,Taylor2013,Taylor2014,Taylor2016,Whittaker2017}, molecular~\cite{Pototschnig2011} and  atomic systems~\cite{Eckert2007,Wolfgramm2012}. For the latter, the maximal  $N_{\rm abs}$ that the sample can tolerate can  be a few photons, as  quantum states such as the GHZ state can be extremely susceptible to noise. 
%As an example, $N_{\rm abs}=1$ is enough to }
%\textcolor{blue}{[I put GHZ state in explicitly. I think it's clear why it is sensitive to noise, otherwise we can cite something, e.g.,, the first paper with a GHZ state?]}. 
%As an example, one could be interested in measuring the total number of excitations in an atomic system without destroying the coherence. In such a setup, even a single absorption event  ($N_{\mathrm{abs}}=1$) can destroy the coherence and change the measurement result \textcolor{blue}{[I agree a citation would be nice, but I just wrote this down thinking about a W state, where you can either (i) detect where the excitations are, which destroys the coherence, or change the number of excitations, which obviously changes the measuremnt result. We could mention this explicitly and cite the first paper on W state?]}. \textcolor{red}{Since we are not really workiing with atomic delicate samples for the rest of the paper, I would either remove it or not expand  on it. }}

The number of absorbed photons $N_{\mathrm{abs}}$ is  not only  determined by the $N$ but also in general depends on  the time that the  sample is exposed to  light and its intrinsic properties. Still, for interferometric measurements~\cite{Wolfgramm2012,Ono2013,Taylor2013,Taylor2014},  placing a constraint in $N_{\rm abs}$ becomes essentially equivalent to fixing the total number of photons. Instead, in this work we will consider frequency measurements in which photons are allowed to interact continuously with the sample. In this case, we will show that  constraining $N_{\rm abs}$ instead of $N$ leads to qualitatively new results. 
 %in frequency measurements in which photons are allowed to interact continuously with the sample.   

%\ma{The measurement of delicate atomic samples } While in interferometric measurements this constraint is equivalent to fixing the total number of photons, we will show here it leads to qualitatively new results in frequency measurements in which photons are allowed to interact continuously with the sample. 

%\textcolor{blue}{here discuss better previous literature, including~\cite{Bollinger1996,Huelga1997,Shaji2007,Hayes2018}. }

In frequency measurements, both the  probe size (e.g. the number of photons $N$) and the total time $T$ are resources for quantum metrology~\cite{Bollinger1996,Huelga1997,Shaji2007,Hayes2018}, and interesting interplays between both appear when the probe is subject to noise or decoherence~\cite{Huelga1997,Shaji2007,Hayes2018,Matsuzaki2011,Chin2012,Chaves2013,Brask2015,Smirne2016}. Optimal  metrological protocols for a fixed $N$ and $T$ have  been considered for atomic  measurements~\cite{Huelga1997,Shaji2007,Hayes2018,Matsuzaki2011,Chin2012,Chaves2013,Brask2015,Smirne2016},  non-linear photonic systems~\cite{Woolley2008}, 
many-body  open quantum systems~\cite{Beau2017}, and time-dependent Hamiltonians~\cite{Naghiloo2017,Pang2017,sun2020exponentially}.  Such time-optimized  protocols can lead to a supraclassical scaling with $N$ of the measurement precision for specific types of correlated or directional noise~\cite{Matsuzaki2011,Chin2012,Dorner2012,Chaves2013,Dur2014,Kessler2014,Arrad2014,Brask2015,Smirne2016},  and to a constant  advantage  for  uncorrelated noise~\cite{Huelga1997,escher2011general,DemkowiczDobrzaski2012,Koodyski2013,knysh2014true}.

%, and  non-linear photonic systems~\cite{}. 
%The importance of time delays in the preparation of measurement of quantum states has also been stressed~\cite{Hayes2018}. 
%\ma{Ultimately, the reason for restricting the total time $T$ is that there always is a limit on the time available to perform a certain experiment. On the other hand, the limit on the number of photons $N$ may arise due to fragile samples or quantum states, as is commonly argued [citations]. As an example, one could be interested in measuring the total number of excitations in an atomic system without destroying the coherence. In such a setup, even a single absorption event can destroy the coherence and change the measurement result [citation?].}

Here, we investigate frequency measurements in a two-mode photonic system, where one wishes to estimate the strength of a beamsplitter-like interaction. This a common task when characterizing the coupling strength in devices such as coupled cavity modes, nanophotonic lattices, superconducting resonators or micromechanical resonators. 
%\textcolor{blue}{\emph{[I'm not sure if we need a citation. It's pretty obvious that if you build a system of coupled resonators, you would like to know how strong the coupling is. This done in essentially every paper, so I'm not sure what to cite here. I don't think \cite{Woolley2008} is a good reference for that reason.]}}. 
In contrast to previous works (see, for example, \cite{Woolley2008}),  we shall not put the constraint on $N$, but on the energy that the system can absorb ($N_{\rm abs}$). This subtle difference turns out to have important consequences:  quantum metrological protocols involving  a small amount $N\approx 5-10$ of photons  %(prepared in NOON, squeezed or  twin-Fock states)  
and a finite amount of samples,  can overcome classical strategies using coherent states of (potentially) arbitrary intensity and an arbitrarily large number of samples. We quantify the quantum advantage as a function of $N_{\rm abs}$ and $T$, and also discuss robustness to imperfections (non-ideal detectors, finite preparation and measurement times). 

In a second part of the article, 
we develop a proposal for realising  such enhancements in state-of-the-art technologies using a cavity or circuit QED set-up~\cite{Haroche2006,Blais2020}.  The key idea is to couple each cavity to an atomic ensemble, and engineer the  dynamics to  both prepare and measure  Fock states in the cavity. The collective coupling between the atoms and the cavity leads to the well-known phenomenon of Dicke superradiance~\cite{dicke54}, which has received renewed attention in the last years  due to the experimental progress in cavity QED  \cite{Haas2014,Norcia2016,Hosseini2017,Kim2018,Kockum2019a} and waveguide QED set-ups \cite{mlynek14,goban15,Solano2017}.  %\textcolor{blue}{Daniel, does this make sense? other relevant citations? I think it does. I added the citation to Solano0217 and one to a review (Kockum et al 2019)}. 
%Building on theoretical ideas developed in the context of waveguide QED~\cite{Romero2009,Peropadre2011,gonzalez-tudela17,Perarnau-Llobet2019,PerarnauLlobet2020,malz2019number},
On the theoretical side, superradiance has  been  exploited for several applications of quantum photonic technologies \cite{demkowicz-dobrzanski15,Dowling2015,polino2020photonic,OBrien2009}, including
%in the context of waveguide QED
  preparation of photonic states \cite{gonzalez-tudela17,Paulisch2018,Uria2020,groiseau2020deterministic},  quantum number-resolved measurements~\cite{Romero2009,Peropadre2011,malz2019number}, and quantum metrology~\cite{Paulisch2019,PerarnauLlobet2020}.  Here, we explore similar applications in the context of cavity or circuit QED, and 
we argue that the collective coupling can be exploited to increase the efficiency and speed of both the measurement and preparation of Fock states.

\section{Framework and physical setup}

%\emph{Framework and physical setup}.  
We model the ``delicate sample'' as a two-mode photonic system, $H=H_0+g H_{\rm int}$ with $H_0=\omega(a^{\dagger}a + b^\dagger b)$
and $H_{\rm int}=  (a^{\dagger}b+ab^{\dagger})/2$.  We wish to estimate the interaction strength $g$ between modes ($g$ has units of frequency and we set $\hbar=1$). %is some unknown Hamiltonian parameter that we wish to estimate.
The inference process  is mediated by photons, which
%are measured after some interaction time. These photons
can also be absorbed by the sample leading to its degradation. The latter process is modelled as a  Markovian dissipative process, so that the  time evolution  of the  photonic state $\rho$ reads:
\begin{align}
\label{evrho}
\dot{\rho}= -i[H_0+g H_{\rm int},\rho]+ \gamma (\mathcal{D}_a(\rho)+\mathcal{D}_{b}(\rho))
\end{align}
with
$\mathcal{D}_x(A)= xAx^{\dagger} -\frac{1}{2}\{x^{\dagger} x ,A \}$. With time, information on $g$ is encoded in $\rho$, but also photons are lost (absorbed).
These two dynamical processes decouple%--i.e., the unitary and dissipative parts of the dynamics commute--
, so that   the evolution can be described as two independent channels:
$\rho(t)= \mathcal{C}_{ g,t}^{\rm uni} \circ \mathcal{C}_{\gamma,t}^{\rm diss}[\rho_0]$,
with $C_{\gamma,t}^{\rm diss}[\rho]= e^{-\gamma t (\mathcal{D}_a+\mathcal{D}_{b})} [\rho]$ and $C_{g,t}^{\rm uni} [\rho]= e^{-i  (H_0+g H_{\rm int}) t} \rho  e^{i   (H_0+g H_{\rm int}) t} $. %Since $[H_0,H_{\rm int}]=0$, $H_0$ will play no role in the estimation of $g$, and will be omitted  in what follows.

If the initial state $\rho_0$ contains $N$ photons, the (average) number of absorbed photons is given by
\begin{align}
  N_{\rm abs}(t) =\tr[\rho(t)\hat N]= N (1-e^{-\gamma t}),
    \label{Nabs}
\end{align}
where $\hat N=a^{\dagger}a+b^{\dagger}b$ counts the total number of photons.
We shall assume that the delicate sample is destroyed  when $N_{\rm abs}(t)=N_{\rm abs}$, which sets a maximum available time for information acquisition. This constrains the possible estimation schemes and, to some extent,  $N_{\rm abs}$ can be seen as a  resource: a larger $N_{\rm abs}$ enables us to obtain more information (and/or in a faster manner) from a single sample.

In order to estimate $g$, we consider the following scenario: We have at our disposal a large number of delicate samples that we can measure sequentially for a total time $T$. The same estimation scheme, or test, is applied to each sample. In each test, a photonic state $\rho_0$ is prepared, it evolves for some time $t$ according to \eqref{evrho} reaching $\rho_{g}$, and is measured by an observable $\Pi$, which yields the outcome $s_j$ with probability $p_j$.
The whole experiment consists of $\nu$ tests, with each test taking a time
$t=T/\nu$, which must satisfy $N_{\rm abs}(t)\leq N_{\rm abs}$. %Note that no constraints are put  on the photon number  $N$ of $\rho_0$  or on  $\nu$; only $N_{\rm abs}$ and $T$ are limited.
%When $\nu \gg 1$ (the regime of interest in this paper)
Then, the uncertainty $\Delta
^2g
$ of   an unbiased estimator  of $g$  satisfies~\cite{helstrom76,holevo82,Braunstein1994}:
\begin{align}
   \Delta^2 g  \geq \frac{1}{\nu \mathcal{F}_{\rm C}} \geq \frac{1}{\nu \mathcal{F}_{\rm Q}}.
   \label{uncertainty}
\end{align}
Here, $\mathcal{F}_{\rm C}= \sum_{j} p_j^{-1} \left( \partial p_j/\partial g \right)^2$ is the Classical Fisher Information (CFI)
%\begin{align}
%$\mathcal{C}= \sum_{j} p_j^{-1} \left( \partial p_j/\partial g \right)^2$,
%\label{cfi}
%\end{align}
and $\mathcal{F}_{\rm Q}$  is the quantum Fisher Information (QFI),
%, which can be formally defined as: $
%    \mathcal{Q}(\rho_{g})=-2 \lim_{x\rightarrow 0} \partial \mathbb{F}(\rho_{g},\rho_{g+x})/\partial x^2  $,
%where $\mathbb{F}(\rho,\sigma)=(\Tr(\sqrt{\sqrt{\rho}\sigma \sqrt{\rho}}))^2$ is  the Uhlmann fidelity between $\rho$  and $\sigma$. The QFI
which will be defined later for each case of interest, and characterizes the potential of $\rho_{g}$ for estimating $g$ with an optimal measurement~\cite{Braunstein1994}.
Our goal is to minimize \eqref{uncertainty} for both coherent and quantum (e.g. Fock, NOON, or squeezed) states. In both cases, we set the two constraints:
\begin{itemize}
    \item the total time $T$ of the experiment.
    \item the maximal number of photons $N_{\rm abs}$ that a single sample can absorb. 
\end{itemize}
For coherent states, we shall put no more constraints, so that the photon number $N$ and the number of samples $\nu$ can be made arbitrarily large, which contrasts with previous works, see e.g.~\cite{Shaji2007}. On the other hand, for quantum states, in Sec.~\ref{Sec:QuantAdvRealScen} we will  also constrain $N$ and the measurement efficiency, and consider a finite rate of preparation and measurement of quantum states. 

%setting as constraints $N_{\rm abs}$ and $T$. Hence, we stress that no constraint is put on $N$, in contrast to previous works on related set-ups, see e.g.~\cite{Shaji2007}. 

%and to find the corresponding optimal strategies,  As discussed in the introduction, and in contrast to previous works  %It is worth pointing  out that the problem at hand can be formally mapped to  standard Mach-Zender interferometry with symmetric (time-dependent) photon loss with phase $\varphi=gt$, %due to the commutativity between the unitary and dissipative contributions in \eqref{evrho},

It is  worth pointing out that,  defining $a_{\pm}\equiv (a\pm b^\dagger)/\sqrt{2}$, we can rewrite $H$ as
%\begin{align}
    $H= (\omega +g) a^{\dagger}_+ a_+ + (\omega -g) a^{\dagger}_- a_-$,
%\end{align}
so an alternative interpretation of this framework
is that we are estimating the frequency difference between two modes. In terms of the original modes $a$ and $b$, both pictures are physically equivalent if a balanced beamsplitter is applied to the input (and output) light. With this transformation, it is also clear that the problem at hand can be formally mapped to  standard Mach-Zehnder interferometry with $\varphi=gt$, where photon absorption corresponds to a  symmetric (time-dependent) photon loss. % with phase $\varphi=gt$, which 
This map naturally enables us to use techniques and results developed in this context~\cite{Giovannetti2011,Tth2014,demkowicz-dobrzanski15}.
%Yet, as motivated above, here the natural resources are both $N_{\rm abs}$ and $T$ (instead of the  number of photons), which will lead to  new insights on non-invasive quantum metrology.

%\section{Optimal classical strategy}
%\dm{[I noticed that you never consider unequal coherent states. We should maybe comment and point out why they're definitely not ideal? We can also just say that the argument is easily extended to the unequal case and that it's worse.]}

%\emph{Optimal classical strategy}. 

%\section{Optimal classical strategy}

\section{Optimal classical strategy}
Let us first analyze the optimal strategy using coherent states. We consider as an input state $\rho_0 = \ket{g_0}\bra{g_0}$ with $\ket{g_0}= \ket{\cos(\theta)\sqrt{ N},\sin(\theta)\sqrt{N}}$. The dissipative channel acts upon the initial state as
$
\ket{\phi}=\mathcal{C}_{\gamma,t}^{\rm diss}[g_0]= \ket{\cos(\theta)\sqrt{e^{-\gamma t}  N},\sin(\theta)\sqrt{e^{-\gamma t}N} }
$. Since the state remains pure, we can  quantify the QFI of $\rho_g = \mathcal{C}^{\rm uni}_{g,t}[\ket{\phi} \bra{\phi}]$ as $\mathcal{F}_{\rm Q}=4 t^2 (\bra{\phi} H_{\rm int}^2 \ket{\phi} -(\bra{\phi} H_{\rm int} \ket{\phi})^2)$~\cite{Braunstein1994}, obtaining:
\begin{align}
\mathcal{F}_{\rm Q}^{\rm coh}=  t^2 N e^{-\gamma t},
\end{align}
which  is  the shot-noise-limit. Using $N_{\rm abs}= N (1-e^{-\gamma t})$ and $t=T/\nu$,  the accumulated QFI in a time $T$ reads:
%\begin{align}
%\xi_{\rm coh} \equiv
$\mathcal{\nu}\mathcal{F}_{\rm Q}^{\rm coh}= T^2 N_{\rm abs} e^{-\gamma T/\nu}/(1-e^{-\gamma T/\nu})\nu.$
%\label{derphasecl}
%\end{align}
%which provides an upper bound on $(\Deltag)^{-2}$.
This expression increases monotonically with $\nu$, so its maximum $\left(\Delta g\right)_{\rm coh}^{-2}\equiv \max_\nu (\mathcal{\nu}\mathcal{F}_{\rm Q}^{\rm coh})$ is obtained for $\nu \rightarrow \infty$:
\begin{align}
    \left(\Delta g\right)_{\rm coh}^{-2}  =  \frac{T}{\gamma} N_{\rm abs}.
    \label{classicalBound}
\end{align}
 Strictly speaking, this bound can be saturated by testing $\nu \rightarrow \infty$ samples, each for an infinitesimally small time $t\rightarrow 0$ using  a number $N\rightarrow \infty$ of photons, in such a way that both $N_{\rm abs}=\gamma t N$ and  $T=\nu t$ remain constant. We  call this theoretical limit ($N, \nu \rightarrow \infty$ and $t \rightarrow 0$, with both $N_{\rm abs}=\gamma t N$ and $T=\nu t$ fixed) the \emph{Poisson limit}, which will play a relevant role in this work.
The importance of the classical bound \eqref{classicalBound} is that it remains finite --being limited by both $T$ and $N_{\rm abs}$-- despite using an unbounded photon number (at each test) and an arbitrarily large number of tests. We also note that optimality of fast measurements  has also been found in  frequency measurements of atomic ensembles~\cite{Huelga1997,Matsuzaki2011,Chin2012,Chaves2013,Brask2015,Smirne2016}.

%We also note that optimality of fast measurements has been fouin frequency estimation of atomic ensembles,  one commonly finds that fast measurement schemes (where the interrogation time of the atoms decays to zero with $N$) are optimal~\cite{Huelga1997,Matsuzaki2011,Chin2012,Chaves2013,Brask2015,Smirne2016}.

%\section{Quantum strategies: Twin Fock states}

\section{Quantum strategies} %: Twin Fock states}. 

We now move to quantum resources. We begin by deriving an upper bound on $\mathcal{F}_Q$ given arbitrary quantum states. We use the results of Refs~\cite{Fujiwara2008,Knysh2011,escher2011general,DemkowiczDobrzaski2012,Koodyski2013}, where the upper bound $\mathcal{F}_{\rm Q}\leq N(1-\mu)/\mu$ is obtained for  an optical interferometer with symmetric photon loss in both arms, as quantified by $\mu$ (the probability of photon loss in each arm of the interferometer). In the context studied here this translates as: 
\begin{align}
    \mathcal{F}_{\rm Q}\leq t^2 N(1-\mu)/\mu
\end{align}
with  $\mu=1-e^{-\gamma t}$ and $N_{\rm abs}=\mu N$. In analogy with the classical derivation, an upper bound on  the accumulated QFI for a given $N_{\rm abs}$ and  $T$ can now be easily obtained using $T=t\nu$:  
\begin{align}
\nu \mathcal{F}_{\rm Q}\leq \frac{T^2 N_{\rm abs}  e^{-\gamma T/\nu}}{\nu (1-e^{-\gamma T/\nu})^2}.
\end{align}  
This expression also increases monotonically with $\nu$, and for  $\nu \gg \gamma T$ it simplifies to: \begin{align}
\nu \mathcal{F}_{\rm Q} \leq \nu \frac{N_{\rm abs}}{\gamma^2 }.
\label{upperbound}
\end{align}  
Approximating $N_{\rm abs}=N(1-e^{-\gamma T/\nu})$ as $N_{\rm abs} \approx \gamma T N /\nu $ for $\nu \gg \gamma T$, we can express \eqref{upperbound} in terms of $N$ as:  
 \begin{align}
\nu \mathcal{F}_{\rm Q} \leq \frac{T}{\gamma}N.
\label{QuantumUpperBound}
\end{align}  
Comparing \eqref{QuantumUpperBound} with  \eqref{classicalBound}, we note that in principle an arbitrarily large advantage over the classical bound  can  be obtained in the Poisson limit (where $N\rightarrow \infty$ and $\nu \rightarrow \infty$ with $T$ and $N_{\rm abs}$ constant). For finite but large~$N$, the quantum advantage scales as $N/N_{\rm abs}$. To understand intuitively this large advantage, we note that $\mathcal{F}_Q \sim t^2 N^2$ for states featuring Heisenberg scaling, which remains constant in the Poisson limit\footnote{In the Poisson limit, Heisenberg scaling may be preserved as $N_{\rm abs}/N \rightarrow 0$ when $N\rightarrow \infty$.   This observation is not in contradiction with the well-known fact that  quantum states lose Heisenberg scaling as $N\rightarrow \infty$ for $\mu>0$ constant~\cite{Fujiwara2008,Knysh2011,escher2011general,DemkowiczDobrzaski2012,Koodyski2013}, as  in this case $N_{\rm abs}/N$ remains finite as $N\rightarrow \infty$. These heuristic considerations will be made more precise for specific states in what follows. }.   
%the Poisson limit both $N\rightarrow \infty $  and $\mu \rightarrow 0$ simultaneously. Hence,   Heisenberg limit may be preserved even for arbitrarily large $N$ as $N_{\rm abs}/N \rightarrow 0$, at least in principle. 

While the previous observations illustrate that large quantum advantages are possible in the Poisson limit, two very relevant questions appear: (i) how close specific quantum states can get to the upper bound Eq~\eqref{upperbound}, and, more importantly: (ii) whether the classical bound~\eqref{classicalBound} can be overcome  in realistic scenarios where $N$ is small   and imperfections are present (non-ideal measurements, finite time of preparation of quantum states).   In what follows, we address (i) by computing  $\mathcal{F}_Q$ for  relevant quantum states in the Poisson limit. Question (ii) will be addressed in  Sec. \ref{Sec:QuantAdvRealScen}.

%Taking the Poisson limit  ($t\rightarrow 0$, $N\rightarrow \infty$ and $N_{\rm abs}$ constant) yields
%\begin{align}
%\mathcal{F}_{\rm Q} \leq \frac{N_{\rm abs}}{\gamma^2}.
%\label{upperbound}
%\end{align}
%Hence, in principle a finite amount of information can be obtained from a single measurement  as $t\rightarrow 0$, which implies  $(\Delta g)^2 = (\nu \mathcal{F}_{\rm Q})^{-1} \rightarrow 0$ for any $T, N_{\rm abs}>0$, i.e.,  an arbitrarily large advantage over the classical bound \eqref{classicalBound}. Intuitively, this  can be understood by noting that $\mathcal{F}_Q \sim t^2 N^2$ for states featuring Heisenberg scaling, which remains constant in the Poisson limit. %In turn, this implies that it is theoretically possible that $(\Delta g)^2 = (\nu \mathcal{F}_{\rm Q})^{-1} \rightarrow 0$ given any $T, N_{\rm abs}>0$,
%, so that $g$ can be estimated with arbitrary precision.  which represents an arbitrarily large advantage over the classical bound \eqref{classicalBound}. 

\begin{figure}[t]
  \centering
  \includegraphics[width=1\columnwidth]{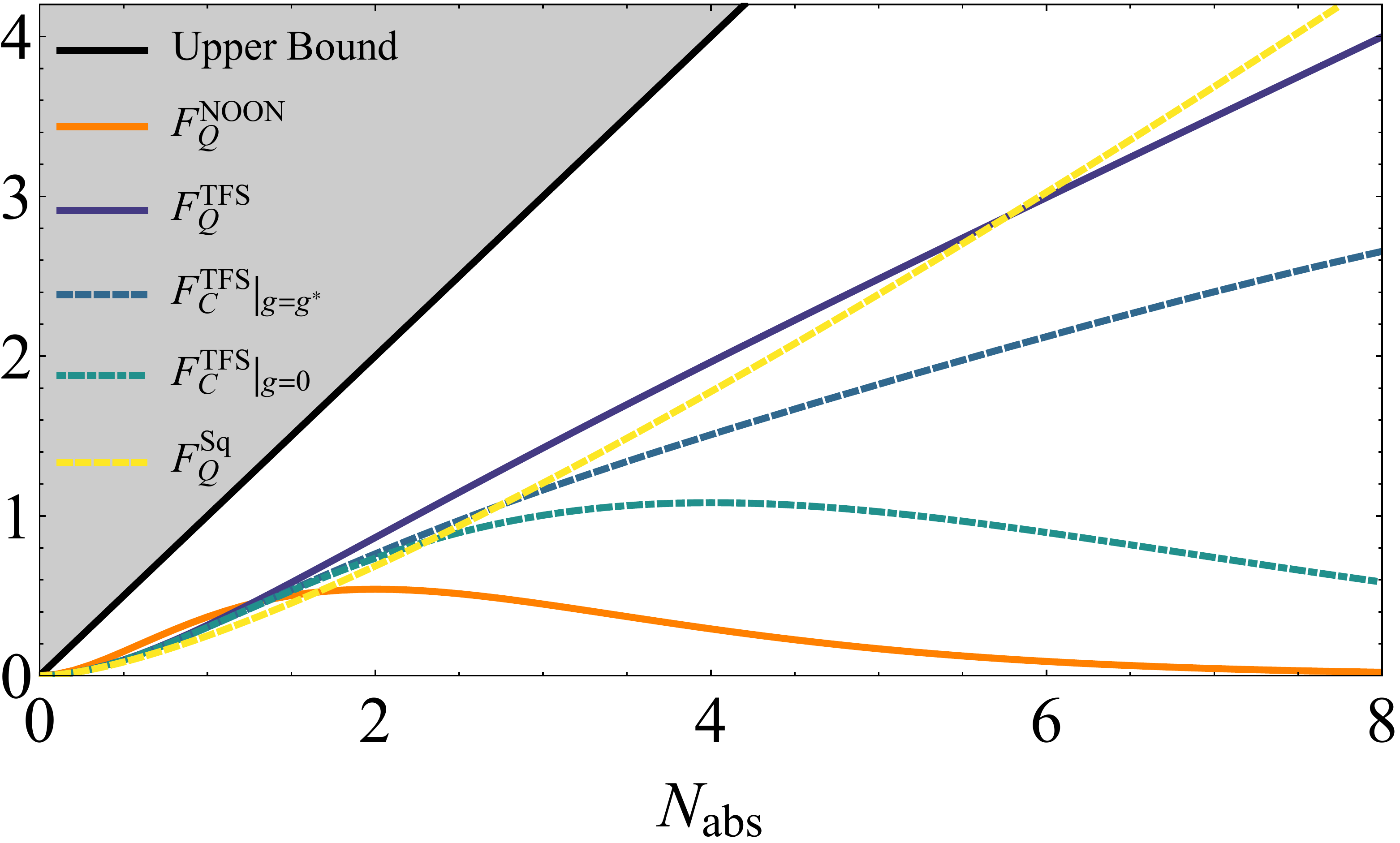}
	    \caption{$\mathcal{F}_{Q}$ for TFS and NOON states (Eqs. \eqref{QpoisTFS} and \eqref{FQnoon}), and $\mathcal{F}_{C}$ for TFS at $g=0$ and $g=g*$ using photon number resolved measurements,  $\mathcal{F}_{C}$ for optimised two mode squeezed states with number-resolved measurements (details in Sec.~\ref{secap:squeezedstates} of the Appendix). All quantities are computed in the Poisson limit ($N\rightarrow \infty$, $t\rightarrow 0$), so that only states with Heisenberg scaling can yield non-zero information in this plot. Parameters: $\gamma=1$. }
	    %\eqref{QFIsqueezed} with optimised $\beta_r, \beta_s$.
		\label{fig:fig1}
\end{figure}

We  start by  the paradigmatic case of twin-Fock states (TFS)~\cite{holland93,campos03,Datta2011}. %, which are a well-known resource in quantum optical estimation~\cite{holland93,campos03}. 
%As in the previous case, we wish to maximize $(\Delta g)^{-2}_{\rm TFS} \equiv \max_{\nu}(\nu \mathcal{F}_{\rm Q}) $ given some $T$ and $N_{\rm abs}$. 
 In this case, we have  $\rho_0=\ket{\phi_0}\bra{\phi_0}$ with $\ket{\phi_0}=\ket{n,n}$ and $N=2n$. This state  evolves for a time $t$ according to \eqref{evrho}, which we recall can be split in two independent channels: a unitary one which encodes $g$, and a dissipative one where photons are lost with probability $\mu=1-e^{-\gamma t}$.  The state after a time $t$ can then be described as:  
\begin{align}
\rho_g = \sum_{k,j} p_k p_j \ket{\phi_{k,j}(t)} \bra{\phi_{k,j}(t)} %\ket{n-k,n-j}\bra{n-k,n-j}
\label{fockStateII}
\end{align}
where $\ket{\phi_{k,j}(t)}=e^{- itg H_{\rm int}} \ket{n-k,n-j}$, and $p_k$ is the probability of absorving $k$ photons after a time $t$, which is given by: 
$p_k = \binom{n}{k}\mu^k (1-\mu)^{n-k} $. To compute the QFI of this state, it is convenient to note that  given $\rho_{g} = \sum_s q_s \ket{\phi_s(t)} \bra{\phi_s(t)}$, with $\ket{\phi_s(t)} =  e^{- itg H_{\rm int}} \ket{\phi_s}$ and $\langle \phi_s| \phi_l \rangle =\delta_{sl}$; the QFI reads
%\begin{align}
$\mathcal{F}_{\rm Q}= 4t^2 [ \sum_i q_i (\Delta H_{i})^2- \sum_{i\neq j} 2q_i q_j  |H_{ij}|^2 (q_i+q_j)^{-1}],$
%\label{FQexact}
%\end{align}
with $(\Delta H_{i})^2 = \bra{\phi_i} H^2_{\rm int} \ket{\phi_i}-(\bra{\phi_i} H_{\rm int} \ket{\phi_i})^2$ and
  $|H_{ij}|^2 = |\bra{\phi_i} H_{\rm int} \ket{\phi_j}|^2$~\cite{Braunstein1994,PARIS2009}. For TFS, we hence obtain,
\begin{align}
\hspace{-1mm}\mathcal{F}_{\rm Q}^{\rm TFS}\hspace{-0.5mm}=\hspace{-0.5mm} 4t^2 \hspace{-1mm}\sum_{ij}^n p_i p_j (n-i) (n-j+1) \hspace{-1mm}\left(\hspace{-0.5mm}\frac{1}{2}- \frac{1}{1+\frac{(n-j+1)(i+1)}{j(n-i)}} \hspace{-0.5mm}\right)\hspace{-1mm}.
%\label{Qexact}
\label{FQexact}
\end{align}
Now, taking the Poisson limit  of \eqref{FQexact} ($t\rightarrow 0 $ with $2 n \gamma t=N_{\rm abs}$), yields
\begin{align}
\mathcal{F}_{\rm Q}^{\rm TFS}= \frac{N_{\rm abs}^2}{\gamma^2}  \sum_{ij}^\infty p_i p_j \left(\frac{1}{2}-\frac{j}{i+1+j} \right) \equiv \frac{\mathcal{G}(N_{\rm abs})}{\gamma^2},
\label{QpoisTFS}
\end{align}
with
\begin{align}
p_k=\frac{(N_{\rm abs}/2)^k e^{-N_{\rm abs}/2}}{k!}.
\label{p_kpois}
\end{align}
For $N_{\mathrm{abs}}\gg1$, the Poisson distribution \eqref{p_kpois} is strongly peaked around $k=N_{\mathrm{abs}}/2$, such that the sum in~\eqref{QpoisTFS} is dominated by $i=j=N_{\mathrm{abs}}/2$. In this regime, $\mathcal{G}(N_{\mathrm{abs}})\approx N_{\mathrm{abs}}/2$, which is  half of the upper bound~\eqref{upperbound}.
In contrast, as $N_{\mathrm{abs}}\to 0$, $\mathcal{G} (N_{\mathrm{abs}})\approx N_{\mathrm{abs}}^2/2$. In general,  a global measurement on both modes is needed to saturate   the QFI \eqref{QpoisTFS}~\cite{PARIS2009}. The $\mathcal{F}_C$ has been considered for more realistic photon number-resolved measurements (NRM)~\cite{Taesoo1998,Datta2011,Pezze2013,Zhong2017}, and we study it in the Poisson limit in Sec.~\ref{secapp:numberres} of the Appendix.  For $g\approx 0$, we find $\mathcal{F}_C^{\rm NRM} |_{g=0} = N_{\rm abs}^2 e^{-N_{\rm abs}/2}/2\gamma^2$. The exponential decay with $N_{\rm abs}$ can be avoided by suitably optimising over the  interaction strength $g$,  leading to a quasi-linear growth of  $\mathcal{F}_C|_{g=g*}$ with~$N_{\rm abs}$ (see Fig.~\ref{fig:fig1}).  %Results are shown  Fig. XX, both at $g\approx 0$ and at the optimal $g^*$ where $\mathcal{F}_C$ is maximised (see also Appendix).

We now consider NOON states. In particular, we take as an initial state  $\rho_0 = \ketbra{\rm NOON}{\rm NOON}$ with $\ket{\rm NOON}= (\ket{0,N} + \ket{N,0})/\sqrt{2}$. In this case, $\mathcal{F}_{Q}^{\rm NOON}$ is given by $N^2 t^2$ times the probability of not losing any photon which,  in the Poisson limit, is given by $p_0^2$ in \eqref{p_kpois}. Hence, %in this limit, %. In that case, $\mathcal{F}_Q^{\rm NOON}=t^2 N^2$, and taking the Poisson limit, we obtain:
\begin{align}
    \mathcal{F}_Q^{\rm NOON}= \frac{N^2_{\rm abs}}{\gamma^2} e^{-N_{\rm abs}}. 
    \label{FQnoon}
\end{align}
For $N_{\rm abs}\ll 1$, NOON perform twice as good as TFS, but this advantage is exponentially suppressed for larger~$N_{\rm abs}$. In fact, NOON states are optimal for~$N_{\rm abs}\ll 1$, which means that the bound~\eqref{upperbound} can not be saturated in this regime. 
%We also note that \eqref{FQnoon} can be saturated by a NRM measurement~\cite{}.  
In the regime $N_{\rm abs}\gg 1$, the upper bound~\eqref{upperbound} can be saturated using two mode squeezed states and number-resolved measurements~\cite{Yurke1986,Olivares2007}, which we show in Sec.~\ref{secap:squeezedstates} of the Appendix. The performance of TFS, squeezed and NOON states in the Poisson limit is summarised in Fig.~\ref{fig:fig1}. In the regime~$N_{\rm abs} = \mathcal{O}(1)$, the performance can be improved by considering more general states, such as the optimal  states discussed in~\cite{Fuentes2016,Huver2008,Dorner2009,Demkowicz2009,Knysh2011,Cable2010,Calsamiglia2016,Matsubara2019}. %, can improve the performance. 
%better for $N_{\rm abs} = \mathcal{O}(1)$. 
%It remains an interesting open question to explore whether  \eqref{upperbound} can be saturated for $N_{\rm abs} = \mathcal{O}(1)$ for more general states, such as~\cite{Fuentes2016,Matsubara2019}, thus complementing previous  results for $N_{\rm abs} \propto N$ ~\cite{Huver2008,Dorner2009,Demkowicz2009,Knysh2011,Cable2010,Calsamiglia2016}.

Summarising, quantum photonic states (TFS, squeezed or NOON states) can provide an arbitrarily large advantage over the classical bound \eqref{classicalBound} in the Poisson limit, which however requires $N\rightarrow \infty$ and $t \rightarrow 0$ (i.e. arbitrarily fast measurements). In the next section, we explore quantum advantages in realistic scenarios with small $N$ and finite $t$, and also consider the presence of non-idealities in the preparation and measurement steps.

\begin{figure}[t]
  \centering
  \includegraphics[width=1\linewidth]{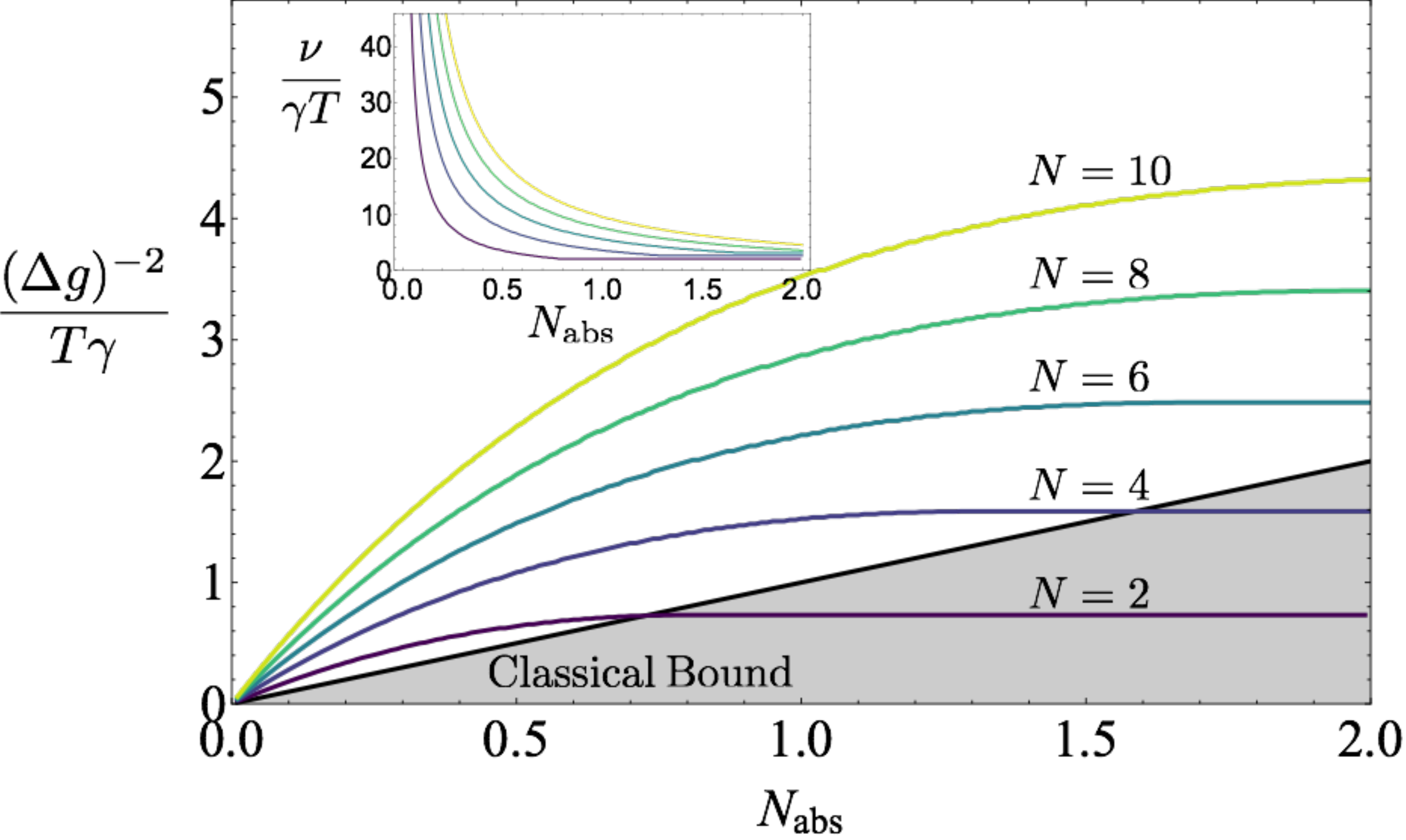}
  \caption{$(\Delta g)^{-2} \equiv \nu \mathcal{F}_{\rm Q}$ per unit of time as a function of $N_{\rm abs}$ for twin Fock states (TFS) of low $N$, together with the classical bound \eqref{classicalBound} for coherent states corresponding to the limit $N\rightarrow \infty$, $\nu \rightarrow \infty$. The numerical results for TFS are obtained by computing $\max_\nu (\nu \mathcal{F}_{\rm Q})$ given the constraints $t\nu =T$ and  $N_{\rm abs}(t)\leq N_{\rm abs}$ with $N_{\rm abs}(t)$  in \eqref{Nabs}. The optimal $\nu$ for TFS, which satisfies $\nu  /T\gamma \approx N/N_{\rm abs}$, is shown in the inset. %Overall, we find good qualitative agreement between the numerical results obtained for low $N$ and  the analytical considerations  where  $N\gg 1$ is used. 
   These results confirm quantum advantages using low-photon entangled states over coherent states of arbitrary intensity. Parameters:  $T=10$ and $\gamma=1$, but essentially the same results are obtained for larger $T$. Note that the bound \eqref{uncertainty}  becomes more accurate as $T$, and hence $\nu$, increases. 
  }
  \label{fig:qficomp}
\end{figure}

\section{Quantum advantage in realistic scenarios}
\label{Sec:QuantAdvRealScen}

We first explore the presence of finite $N$. First, when $N$ is finite with $N\gg 1$, we can use $N_{\rm abs} \approx N \gamma t$ and $\nu =T/t$ to approximate $(\Delta g)^{-2} = \nu \mathcal{F}$ as
\begin{align}
  (\Delta g)^{-2} \approx (\Delta g)^{-2}_{\rm coh}  \frac{N}{N_{\rm abs}^2} \mathcal{F},  
  \label{QFIquant}
\end{align}
where  $\mathcal{F}$  is any of the CFI or QFI shown in Fig.~\ref{fig:fig1}. This expression can be compared with \eqref{classicalBound}, where we note  a linear increase of the quantum advantage  with $N$. Interestingly, quantum advantages are also observed when $N$ is small, so that Eq. \eqref{QFIquant} is not necessarily valid. This is shown in Fig. \ref{fig:qficomp}, where we compute the QFI of TFS exactly using \eqref{FQexact} for low $N<10$. Notably, we see that even quantum states with $N=2$ can outperform arbitrary classical strategies for sufficiently small $N_{\rm abs}$. Importantly, the number of repetitions of the experiment $\nu/\gamma T \approx N/N_{\rm abs}$ is also small in such scenarios (see inset of Fig. \ref{fig:qficomp} ). While these considerations are promising, the results of Fig. \ref{fig:qficomp} are obtained assuming   
 perfect preparation and measurement devices, which motivates the  analysis of noise  robustness  in what follows. 

Let us  consider the  main practical limitations:  measurement devices with efficiency $\eta <1$, and an extra time $t_{\rm ext}$ to prepare and measure  quantum states~\cite{Hayes2018}. 
More precisely, we model photon loss in the measurement process through a beamsplitter interaction with vacuum of each mode $a$, $b$:
\begin{align}
\label{modelnoise}
    &a \rightarrow \sqrt{\eta} a + \sqrt{1-\eta} e_1
    \nonumber\\
    &b \rightarrow \sqrt{\eta} b + \sqrt{1-\eta} e_2
\end{align}
so that there is a probability $1-\eta$ of losing a photon. Since $\eta$ is the same in both modes, this beam splitter interaction commutes with $H_{\rm int}$, so that this operation can be applied either before (imperfect state preparation) or after (imperfect detector)  the acquisition of information of $g$.    In particular, we can generalise $\mu$ as:
    $\mu = 1-e^{-\gamma t}\eta$,
where $\eta$ is independent of the time of the process, and where we recall that $N_{\rm abs}=N(1-e^{-\gamma t})$ (i.e. the photons loss in the vacuum do not damage the sample). Indeed, $\eta$ quantifies the efficiency of the detector. On the other hand, we also assume the existence of an extra time $t_{\rm ext}$ to prepare and measure quantum states of light. The total time of the process is hence generalised to: 
\begin{align}
T=\nu (t+ t_{\rm ext}), 
\end{align}
where  $t$ is the time used in the interference process.

\begin{figure*}
		\centering
		\includegraphics[width=0.47\linewidth]{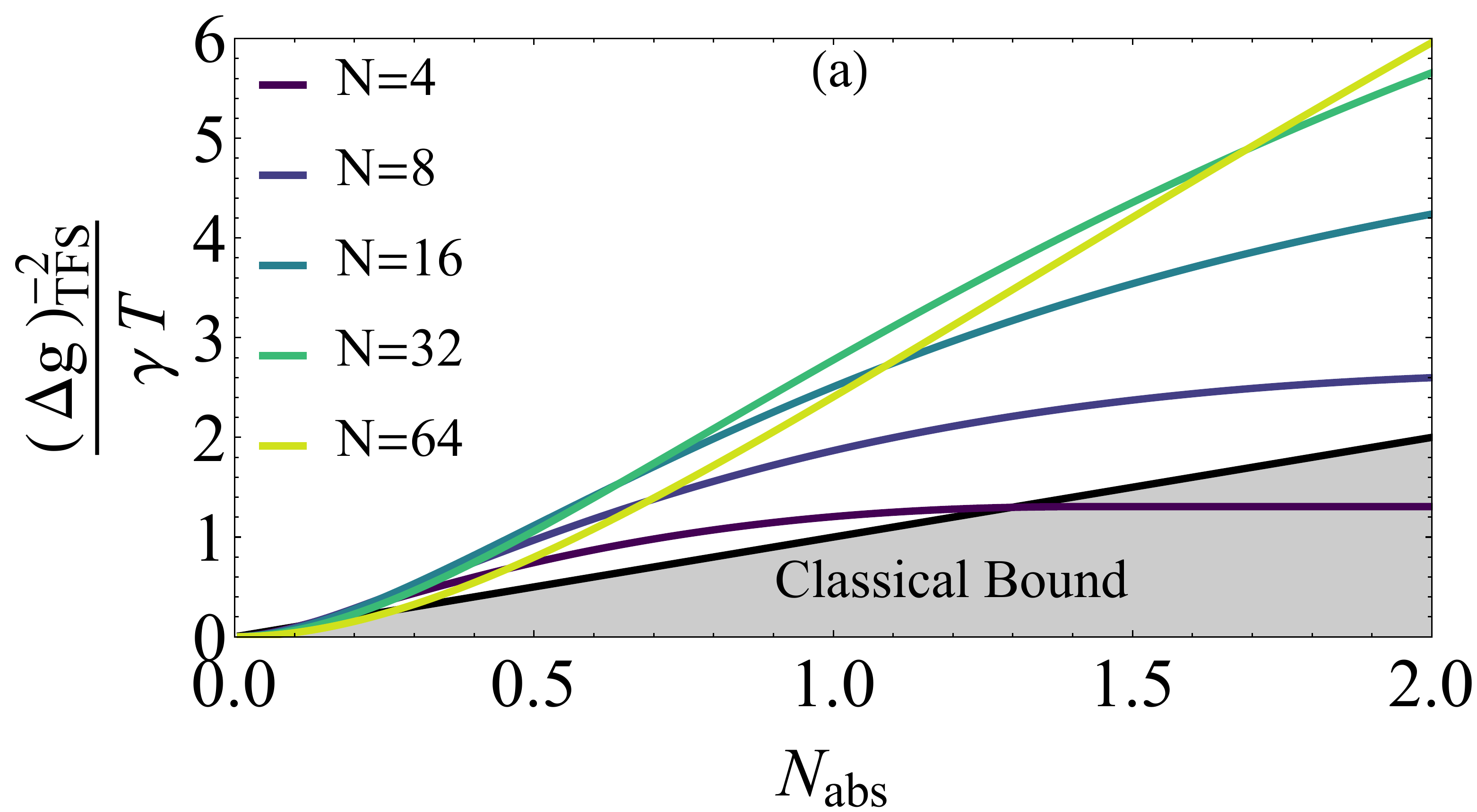} \hspace{5mm}
	    \includegraphics[width=0.47\linewidth]{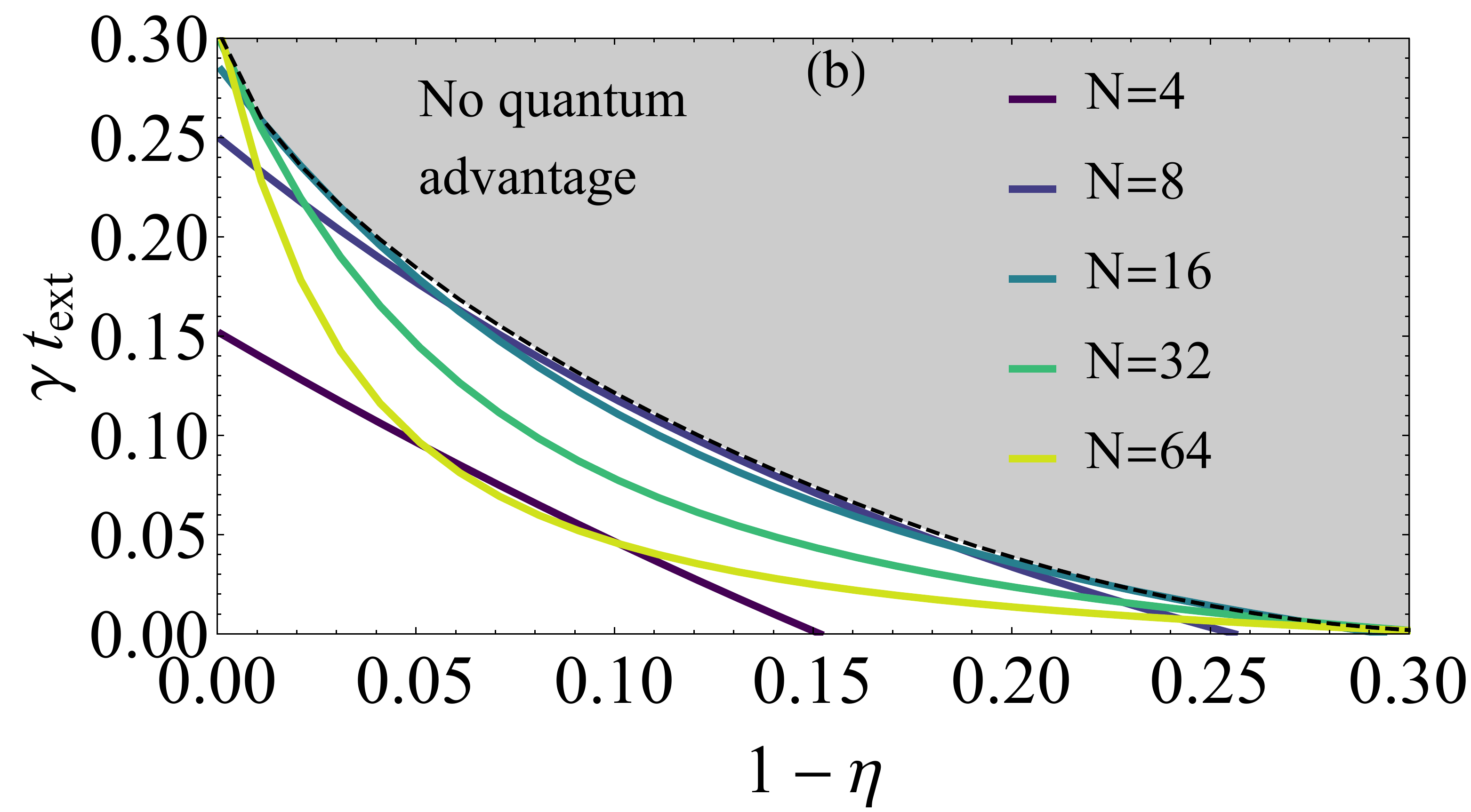} %\hspace{0.1mm}
	    \caption{(a) $(\Delta g)^{-2}/\gamma T$  as a function of $N_{\rm abs}$ for TFS of different $N$ with $\eta=0.96$ and $\gamma t_{\rm ext}=0.04$, together with the classical bound \eqref{classicalBound}. % for coherent states corresponding to the limit $N\rightarrow \infty$, $\nu \rightarrow \infty$. 
	    The numerical results for TFS are obtained by computing  $\max_\nu (\nu \mathcal{F}_{\rm Q}^{\rm TFS})$ (using the exact expression \eqref{FQexact}) given the constraints $t\nu =T$ and  $N_{\rm abs}(t)\leq N_{\rm abs}$ with $N_{\rm abs}(t)$  in \eqref{Nabs}. The optimal $\nu$ is  well approximated by $\nu /T\gamma \approx N/N_{\rm abs}$. (b) In grey, region where quantum advantages  are  not possible with TFS of any $N$ and with $N_{\rm abs}=1$.   The curves are implicitly defined by  $(\Delta g)^{2}_{\rm TFS}/(\Delta g)^{2}_{\rm coh} =1 $ given TFS of different photon number. That is, they are defined by the values of $\{t_{\rm ext}, 1-\eta \}$ for which $(\Delta g)^{2}_{\rm TFS}/(\Delta g)^{2}_{\rm coh} =1 $, and hence delimit regions where quantum advantages are possible given TFS of different $N$ (on the left of the curves). In the grey region, no quantum advantages are possible, regardless of the value of $N$. }
		\label{fig:fig2}
	\end{figure*}

Given this model, we first illustrate in Fig.~\ref{fig:fig2} (a)  that  TFS with low $N$ can  overcome the bound \eqref{classicalBound} when $N_{\rm abs}$ is small, even in the presence of imperfections (in the figure, we take $\eta=0.96$ and $\gamma t_{\rm ext}=0.04$). These results can be compared with the ideal scenario ($\eta=1$ and $\gamma t_{\rm ext}=0$) shown in Fig. \ref{fig:qficomp}, which illustrates that imperfections become most relevant as $N$ increases. Similar results are found for NRM (see Sec.~\ref{secapp:imperfect} of the Appendix). 
Importantly, in both cases the corresponding protocol requires measuring a finite number of samples, $\nu /T\gamma \approx N/N_{\rm abs}$, making it feasible in practice.

Second, the robustness of such quantum advantages to finite detector efficiency ($\eta$)  and extra times ($t_{\rm ext}$) is analysed in Fig.~\ref{fig:fig2} (b) for TFS and $N_{\rm abs}=1$. In particular, we plot the values of $\{\gamma t_{\rm ext}, 1-\eta \}$ for which $(\Delta g)^{2}_{\rm TFS}/(\Delta g)^{2}_{\rm coh} =1 $ for different $N$. That is, on the left of the curves of Fig.~\ref{fig:fig2} (b),  quantum advantages are possible (i.e., $(\Delta g)^{2}_{\rm TFS}/(\Delta g)^{2}_{\rm coh} \geq 1 $), and on the right, they are not. 
In the grey region, no quantum advantages are possible regardless of $N$. 
%Roughly speaking, one requires at least $\eta >0.9$ and $\gamma t_{\rm ext}<0.1$, which also depends on $N$. 
Whereas for $t_{\rm ext}=0$ or $\eta=1$ increasing $N$ is generally favourable, we find that TFS of low $N$ ($N\approx 8-20$) are most robust to noise  when both $t_{\rm ext}>0$ and $1-\eta$ are non-negligible.
In particular, for TFS of $N=8$, the quantum advantage prevails for $\eta >0.9$ and $\gamma t_{\rm ext}<0.12$. 

In Sec.~\ref{secapp:imperfect} of the Appendix, we provide an extended discussion of noise robustness, including upper bounds on general states.
We find that, while the presence of $\eta$ and $t_{\rm ext}$ severely restricts the maximal possible quantum advantage, the main practical conclusion of our paper remains: In frequency measurements of delicate samples, few-photon quantum states (TFS, NOON, or squeezed) can overcome classical strategies using arbitrarily intense coherent states and an arbitrarily large number of samples when $N_{\rm abs}$ is low. %(i.e. the bound Eq.~\eqref{classicalBound}). 

\section{Implementation in Cavity QED}
Although  quantum advantages can be obtained already for low $N$ and can  tolerate imperfections (see Fig.~\ref{fig:fig2} and Sec.~\ref{secapp:imperfect} of the Appendix), it is crucial to devise implementations where quantum states are prepared and measured fast, and with high efficiency. Recent examples include the preparation of Fock states in the motional states of trapped ions \cite{Wolf2019}. 
%Perhaps the greatest challenge in observing quantum-enhanced metrology is producing and measuring quantum states of light, such as TFS, GHZ and NOON states.
Here, we take an alternative route and  outline a protocol based on cavity QED~\cite{Haroche2006} or circuit QED~\cite{Blais2020}
capable of generating and observing Fock states of light in cavities, inspired by recent progress on metrology and number-resolved measurements in waveguide QED~\cite{gonzalez-tudela17,paulisch18b,PerarnauLlobet2020,malz2019number}.
The key idea is to engineer an interaction of the cavity with a set of resonant qubits/atoms.  More precisely,   in addition to Eq.~\eqref{evrho}, the two cavities are each coupled to a collection of $N_{\rm at}$ resonant atoms via
\begin{equation}
  H_{\mathrm{Dicke}}(t)=\sum_{\alpha=a,b}
  \sum_i^{N_{\rm at}}\left[\omega\sigma^z_{i,\alpha}
  +J(t)(\sigma^-_{i,\alpha}\alpha^\dagger+\mathrm{H.c.})\right],
  \label{eq:Hatoms}
\end{equation}
such that the total Hamiltonian of the system reads $H(t)=H_0+H_{\mathrm{int}}+H_{\mathrm{Dicke}}$.
In our protocol we require that the collectively enhanced atom--cavity coupling $J(t)$ is tuneable and much faster than the cavity decay rate. Strong coupling, together with control of the atom--cavity interaction is routinely achieved in experiment~\cite{Walther2006,Kockum2019a}, in particular to generate or absorb individual photons~\cite{Hennrich2000}.
Extending this level of control to ensembles of atoms the cavity and combining this with sufficiently high quality readout is challenging, but actively investigated~\cite{Haas2014,Mivehvar2021}.
%collective spin of the atoms $S_\alpha^z=\sum_i\sigma^z_{\alpha,i}$. 
%\textcolor{blue}{
%We consider here the linear regime, in which the number of photons in the cavity is  smaller than the number of atoms $n\ll N_{\rm at}$, 
%, and thus we can approximate $S^-_\alpha\approx \sqrt{N_{\rm at}}c_\alpha$, where $c_\alpha$ is a bosonic annihilation operator, . The nonlinear regime where $N_{\rm at}\approx n$ can be treated with exact diagonalisation of \eqref{eq:Hatoms}.
%}
%The this set-up  can be used to prepare and measure Fock states as we describe below.

Initially, all $2N_{\rm at}$ atoms  %$N_{\rm at}$ ($\alpha=a,b$)
are excited, and each cavity is set to the ground state: $n=0$,  where $n$ is the number photons in each cavity. Through the atom-cavity (AC) Dicke interaction \eqref{eq:Hatoms}, the excitations are transformed into photons in the cavity. In general, we can diagonalize exactly Eq.~\eqref{eq:Hatoms} to find the state evolution (see Appendix \ref{secap:implQED}). Yet, here we provide an analytical solution in the linear regime where $n \ll N_{\rm at}$ along the evolution, which turns out to also be relevant for the implementation. In this case, we can  linearize the spin operator around the state with all atoms excited, obtaining $\sum_i\sigma^-_\alpha\simeq\sqrt{N_{\rm at}}c_\alpha^\dagger$, where $c_\alpha$ is a bosonic annihilation operator. Thus, Eq.~\eqref{eq:Hatoms} transforms into
\begin{equation}
    H_\mathrm{Dicke}\simeq \sum_{\alpha=a,b}\sum_{i=1}^{N_{\rm at}}[ \omega\sigma_{i,\alpha}^z
    + (J(t)\sqrt{N_{\rm at}}c^\dagger_\alpha \alpha^\dagger+\mathrm{H.c.})].
    \label{eq:Hlin}
\end{equation}
The time evolution is therefore a two-mode squeezing operation
%\begin{equation}
    $e^{-iHt}=e^{-\zeta\alpha^\dagger c^\dagger+\zeta^*\alpha c}$,  with $\zeta = i\sqrt{N_{\rm at}}Jt = re^{i\phi}$.
%\end{equation}
Under this evolution,
\begin{equation}
    \begin{aligned}
        &\langle \alpha^\dagger\alpha\rangle = \sinh^2r = \sinh^2(\sqrt{N_{\rm at}}Jt) \\
        &\langle\alpha^\dagger\alpha^\dagger\alpha\alpha\rangle
        =\sinh^2(r)[\cosh(2r)-\sinh^2(r)].
    \end{aligned}
\end{equation}
with $\alpha= a,b$. 
In particular, the optimal time scale to obtain on average $n$ photons in each cavity is 
\begin{equation}
    t_\mathrm{prep} = \frac{\sinh^{-1}(\sqrt{n})}{\sqrt{N_{\rm at}} J}.
    \label{eq:apptopt}
\end{equation}
On the other hand,  the probability of obtaining $k_{\alpha}$ photons in one cavity at time $t$ is given by: $q(k_\alpha)= (\tanh \sqrt{N_{\rm \alpha}} J t)^{2k_{\alpha}}/(\cosh \sqrt{N_{\rm \alpha}} J t)^2$. For the  time \eqref{eq:apptopt}, it reads:
\begin{align}
q(k_\alpha) \approx  \frac{n^{k_\alpha}}{(n+1)^{k_\alpha+1}},
\label{probqalpha}
\end{align}
which we recall holds in the regime $N_{\rm at} \gg n$.

Given these considerations, for a given target photon number $n$, the state preparation protocol consists of the following steps.
%\begin{enumerate}
%    \item Let cavities and atomic arrays interact for $t_{\rm prep}$.
%    \item Measure the number of excited atoms in each atomic array, which yields the result $N_{\rm at} -k_\alpha$ with probability $q(k_\alpha)$ ($\alpha=a,b$).
%    \item The measurement heralds a Fock state in the cavity with known photon number $k_\alpha$. This Fock state is the initial state of the measurement.
%\end{enumerate}
%Starting from the ground state, a photonic Fock state in each cavity is produced by exciting the atoms, allowing them to interact with the cavity, and in the end heralding how many atoms have decayed.  For a given target photon number $n_{\rm ph}$, the protocol consists on:
\begin{enumerate}
\item~Starting from an initial state of all atoms and both cavities in the ground state, all atoms are excited.
\item~The cavity-atom interaction is turned on, $J(t)=J$, for a time $t_\mathrm{prep}$. 
\item~The cavity-atom interaction is turned off and the number of remaining excited atoms is measured.
\end{enumerate}
    Conditional on the measurement outcome $N_{\rm at}-k_\alpha$, the cavity is projected in a Fock state with $k_\alpha$ photons, which takes place with probability~$q(k_{\alpha})$ given in Eq.~\eqref{probqalpha}. This process can be done in parallel for both cavities, and on average $N=2n$ photons are created. 
%For $n_{\rm ph} \ll N_{\rm at}$ (linear regime), we find that $t_\mathrm{prep} \approx \sqrt{n_{\rm ph}/N_{\rm at} J^2}$ and $q(k_{\alpha}) \approx n_{\rm ph}^{k_\alpha}/(n_{\rm ph}+1)^{k_\alpha+1}$; whereas  for  $n_{\rm ph} \approx  \mathcal{O}(N_{\rm at})$ (non-linear regime)  both $t_{\rm prep}$ and $q(k_\alpha)$ can be determined numerically by solving  \eqref{eq:Hatoms} (see Sec.~\ref{secap:implQED} of the Appendix for details). %shown in Fig.~\ref{fig:fig3}b. [ NEED TO UPDATE FIGURE, variance is in the SI ]
%This can be done in parallel for both cavities.

%Starting from the ground state, a photonic Fock state in each cavity is produced following the following three steps. 
%(i) All atoms are excited.
%(ii) The cavity-atom interaction  is turned on, $J(t)=J$, for a time $t_{\mathrm{opt}}\approx \log(4N_{\rm at})/(2J\sqrt{N_{\rm at}})$.
%As we show in the SM and illustrate in Fig.~\ref{fig:fig3}a, this is the optimal waiting time for maximizing the photon number $n_{\rm ph}$, leading to $n_{\rm ph}/N_{\rm at} \approx 0.8$.
%(iii) The cavity-atom interaction is turned off and $S_\alpha^z$ is measured.
%Conditional on the measurement outcome $k_\alpha$, the cavity is projected in a Fock state with $N_{\rm at}-k_\alpha$ photons, with a probability distribution shown in Fig.~\ref{fig:fig3}b. 
%This can be done in parallel for both cavities.

\begin{figure*}
		\centering
		\includegraphics[width=0.47\linewidth]{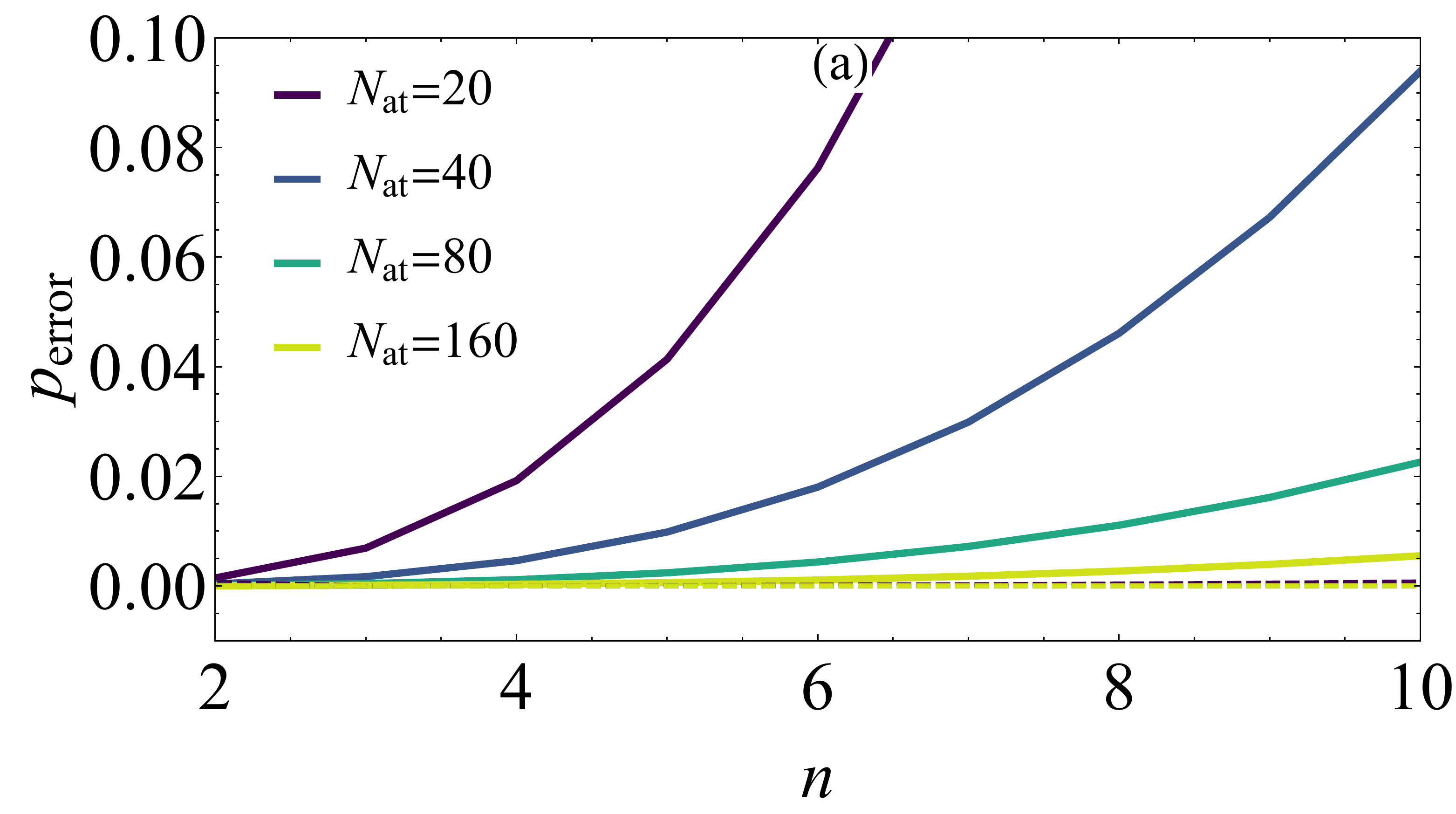} \hspace{5mm}
	    \includegraphics[width=0.47\linewidth]{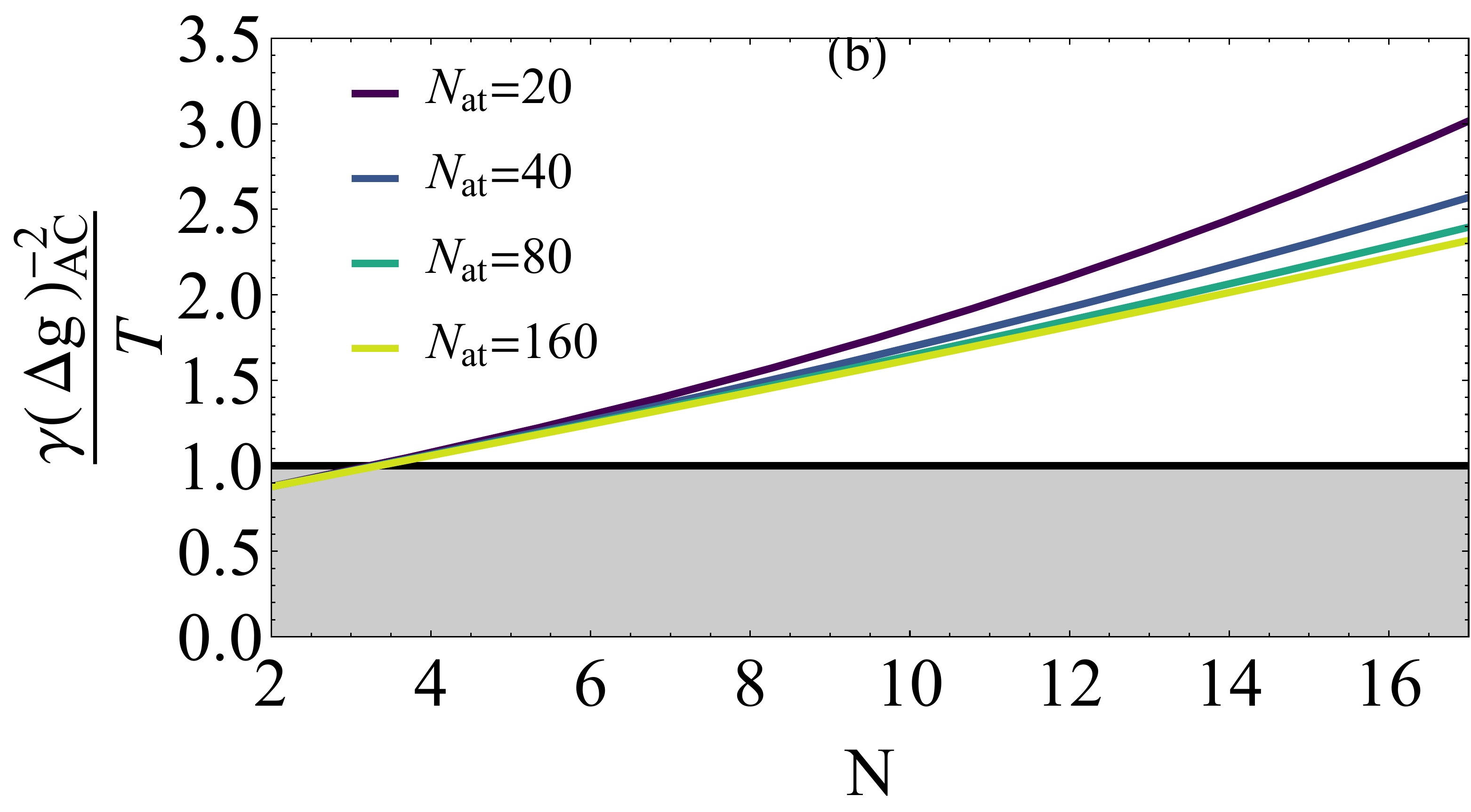} %\hspace{0.1mm}
	    \caption{(a) Error probability in the absorption process, i.e., probability that the photons are not absorbed by the atoms. We take as an initial state for the cavity a Fock state with $n$, which interacts with the atomic array   (the atoms start in the ground state) through \eqref{eq:Hatoms} for a time $t_{\rm meas}$. In dashed lines, we show the same for two repetitions of the absorption process, leading to $p_{\rm error}\approx 0$  (b) Measurement precision $(\Delta g)^{-2}_{\rm AC}$ for probabilistic TFS prepared through the atom-cavity coupling at $g\approx 0$,  using NRM and with $N_{\rm abs}=1$. Details on the calculation of $(\Delta g)^{-2}_{\rm AC}$ are provided in Sec.~\ref{secap:implQED}  of the Appendix. Parameters for both plots: $J=1$ and $\omega=1$. }
	\label{fig:fig3}
	\end{figure*}

After this preparation stage, we let the photonic system evolve according to Eq.~\eqref{evrho} for some time $t$, which is the time relevant for our discussion in the rest of the article. The number of photons involved in the process is $N$.
In the mean time, the remaining $N_{\rm at}-k_\alpha$ atoms need to be reset to the ground state in order to prepare them for the final measurement.

To measure the final photon distribution in the cavity, we again turn on the atom-cavity coupling $J(t)$, which allows the atoms to reabsorb photons from the cavity.
In contrast to the preparation stage, the atoms are now prepared in the ground state, whereas the photonic state contains at most $2n$ photons. As in the preparation stage, we can linearize the spin operator (now around the ground state)
obtaining $\sum_i\sigma^-_\alpha\simeq\sqrt{N_{\rm at}}c_\alpha$, where $c_\alpha$ is a bosonic annihilation operator.
 Thus, Eq.~\eqref{eq:Hatoms} can now be transformed as
\begin{equation}
    H_\mathrm{Dicke}\simeq \sum_{\alpha=a,b}\sum_{i=1}^{N_{\rm at}}[ \omega\sigma_{i,\alpha}^z
    + (J(t)\sqrt{N_{\rm at}}c_\alpha \alpha^\dagger+\mathrm{H.c.})].
\end{equation}
The time evolution is therefore a standard beam splitter interaction with oscillations with period $T= \pi /J\sqrt{N_{\rm at}}$. We take $t_{\rm meas}=T/2$, i.e.
\begin{align}
t_{\rm meas}= \frac{\pi }{2J\sqrt{N_{\rm at}}},
\label{tmeas}
\end{align}
to maximise the probability that all photons are absorbed. 
Given a $n$-photon state in the cavity as an initial state, this absorption process has an error probability $p_{\rm error} = \mathcal{O}(n/N_{\rm at})^2$, where $1-p_{\rm error}$ is the probability that all photons are absorbed. In order to relate it to the efficiency $\eta$ of the detector, %where $1-\eta$  quantifies the probability of losing a photon, 
we can identify:
\begin{align}
    1-p_{\rm error} \equiv \eta^n.
    \label{perror}
\end{align}
Note that, strictly speaking, the $\eta$ introduced here is not the same as the one  introduced in \eqref{modelnoise} (in the sense that they arise through different physical processes), but their meaning is essentially the same: $1-\eta$ the probability of not losing a photon in the detection process. In particular, a small $p_{\rm error}$ guarantees $\eta\approx 1$, as desired. %Hence the detector efficiency satisfies $\eta = $
%, i.e., the probability of not measuring a photon is of order  $\mathcal{O}((n/N_{\rm at})^2)$. 
The error probability $p_{\rm error}$  is shown in  Fig.~\ref{fig:fig3} for different $N_{\rm at}$. From the figure, it becomes clear that for $N_{\rm at} >20$ already very high efficiencies can be obtained, so that quantum advantages are possible according to Fig.~\ref{fig:fig2}.
Furthermore, we note that, if $p_{\rm error}$ is non-negligible,
the absorption process can be repeated, which substantially reduces $p_{\rm error}$, as also shown in Fig.~\ref{fig:fig3}. In this case, adding up the total number of excitations observed in the measurements yields the measurement result, thus implementing a number-resolved measurement (NRM) in each cavity.

%This is illustrated in Fig. 

%As discussed in the main text, this has an error probability $p_{\rm error}$ of order $\mathcal{O}((n_{\rm ph}/N_{\rm at})^2)$. 

%After a time $t_{\mathrm{meas}}=\pi /2J \sqrt{N_{\rm at}}$, the photons are re-absorbed with an error $p_{\rm error}$ of order $\mathcal{O}((n_{\rm ph}/N_{\rm at})^2)$, see Fig.~\ref{fig:fig3} (a).  If $p_{\rm error}$ is non-negligible, this process can be repeated  which substantially reduces $p_{\rm error}$; this is also illustrated in Fig.~\ref{fig:fig3}.  Adding up the total number of excitations observed in the measurements yields the measurement result, thus implementing a number-resolved measurement (NRM) in each cavity. %Both the efficiency and speed can be improved by increasing $N_{\rm abs}$, making this scheme  

%Hence, by increasing $N_{\rm at}$,  both $\etat \rightarrow 1$ and $t_{\rm ext}\rightarrow 0$, making this scheme particularly suited for this task (see Fig. XX).  
In practice, the following constraints need to be satisfied for the successful implementation of this scheme: \begin{enumerate}
    \item $p_{\rm error} = \mathcal{O}(n/N_{\rm at})^2 \ll 1$, i.e.,  a high detector efficiency in the NRM, and
    \item  $ t_{\rm ext} =  (t_{\rm prep}+t_{\rm meas}) \propto 1/ J \sqrt{N_{\rm at}} \ll \gamma^{-1}$, so that $t_{\rm ext}$ can be neglected.
\end{enumerate} 
The linear regime where $N_{\rm at} \gg n$ is hence convenient for the implementation of this protocol, and it is also useful to engineer a strong atom-cavity coupling $J\gg \gamma$. Experimental requirements needed for the successful implementation of this proposal are discussed in the Sec.~\ref{secap:implQED} of the Appendix, where we argue  that both cavity QED~\cite{Zhang2012,Hume2013,Haas2014} and circuit QED~\cite{Jeffrey2014,Walter2017,Dassonneville2020} appear as promising platforms.   % and  by taking $N_{\rm at} \gg n_{\rm ph}$, making this scheme particularly suited for this task.
%}
%In ideal conditions ($\eta=1$, $t_{\rm ext}=0$), 

Even when both conditions are met,  the discussed preparation scheme does not produce TFS, but rather Fock states distributed according to $q(k_{\alpha})$. Furthermore, the NRM in each cavity is in general not optimal for metrology, i.e., the corresponding CFI does not saturate $\mathcal{F}_{\rm Q}$ (see  Fig.~\ref{fig:fig1}, Appendix~\ref{secapp:numberres}, and Refs.~\cite{Taesoo1998,Pezze2013,Zhong2017}). While both effects lead to a reduction for the quantum advantages reported in Fig.~\ref{fig:fig2}, we find that  quantum advantages can be obtained for $N = 2n \approx 4-20$, and $N_{\rm at} \approx 20-100$ with the discussed implementation, as shown in Fig.~\ref{fig:fig3} for $N_{\rm abs}=1$. Details on the calculation of $(\Delta g)^{-2}_{\rm AC}$ for the atom-cavity (AC) implementation, as well as an analysis to robustness to non-idealities ($\eta>1$ and $t_{\rm ext}>0$), is provided in Appendix~\ref{secap:implQED}.

%\section{Conclusions}

\section{Conclusions} 
While quantum advantages in photonic metrology are well established when the (average) photon number is fixed~\cite{demkowicz-dobrzanski15,Dowling2015,polino2020photonic}, in specific situations, other constraints might be more meaningful.
For example, when non-linear effects or quantum backaction limit the light intensity, a more sophisticated analysis has to take these effects into account
%it is more subtle whether there can exist a quantum advantage when an arbitrarily powerful coherent state is employed
\cite{Mitchell2017}. 
Here, we study a complimentary situation, where photosensitive samples imply that the number of absorbed photons $N_{\rm abs}$ is limited~\cite{Wolfgramm2012,Taylor2013,Taylor2014,Taylor2016,Whittaker2017}.
While in interferometric measurements this coincides with the traditional constraint on $N$,  we have shown that it leads to qualitatively new results in frequency measurements. In this case, finite quantum resources for metrology (finite photon number, finite number of samples, finite time) can become more powerful than infinite classical resources (unbounded photon number, unlimited number of samples, finite time). %\textcolor{blue}{Such advantages can remain in the presence non-ideal detectors and extra preparation times of the quantum states.} 
We have characterized these advantages as a function of $N_{\rm abs}$ and $T$ for a  model of two coupled oscillators,  but our results suggest similar advantages in frequency measurements of delicate samples where photon absorption can be modelled as a Markovian process. 

In a second part of the article, we have discussed a possible implementation of these advantages in a cavity QED set-up. The key idea is to couple   cavities  to a set of resonant atoms. By suitably controlling  the atoms, we have devised strategies to prepare and measure Fock states in a fast and efficient way. In particular, we showed that superradiance, arising due to
the collective atom-cavity coupling, can be exploited to enhance the quality of the measurement and preparation of photonic states. These results illustrate the potential of    superradiance for applications in quantum metrology and information~\cite{OBrien2009,demkowicz-dobrzanski15,Dowling2015,polino2020photonic}.

\textit{Acknowledgements.} We thank J. Ko{\l}ody{\'{n}}ski and M. W. Mitchell for insightful discussions.    M. P.-L. acknowledges funding from Swiss National Science Foundation (Ambizione  PZ00P2-186067). 
D.M.\ and J.I.C.\ acknowledge funding from ERC Advanced Grant QENOCOBA under the EU Horizon 2020 program (Grant Agreement No. 742102).

\newpage

\appendix

\begin{widetext}

\section{Squeezed states}
\label{secap:squeezedstates}
%\onecolumn

Squeezed states also enable Heisenberg scaling~\cite{Yurke1986,Olivares2007}, and hence can provide a finite amount of information in the Poisson limit $t \rightarrow 0$ and $N \rightarrow \infty$. This is explored in this section.

We  take as an initial state a vacuum-squeezed state together with a coherent-squeezed state: $\ket{\phi_0}= D_a(\alpha) S_a(s) S_b(r) \ket{0}$ with $D_a(\alpha)= e^{-\alpha (a^{\dagger}-a)}$,
 and $S_y(x)= e^{\frac{x}{2}(y^2-(y^\dagger)^2)}$, $x=s,r$ and $y=a,b$; and we take $\alpha, r>0$ and $s<0$. The average total number of photons is:
 \begin{align}
     N=\alpha^2 + \sinh^2 r + \sinh^2 s,
 \end{align}
 and it is convenient to define: $\beta_x = \sinh^2 x/N$, $x=r,s$.
 We consider number-resolved measurements in the output states. Following~\cite{Olivares2007}, the $\mathcal{F}_C$ in the regime $N\gg 1$ is given by (the extra factor $t^2$ comes from the fact that we are measuring a frequency):
 \begin{align}
     \mathcal{F}_C^{\rm sq} = \frac{N^2 t^2 (1-2\beta_r)^2}{N\frac{1-e^{-\gamma t}}{e^{-\gamma t}}+\frac{1}{4}\left(\frac{1}{\beta_r}+\frac{\beta_r}{\beta_s}-3 \right)}.
 \end{align}
Taking the Poisson limit ($t\rightarrow 0$ with $Nt\gamma=N_{\rm abs}$), leads to
\begin{align}
     \mathcal{F}_C^{\rm sq} = \frac{N_{\rm abs}^2}{\gamma^2} \frac{ 4(1-2\beta_r)^2}{4N_{\rm abs}+\left(\beta_r^{-1}+\beta_r 2\beta_s^{-1}-3 \right)}.
\end{align}
At this point one can  maximise this expression with respect to $\beta_x$, with the constraints $0<\beta_x<1$ and $\beta_a+\beta_b \leq 1$, in order to optimise the measurement precision. 
For  $N_{\rm abs} \ll 1$, we find that $\mathcal{F}_C^{\rm sq} \approx  N_{\rm abs}^2/\gamma^2$, hence squeezed states perform close to NOON states in this regime. On the other hand, for $N_{\rm abs} \gg 1$, we have $\mathcal{F}_C^{\rm sq} \approx N_{\rm abs}/\gamma^2$, thus saturating the upper bound presented in the main text. For intermediate values of $N_{\rm abs}$, a numerical optimisation is presented in Fig. 1 of the main text. 
It remains an interesting question whether more generalised Gaussian states and/or measurement schemes~\cite{Fuentes2016,Matsubara2019} can outperform this proposal in this regime.

\section{Classical Fisher Information for number-resolved measurements}
\label{secapp:numberres}

In this section we compute the classical Fisher information,
\begin{align}
\label{cfi}
\mathcal{F}_{\rm C}= \sum_j \frac{1}{q_j} \dot{q}_j^2,
\end{align}
where $q_j$ is the probability to obtain outcome $j$, 
for number resolved measurements in the set-up described in the main text. 
Because of the map between our framework and Mach-Zehnder interferometry with symmetric (time-dependent) photon loss, we can follow similar derivations in that context, see e.g. Refs.~\cite{Taesoo1998,Pezze2013,Zhong2017}.
%In this section we consider number-resolved measurements. 
Let us first compute the probabilities of the photon-number measurements for the initial Fock state $\ket{k,m}$ and unitary $U_{g}=e^{-itH_0}e^{-itH_{\rm int}}$. First of all we note that $e^{-itH_0}$ commutes with the initial state (with a well defined photon number in each arm) and hence can be ignored for the number-resolved statistics. We first compute:
\begin{align}
P(k,m,q,g)\equiv |\bra{k-q,m+q}e^{-ig tH_{\rm int}} \ket{k,m}|^2
\end{align}
with $\ket{k,m}=\frac{1}{\sqrt{k!m!}}(a^\dagger)^k (b^\dagger)^m \ket{0,0}$ and $q\in [-m,k]$. Noting that:
\begin{align}
&a^\dagger_g \equiv e^{-ig tH_{\rm int}} a^\dagger e^{ig tH_{\rm int}}= a^{\dagger} \cos\left( \frac{tg}{2}\right)+ i b^{\dagger} \sin\left( \frac{tg}{2}\right)
\nonumber\\
&b^\dagger_g \equiv e^{-ig tH_{\rm int}} b^\dagger e^{ig tH_{\rm int}}= b^{\dagger} \cos\left( \frac{tg}{2}\right)+ i a^{\dagger} \sin\left( \frac{tg}{2}\right),
\end{align}
we obtain
\begin{align}
P(k,m,q,g)= \frac{(m+q)!(k-q)!}{m!k!} \left(\tan\theta \right)^{2q} \left(\sum_{j=0}^m \binom{m}{j} \binom{k}{j+q}  (-1)^j  \left(\cos\theta\right)^{k+m-2j} \left(\sin\theta \right)^{2j} \right)^2
\end{align}
where we defined $\theta\equiv tg/2$, and $q\in [-m,k]$. This can also be expressed as:
\begin{align}
P(k,m,q,g)=&\frac{(m+q)! (k-q)!}{m! k!} \left(\tan\theta \right)^{2q}
\nonumber\\
& \bigg((-1)^k \binom{m}{k+1} \binom{k}{k+q+1} \frac{(\sin \theta )^{2
   k+2}}{ (\cos \theta )^{2-m-k}} \, _3F_2\left(1,-m+k+1,q+1;k+2,k+q+2;-\tan ^2(\theta
   )\right)\nonumber\\
  & +\binom{k}{q} (\cos \theta )^{m+k} \, _2F_1\left(-m,q-k;q+1;-\tan ^2(\theta
   )\right)\bigg)^2,
   \label{P(k,m,q)}
\end{align}
where $_pF_q(a;b;z)$ is the generalized hypergeometric function.

Next, we compute the probability to obtain outcomes $\{ k,m\}$ for the state:
\begin{align}
\rho = \sum_{k,j} p_k p_j \ket{n-k,n-j}\bra{n-k,n-j}
\label{fockState}
\end{align}
with
\begin{align}
p_k = \mu^k (1-\mu)^{n-k} \binom{n}{k}, \hspace{10mm} \mu=1-e^{-\gamma t}.
\end{align}
It is convenient to introduce:
\begin{align}
&n_{\rm loss}\equiv 2n-(k+m)
\nonumber\\
& q\equiv n-k
\end{align}
Then, we have for \eqref{fockState}
\begin{align}
P^{\rm Fock}(n_{\rm loss},q,g)=\sum_{l=0}^{n_{\rm loss}} p_l p_{n_{\rm loss}-l} P(n-l,n+l-n_{\rm loss},q-l,g)
\end{align}
The only remaining step is to compute the classical Fisher information using \eqref{cfi}:
\begin{align}
\label{cfiF}
\mathcal{F}_{\rm C}^{\rm TFS}(g)=\sum_{n_{\rm loss}=0}^{2n} \quad \sum_{q=n_{\rm loss}-n}^{n} \frac{\left( \partial_g P^{\rm Fock}(n_{\rm loss},q,g)\right)^2}{P^{\rm Fock}(n_{\rm loss},q,g)}
\end{align}
To gain some analytical insight for this expression, we can compute it in the limit $g \rightarrow 0$. Following~\cite{Pezze2013,PerarnauLlobet2020}, we can expand the probabilities around $g\approx 0$ up to order $\mathcal{O}(g^2)$,  and realise that only instances where $n+1$ photons are measured in one of the modes contribute to the CFI. This leads to the following analytical expression for the CFI~\cite{Datta2011}:
\begin{align}
\lim_{g \rightarrow 0} \mathcal{F}_{\rm C}^{\rm TFS}(g) = 2 t^2 (n+1)n (1-\mu)^{n+1}
\label{cfivarphi0},
\end{align}
which shows an exponential decay with $n=N/2$. This is hence far from $\mathcal{F}_{\rm Q}$ for large $N$, while still  better than NOON states. For future convenience, we also give the corresponding expression when the initial state is $\rho = \sum_{k,j} p_k p_j \ket{n-k,m-j}\bra{n-k,m-j}$:
\begin{align}
   \lim_{g \rightarrow 0} \mathcal{F}_{\rm C}^{n,m}(g)=t^2\left( (1-\mu)^{n+1}(n+1)m+(1-\mu)^{m+1}(m+1)n\right).
   \label{eq:genexp}
\end{align}

Finally, we numerically compute \eqref{cfiF} for different $g$, $\mu$ and $N$ for the state \eqref{fockState} (and setting $t=1$). An example is shown in Fig.~\ref{fig:cfi}, from it we see that in general the dependence of $\mathcal{F}_{\rm C}$ with $g$ is non-trivial, and so is the point where it is maximized. The figure also shows that indeed \eqref{cfi} does not saturate $\mathcal{F}_{\rm Q}$ for number-resolved measurements. In Fig.~\ref{fig:cfi} we also show \eqref{cfiF} as a function of $N$, where it becomes apparent the importance of maximizing $\mathcal{F}_{\rm C}$ with respect to $g$ for the number-resolved-measurements as $N$ increases. The numerical results suggest that a systematic improvement with respect to the Shot-Noise-Limit can be obtained for any $N$ with suitably optimized number-resolved-measurements.

\begin{figure}[htb]
  \centering
  \includegraphics[width=.4\linewidth]{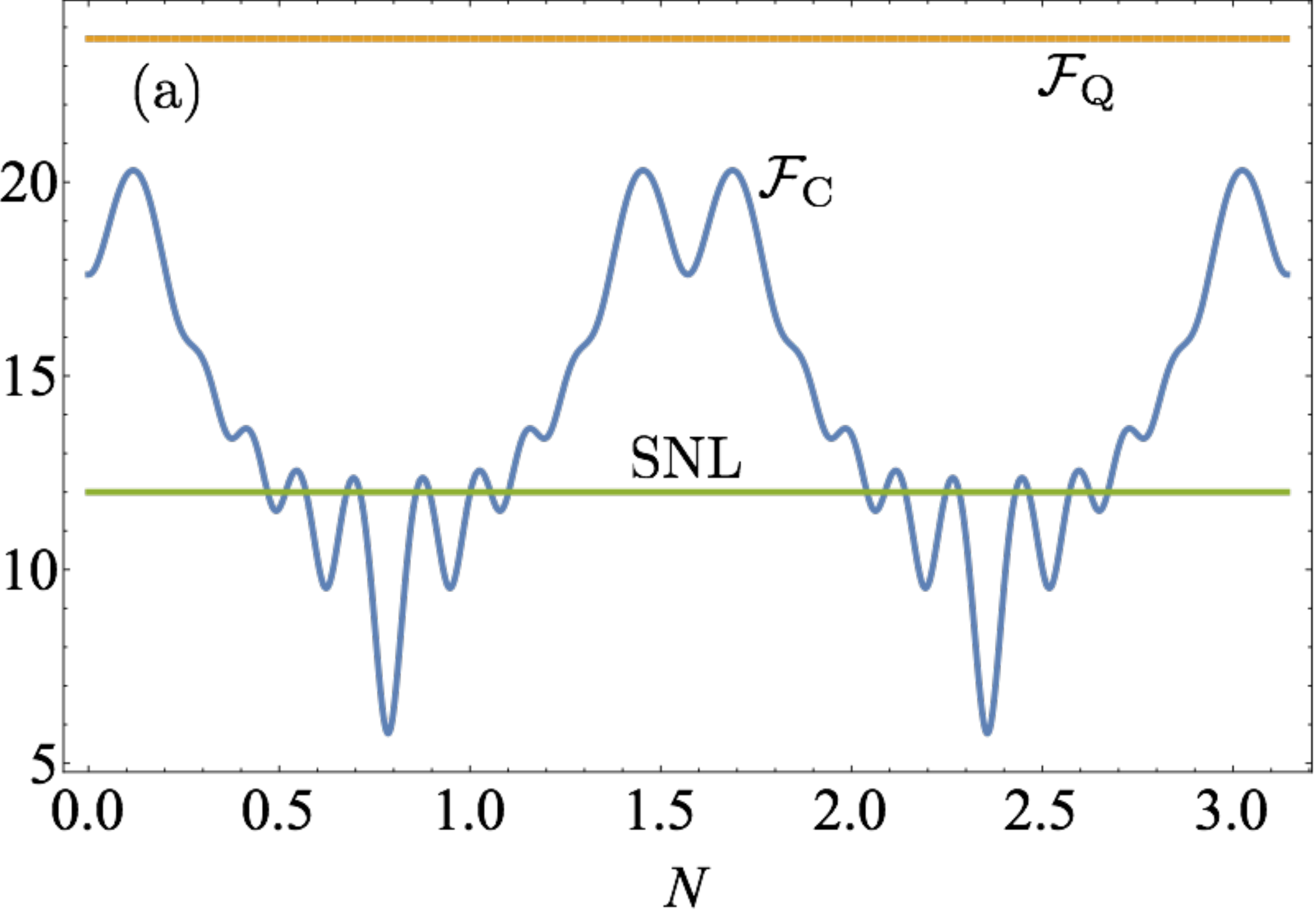} \hspace{10mm}
  \includegraphics[width=.4\linewidth]{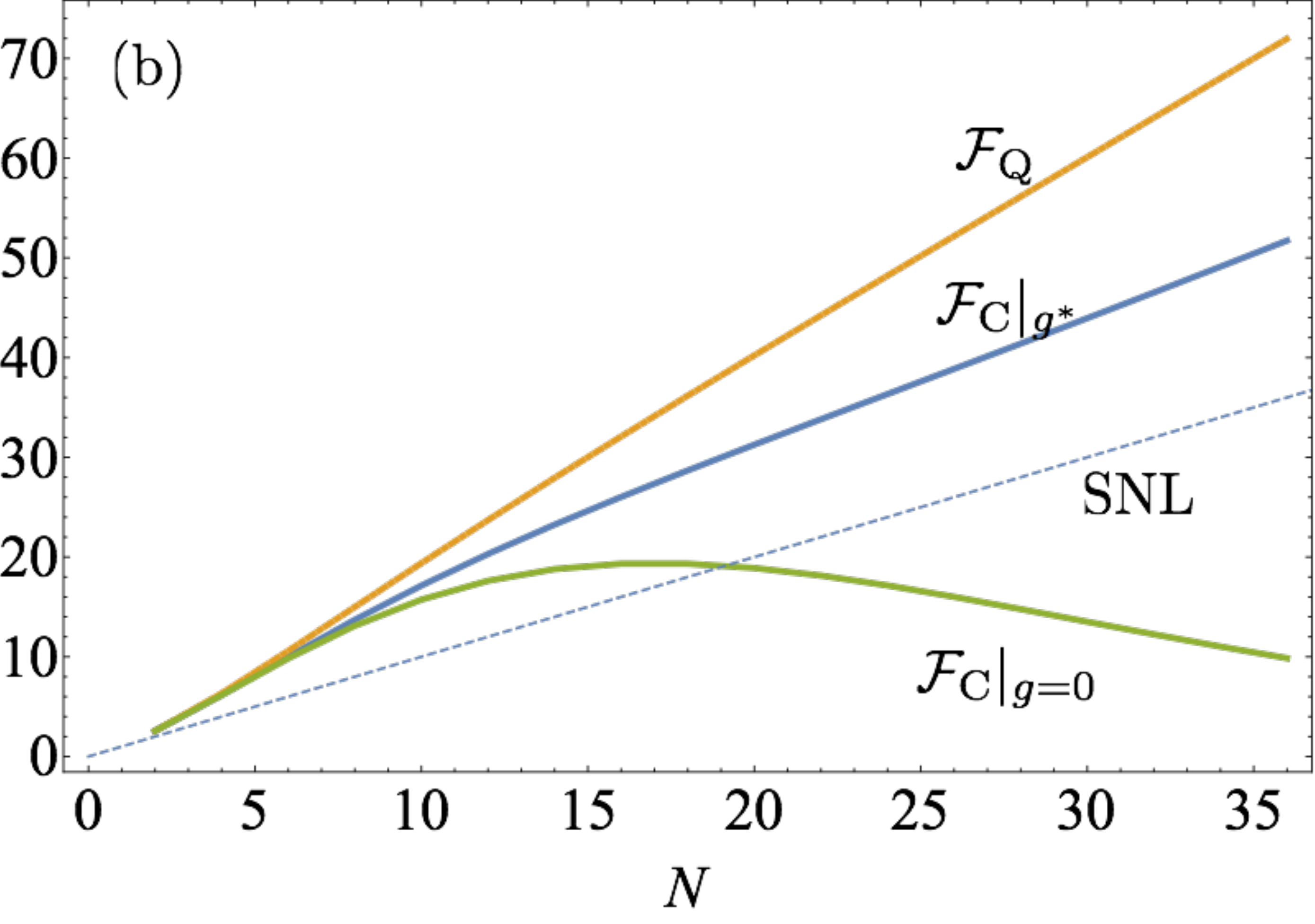}
  \caption{(a) CFI, QFI, and SNL vs $\phi$. Parameters: $N=12$, $\mu =0.2$, and $t=1$. (b) $\mathcal{F}_{\rm Q}$ and $\mathcal{F}_{\rm C}$ vs $N$ for $\mu =0.2$ and $t=1$. }
  \label{fig:cfi}
\end{figure}

Let us now look at the Poisson limit $N\rightarrow \infty$, $t\rightarrow 0$ while satisfying $2 \gamma t n \rightarrow N_{\rm abs}$.
First, from \eqref{cfivarphi0}, we can study the Poisson limit $2 \gamma t n \rightarrow N_{\rm abs}$, where we obtain:
\begin{align}
\lim_{g \rightarrow 0} \mathcal{F}_{\rm C}^{\rm TFS} = \frac{N_{\rm abs}^2}{2\gamma^2} e^{-N_{\rm abs}/2}
\label{cfivarphi0poislim}.
\end{align}
To obtain similar analytical expression for arbitrary $g$ is challenging due to the complexity of obtaining asymptotic expressions for the generalized hypergeometric functions. Yet, we can obtain numerical results for rather large $N \approx 100-1000$. To do this, it is useful to approximate $\mathcal{F}_{\rm C}$ in \eqref{cfiF} as:
\begin{align}
\label{cfiFII}
\mathcal{F}_{\rm C}^{\rm TFS}(g)\approx \sum_{n_{\rm loss}=0}^{N_{\rm loss}^{\rm max}} \quad \sum_{q=-q_{\rm max}}^{q_{\rm max}} \frac{\left( \partial_g P^{\rm Fock}(n_{\rm loss},q,g)\right)^2}{P^{\rm Fock}(n_{\rm loss},q,g)}
\end{align}
since only terms with $q = \mathcal{O}(1)$ and $N^{\rm max}_{\rm loss}=\mathcal{O}(N_{\rm abs})$ contribute to the sum  in the Poisson limit (other terms are exponentially small). In our numerical simulations, we take $q_{\rm max}=10$ and $N^{\rm max}_{\rm loss}=4N_{\rm abs}$, and the results become in practice indistinguishable when these parameters are varied. Numerical results are shown in Fig.  2  of the main text.

\subsubsection{Optimal measurement}

Finally, we note that the optimal measurement $L$ where the classical Fisher info \eqref{cfi} saturates $Q$ is given by~\cite{PARIS2009}:
\begin{align}
L& =i \sum_{k,l} \frac{\tilde{p}_k \tilde{p}_l -\tilde{p}_{k-1} \tilde{p}_{l+1}}{\tilde{p}_k \tilde{p}_l +\tilde{p}_{k-1} \tilde{p}_{l+1}} \sqrt{k(l+1)} \bigg( \ket{k,l} \bra{k-1,l+1} +\ket{l,k} \bra{l+1,k-1}   \bigg)
\nonumber\\
&= i \sum_{k,l} \frac{(n-k+1)(l+1) - k (n-l)}{n(l+k+1) +k(n-l)} \sqrt{k(l+1)} \bigg( \ket{k,l} \bra{k-1,l+1} +\ket{l,k} \bra{l+1,k-1}   \bigg) .
\label{optmeas}
\end{align}
with $\tilde{p}_k=p_{n-k}$. In the presence of photon loss (i.e. $p_0 \neq 1$), this does not correspond to a number resolved measurement of the two modes, but to a global measurement involving both modes.

\section{Robustness to imperfections: Photon loss in the detection process and finite time for preparation and measurement of quantum states}

\label{secapp:imperfect}

In this section, we discuss how  $(\Delta g)^{-2} \equiv \nu \mathcal{F}$ of general quantum states decreases due to: finite $N$, photon loss in the measurement process, and an extra-time $t_{\rm ext}$ of preparation/measurement of quantum resources. We then also discuss the particular case of twin Fock states (TFS), both with optimal measurements  ($\mathcal{F}_{\mathcal{Q}}^{\rm TFS}$) and with photon number-resolved measurements (NRM) ($\mathcal{F}_{\mathcal{C}}^{\rm TFS}$).

%\subsection{QFI of TFS}

We start by  giving general bounds on $\nu \mathcal{F}_{\mathcal{Q}}$ for any quantum state given different constraints:

\begin{figure}[t]
  \centering
  \includegraphics[width=0.45\linewidth]{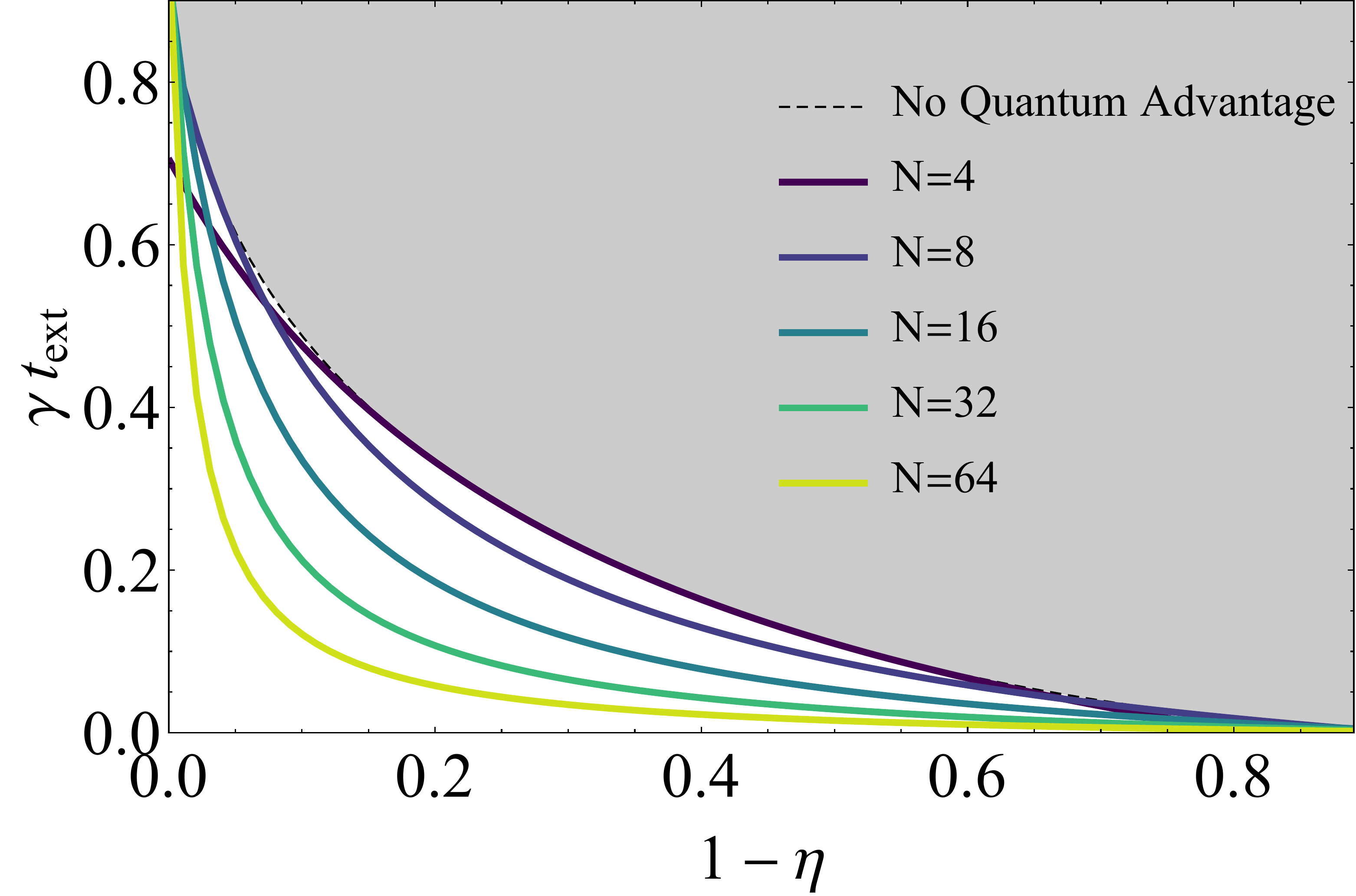}
  \caption{Here we illustrate regions where quantum advantages become possible in principle. For that, we plot the values of $\{t_{\rm ext}, 1-\eta \}$ for which the upper bound $\mathcal{J}(t)$, defined in \eqref{genboundad}, satisfies $\mathcal{J}(t)=1$. Recall that if $\mathcal{J}(t)<1$ then no quantum advantage is possible (using any state). This is shown in grey, by maximising \eqref{genboundad} over any $t$ (and hence $N$). On the other hand, for the different colourful curves $t$ is fixed implicitly by $N_{\rm abs}=N(1-e^{-\gamma t})$, which provide upper bounds for a state with $N$ photons.
  }
  \label{figApp:tfscfiIIa}
\end{figure}

\begin{figure}[t]
  \centering
  \includegraphics[width=0.45\linewidth]{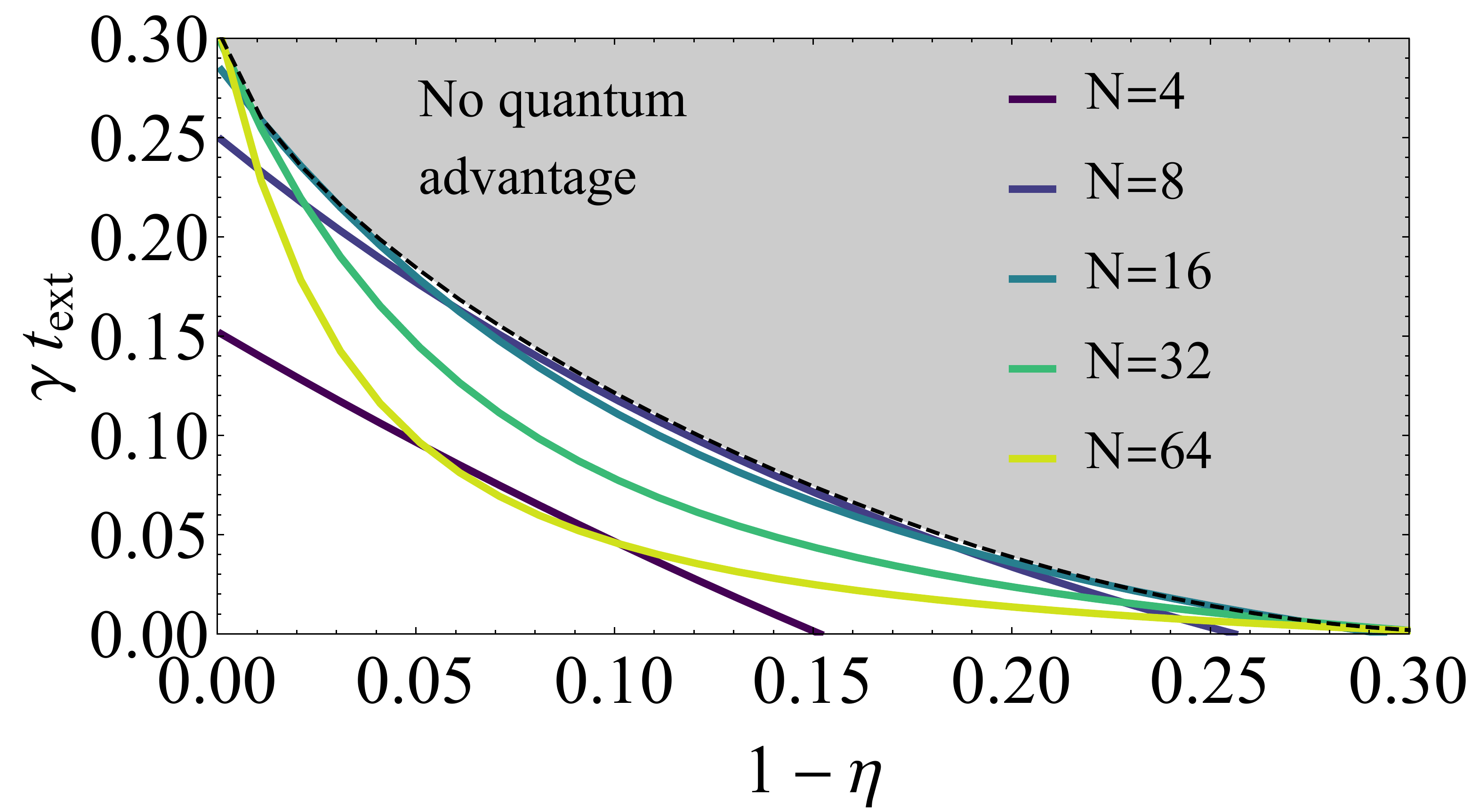}
  \includegraphics[width=0.45\linewidth]{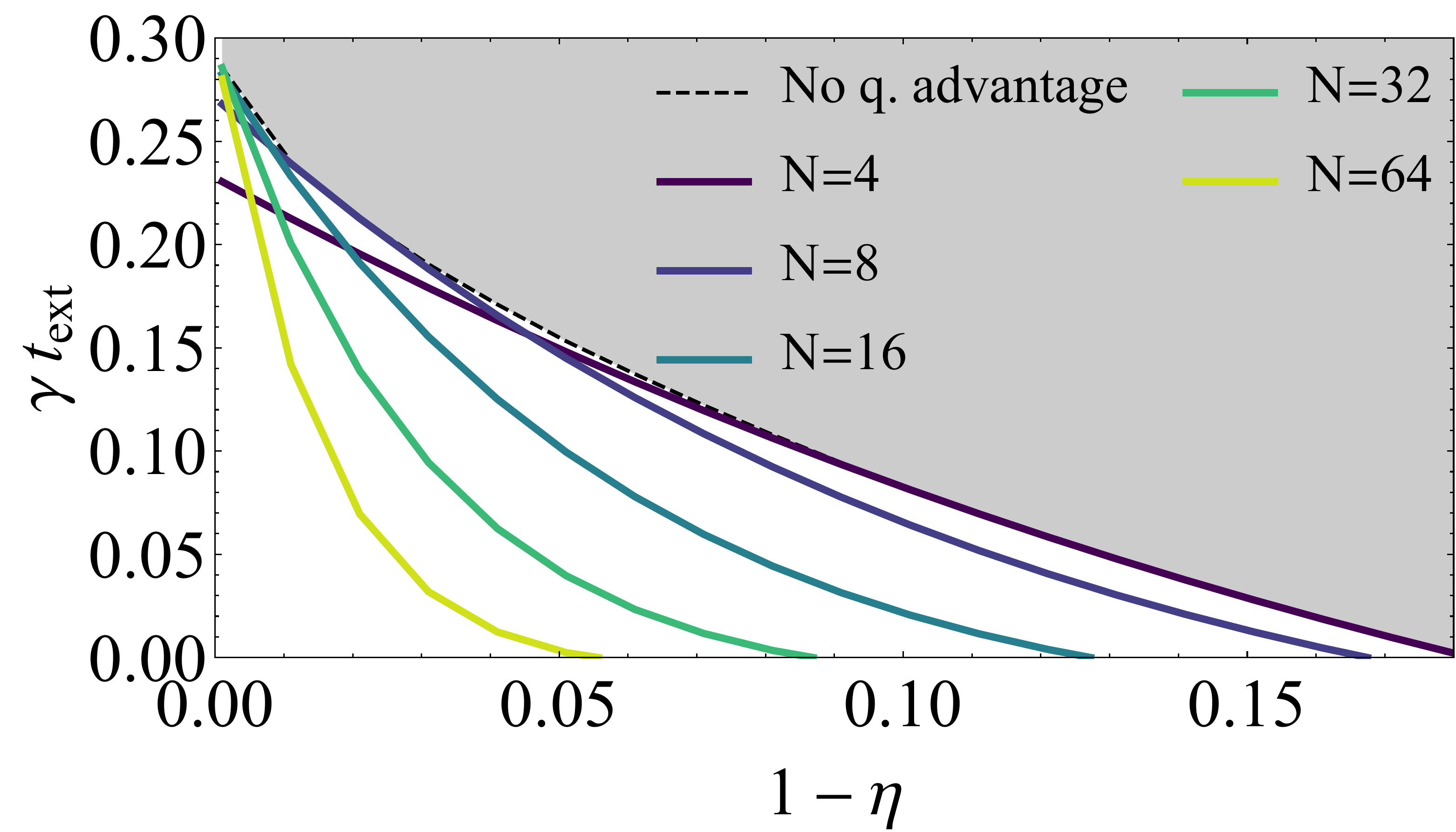}
  \caption{Similarly to Fig.~\ref{figApp:tfscfiIIa}, we plot the values $\{1-\eta, t_{\rm ext} \}$ for which $(\Delta g)^{-2}_{\rm coh}/(\Delta g)^{-2}_{\rm TFS}=1$ (left) or $(\Delta g)^{-2}_{\rm coh}/(\Delta g)^{-2}_{\rm NRM}=1$ (right) for TFS of different $N$ and $N_{\rm abs}=1$. In other words,  lower values of $\{1-\eta, t_{\rm ext} \}$  enable quantum advantages with TFS. For NRM, we consider $g\approx 0$.
  }
  \label{figApp:tfscfiII}
\end{figure}

\begin{itemize}
\item \emph{Finite N}.
Let us first assume finite $N$, but $\eta=1$ and $t_{\rm ext}=0$. Using the upper bound of~\cite{Fujiwara2008,Knysh2011,escher2011general,DemkowiczDobrzaski2012,Koodyski2013},  we have:
\begin{align}
    \mathcal{F}_{\mathcal{Q}} \leq t^2 N \frac{e^{-\gamma t}}{1-e^{-\gamma t}} \leq \frac{tN}{\gamma},  
\end{align}
where the factor $t^2$ comes from the fact that we are measuring a frequency-like quantity. Considering now $\nu=T/t$ measurements, we obtain:
\begin{align}
    (\Delta g)^{-2}_N \equiv \nu \mathcal{F}_{\mathcal{Q}} \leq \frac{TN}{\gamma} = \frac{N}{N_{\rm abs}} (\Delta g)^{-2}_{\rm coh}.
\end{align}
Since any $\eta<1$ or $t_{\rm ext}>0$ can only decrease $(\Delta g)^{-2}_N$, this bound also holds for any $\eta$ and $t_{\rm ext}$. 
\item \emph{Photon loss}. 
Let us now assume $0<\eta<1$. The starting point is now:
$\mathcal{F}_{\mathcal{Q}} \leq t^2 N \eta e^{-\gamma t}/(1-\eta e^{-\gamma t})$,  
which implies for the whole process with $\nu=T/t$ measurements:
\begin{align}
    (\Delta g)^{-2}_\eta \equiv \nu   \mathcal{F}_{\mathcal{Q}} \leq T \eta N_{\rm abs} \frac{t e^{-\gamma t}}{(1-e^{-\gamma t})(1-\eta e^{-\gamma t})}, 
    \label{apetamax}
\end{align}
which we expressed in terms of the constants $T$ and $N_{\rm abs}$. The right hand side of \eqref{apetamax} is maximised for $t \rightarrow 0$ (or equivalently $N \rightarrow \infty$), leading to the bound:
\begin{align}
    (\Delta g)^{-2}_\eta  \leq (\Delta g)^{-2}_{\rm coh} \frac{\eta}{1-\eta}. 
    \label{appmax}
\end{align}
Hence the quantum advantage comes in the form of a better prefactor $\eta/(1-\eta)$. This prefactor advantage takes the same form as the one obtained in noisy photon interferometry  when comparing coherent and quantum states with a fixed $N$~\cite{Fujiwara2008,Knysh2011,escher2011general,DemkowiczDobrzaski2012,Koodyski2013}. %, but note that \eqref{appmax} can be saturated only in the limit $N\rightarrow \infty$. 
\item \emph{Extra time}.
Let us now assume an extra time $t_{\rm ext}$ necessary to prepare and measure using quantum resources. Let us consider the extreme case  $t_{\rm ext}>0$ but $\eta=1$ and no constraints on $N$.  Taking $\eta=1$, we have:
\begin{align}
(\Delta g)^{-2}_{t_{\rm ext}} \equiv \nu   \mathcal{F}_{\mathcal{Q}} \leq T  N_{\rm abs} \frac{t^2 e^{-\gamma t}}{(t+t_{\rm ext})(1- e^{-\gamma t})^2}.
\end{align}
This expression is again optimised in the limit $t\rightarrow 0$ (and hence $N\rightarrow \infty$), yielding: 
\begin{align}
(\Delta g)^{-2}_{t_{\rm ext}}  \leq (\Delta g)^{-2}_{\rm coh} \frac{1}{\gamma t_{\rm ext}}.
\end{align}
\end{itemize}

Let us now consider the general case where $0<\eta<1$ and $t_{\rm ext}>0$. In general we have,
\begin{align}
\frac{(\Delta g)^{-2}_{t_{\rm ext},\eta}}{(\Delta g)^{-2}_{\rm coh} }  \leq     \frac{\gamma t^2 \eta e^{-\gamma t}}{(t+t_{\rm ext})(1- e^{-\gamma t})(1- \eta e^{-\gamma t})} \equiv \mathcal J(t).
\label{genboundad}
\end{align}
This expression is no longer optimised in general at $t\rightarrow 0$, and in fact it does not allow for an analytical solution. We can still solve it numerically for any value of the desired value of $\mathcal{J}(t)$, which provides an upper bound on the quantum advantage. In particular, a quantum advantage is impossible  if $\mathcal{J}(t)=1$ $\forall t$. This condition is plot in Fig.~\ref{figApp:tfscfiIIa}, where we compute the values of $\eta,t_{\rm ext}$ for which $\max_t \mathcal{J}(t)=1$.  In Fig.~\ref{figApp:tfscfiIIa}, we also plot the values $\eta,t_{\rm ext}$ for which $\max_t \mathcal{J}(t)=1$ but by fixing $t$ implicitly by $N_{\rm abs}=N(1-e^{-\gamma t})$, which provide upper bounds for a state with $N$ photons. Interestingly, we find that the upper bound becomes more robust to noise in the regime where both $t_{\rm ext}$ and $1-\eta$ are not close to zero. This suggests that in realistic implementations few-photon quantum states are desirable when $N_{\rm abs}$ is low.

Let us now move to the particular case of TFS by considering $(\Delta g)^{-2}_{\rm TFS} \equiv \max_\nu (\nu \mathcal{F}_{\mathcal{Q}}^{\rm TFS})$ and $(\Delta g)^{-2}_{\rm TFS} \equiv \max_\nu (\mathcal{F}_{\mathcal{C}}^{\rm TFS})$. It is useful to point out that, as long as $N_{\rm abs}\leq N_{\rm abs}^*$ ($N_{\rm abs}^*$ needs to be determined in each case, e.g., numerically), $ \nu \mathcal{F}_{\rm Q}^{\rm TFS}$ (and  $ \nu \mathcal{F}_{\rm C}^{\rm TFS}$) is maximised by testing each sample until it is damaged, that is:
\begin{align}
&T=t \nu
\nonumber\\
&t=\frac{-1}{\gamma} \ln\left(1-\frac{N_{\rm abs}}{N} \right) \approx \frac{N_{\rm abs}}{\gamma N}
\label{choicet}
\end{align}
for $N_{\rm abs}\leq N_{\rm abs}^*$ (otherwise, the choice \eqref{choicet} is taken with $N_{\rm abs}=N_{\rm abs}^*$), and we recall that $N_{\rm abs}^*$  needs to be determined numerically for each case. Given this choice,
 we first analyse the regions where quantum advantages are possible using TFS, which is shown in Fig.~\ref{figApp:tfscfiIIa}. As expected, the regions are more restrictive than for the upper bound shown in Fig.~\ref{figApp:tfscfiII}. 
On the other hand, an example of the quantum advantage as a function of $N_{\rm abs}$ for different TFS and with $\eta=0.95$ and $\gamma t_{\rm ext}=0.05$ is presented in Fig.~\ref{figApp:tfscfi}. Similar results are obtained when considering $\mathcal{F}_{\mathcal{C}}^{\rm TFS}$
 with NRM, which is in general lower than $\mathcal{F}_{\mathcal{C}}^{\rm TFS}$ as expected from Fig.~\ref{fig:cfi}.  This is shown in both Figs. \ref{figApp:tfscfiII} and \ref{figApp:tfscfi}   by considering $g\approx 0$.

% \textcolor{blue}{
% While the presence of $\eta$ and $t_{\rm ext}$ severely restricts the maximal possible quantum advantage, the main practical conclusion of our paper remains: In frequency measurements of delicate samples, few-photon quantum states can overcome classical strategies using arbitrarily intense coherent states and an arbitrarily large number of samples. Notably, this is valid even in the presence of  photon loss in the detectors (e.g. $\eta \approx 0.9$), and extra time to prepare and measure quantum states ($\gamma t_{\rm ext} \approx 0.1$),  and for low $N \approx 8$.}

%\begin{figure}[t]
%  \centering
%  \includegraphics[width=0.5\linewidth]{UpperBounds.pdf}
%  \caption{\textcolor{blue}{In dashed line, we plot the values of $\{\eta,t_{\rm ext}\}$ for which $\max_t \mathcal{J}(t)=1$, which discards a region where quantum advantages are not possible. The other lines correspond to the $\{\eta,t_{\rm ext}\}$ for which $J(t)=1$ with $t$ defined implicitly by: $N_{\rm abs}=N(1-e^{-\gamma t})$ (in the plot we take $N_{\rm abs}=1$, $\gamma=1$).} 
%  }
%  \label{figApp:UpperBound}
%\end{figure}

% Basically we wish to understant for which values of $\eta, t_{\rm ext}$ the RHS of \eqref{genboundad} can become larger than one; this shows when a quantum advantage is not impossible. 

\begin{figure}[t]
  \centering
  \includegraphics[width=0.45\linewidth]{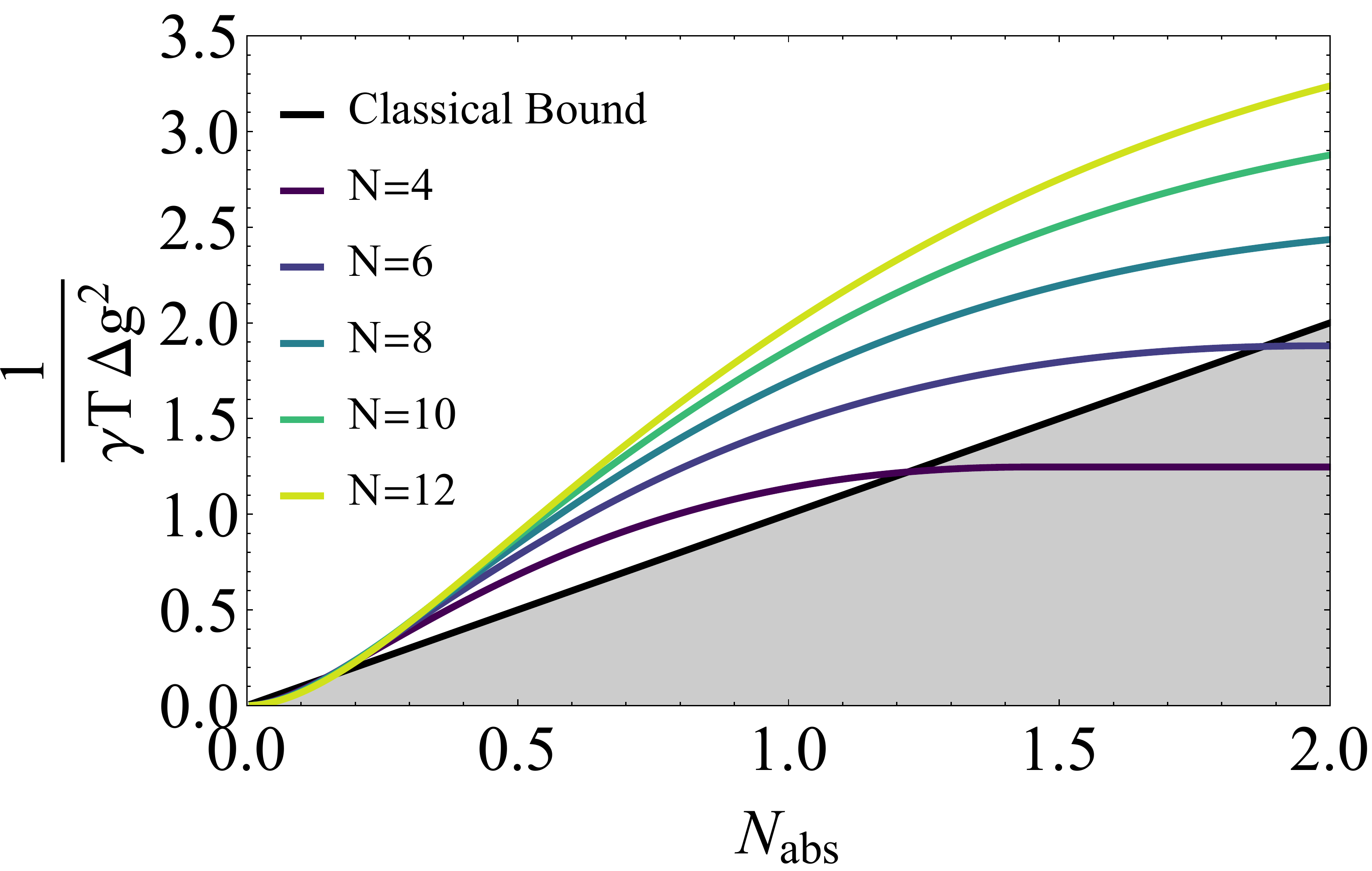}
  \includegraphics[width=0.45\linewidth]{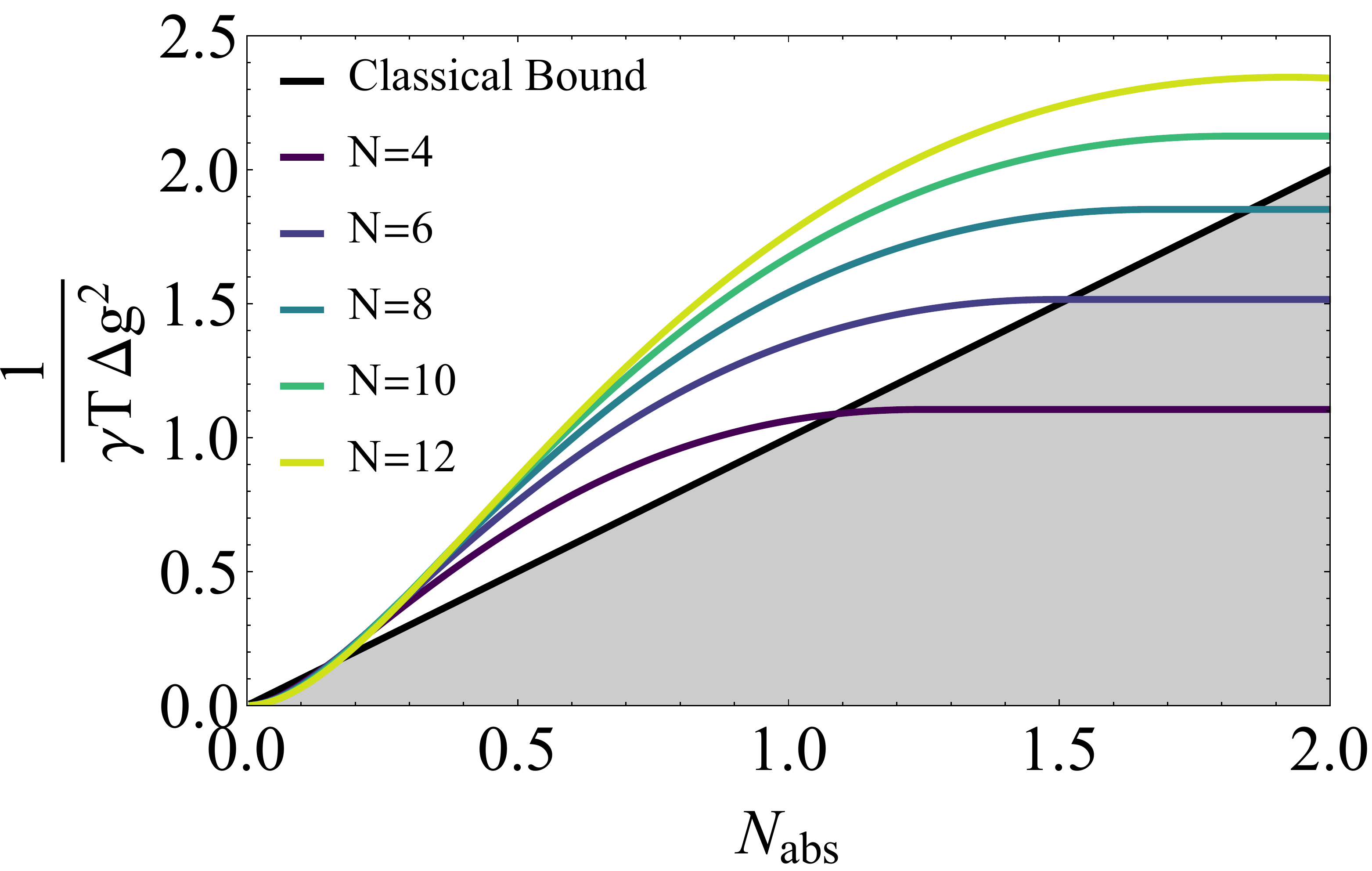}
  \caption{Plot of   $(\Delta g)^{-2}_{\rm TFS}=1$ (left) and $((\Delta g)^{-2}_{\rm NRM}=1$ as a function of $N_{\rm abs}$ for TFS of different $N$. For NRM, we consider $g\approx 0$.  We take $\eta=0.95$ and $\gamma t_{\rm ext}=0.05$}
  \label{figApp:tfscfi}
\end{figure}

\section{Measurement precision in cavity QED implementation}
\label{secap:implQED}
In this section we discuss the implementation in cavity QED, and in particular how to compute the measurement precision (both QFI and CFI).

At each run of the experiment, the preparation returns probabilistically a Fock state $\ket{m,l}$, where the photon numbers $m,l$ are known due to measurement of the remaining excited atoms, and are distributed according to the known distribution $q(k)$ [Eq.~\eqref{probqalpha}].
Given $m,l$, the interaction time $t_{m,l}$ between cavities is taken such that on average $N_{\rm abs}$ photons are absorbed, i.e.:
\begin{align}
    N_{\rm abs}= (m+l)(1-e^{-\gamma t_{m,l}}). 
\end{align}
We then repeat the experiment as often as possible in the given finite maximum time $T$, such that in the end $\sum_{m,l} \nu_{m,l} t_{m,l}\lesssim T$, where we defined $\nu_{m,l}$ which is the number of times a certain combination $m,l$ appeared in the experimental run.
%We assume the total time $T$ and $N_{\rm abs}$ to be limited.
We are interested in the regime $\nu \equiv \sum_{m,l}  \nu_{m,l}  \gg 1$ (i.e. $T/t_{m,l} \gg 1$), where the statistics are dominated by typical sequences in which  $ \nu_{m,l} = q(m) q(l) \nu$ which,  together with $T = \sum_{m,l} \nu_{m,l} t_{m,l}$, provides $\nu$. 
%and  the  interaction time $t$ between cavities are fixed and independent of  $m,l$. The average number of absorbed photon per sample is then given by:
%\begin{align}
%N_{\rm abs} = \sum_{m,l} q(m) q(l) (m+l) (1-e^{-\gamma t}),
%\end{align}
%or equivalently,
%\begin{align}
%N_{\rm abs} = (1-e^{-\gamma t}) 2 \langle m \rangle  ,
%\end{align}
%with $\langle m \rangle =\langle l \rangle  \approx 0.78n$.

%\subsection{Quantum Fisher Information}

\begin{figure}[t]
  \centering
  \includegraphics[width=0.5\linewidth]{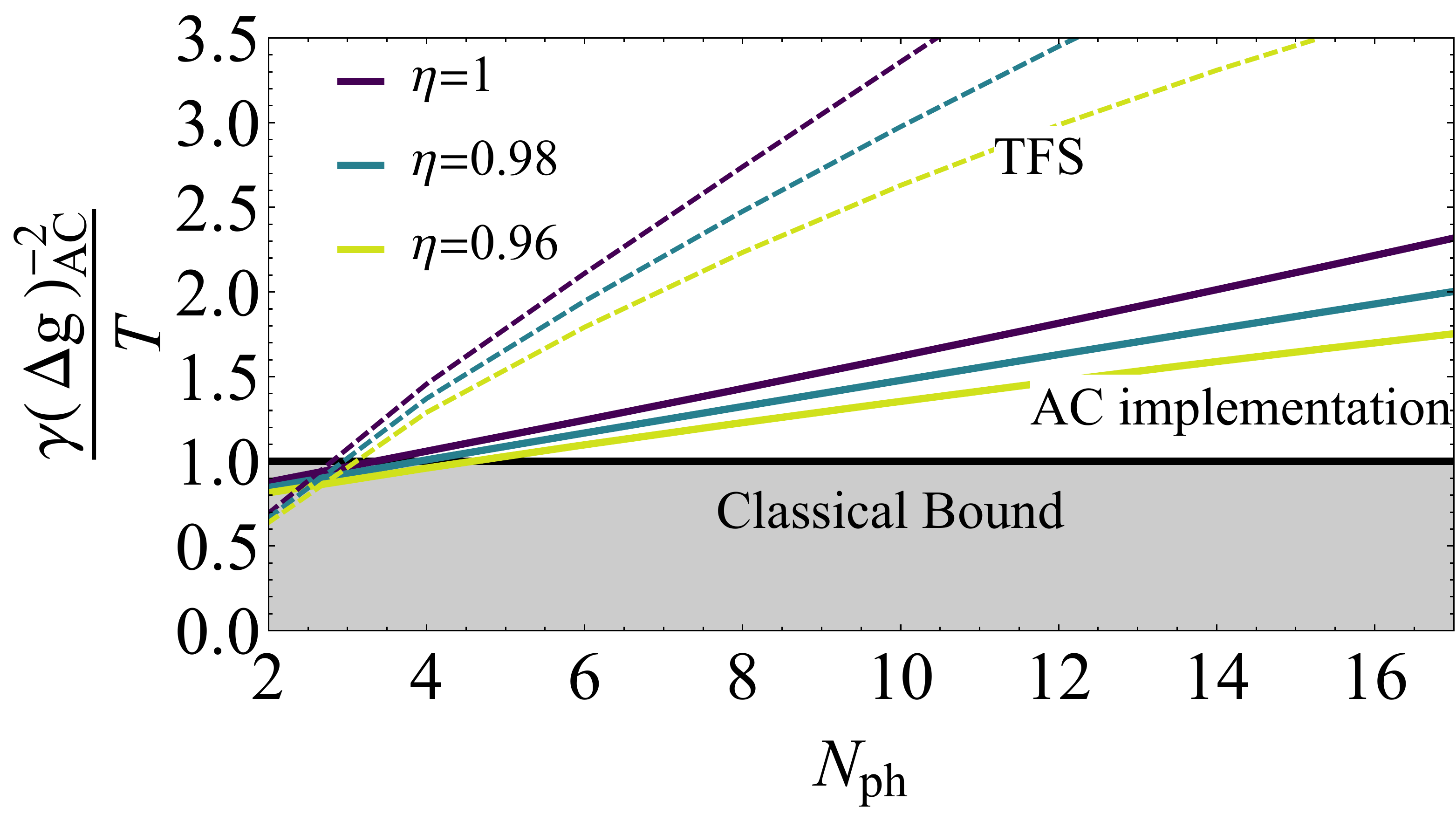}
    \caption{Plot of   $(\Delta g)^{-2}_{\rm AC}=1$ for TFS (dashed) and the atom-cavity (AC) implementation (solid lines)  as a function of $n$, and given different values of $\eta$ (we take $t_{\rm ext}=0$). We take $N_{\rm at}=100$ for the AC implementation, and for both cases we consider NRM with $g\approx 0$.
  }
  \label{figApp:tfscfiIIac}
\end{figure}

Assuming a large number of runs of the experiment $\nu\gg 1$ (so that enough statistics for each $m,l$ are obtained), the measurement uncertainty $(\Delta g)^{-2}$ is bounded by
\begin{align}
(\Delta g)^{-2}_{\rm AC, \mathcal{Q}} \equiv  \sum_{m,l} \nu_{m,l}  \mathcal{F}_{\rm Q}^{m,l} =  \frac{T}{ \sum_{m,l}q(m)q(l)t_{m,l}}\sum_{m,l} q(m) q(l) \mathcal{F}_{\rm Q}^{m,l} %\equiv (\Delta g)^{-2}_{\rm AC}
\label{DeltaVarphiDicke}
\end{align}
with $\mathcal{F}_{\rm Q}^{m,l}$ is the QFI for an initial Fock state $\ket{m,l}$, and it reads (following the same reasoning provided in the main text): 
\begin{equation}
\begin{aligned}
\mathcal{F}_{\rm Q}^{m,l}= 2t^2_{m,l} \sum_{i}^l \sum_j^m p_{i,l} p_{j,m}
&\left[(l-i) (m-j+1) \left(\frac{1}{2}- \frac{1}{1+\frac{(m-j+1)(i+1)}{j(l-i)}} \right)\right.\\
&+\left.(m-j) (l-i+1) \left(\frac{1}{2}- \frac{1}{1+\frac{(l-i+1)(j+1)}{i(m-j)}} \right) \right].
\label{Qfockml}
\end{aligned}
\end{equation}
with $p_{j,m}=\binom{m}{j} \mu^j (1-\mu)^{n-j}$ and $\mu=1-e^{-\gamma t}$.

We can compute $(\Delta g)^{-2}_{\rm NRM} \equiv \nu \mathcal{F}_{\mathcal{C}}$ using NRM in a similar fashion. For simplicity, we focus on the case $g\approx 0$, which can be solved analytically, but the case of general $g$ can be straightforwardly treated following our previous considerations for TFS. Analogously  with our previous considerations, we have,
 \begin{align}
(\Delta g)^{-2}_{\rm AC} =   \frac{T}{ \sum_{m,l}q(m)q(l)t_{m,l}}\sum_{m,l} q(m) q(l) \mathcal{F}_{\rm C}^{m,l}
\label{DeltaVarphiDickeII}
\end{align}
with (see Eq.~\eqref{eq:genexp}),
\begin{align}
    \mathcal{F}_{\rm C}^{m,l}= t^2_{m,l}\left( (1-\mu)^{l+1}(l+1)m+(1-\mu)^{m+1}(m+1)l\right)
\end{align}
with $\mu=1-e^{-\gamma t}\eta$.  
%In Fig. XX, we also compare 

We show $(\Delta g)^{-2}_{\rm AC}$ in Fig.~\ref{figApp:tfscfiIIac}, where it is compared with TFS. In our simulations, we compute $q(k)$ exactly by solving the Hamiltonian Eq. (12) of the main text. As expected, the probabilistic preparation of Fock states in our implementation diminishes its performance with respect to ideal TFS. Yet, we see that already at $n \approx 6$, quantum advantages appear possible, even in the presence of detector inefficiencies ($\eta<1$). The tolerance to imperfections (both $\eta$ and $t_{\rm ext}$ of $(\Delta g)^{-2}_{\rm AC}$ is shown in Fig.~\ref{figApp:tfscfiIIcas}.  

%In Figs. , we compare... 

\begin{figure}[t]
  \centering
  \includegraphics[width=0.5\linewidth]{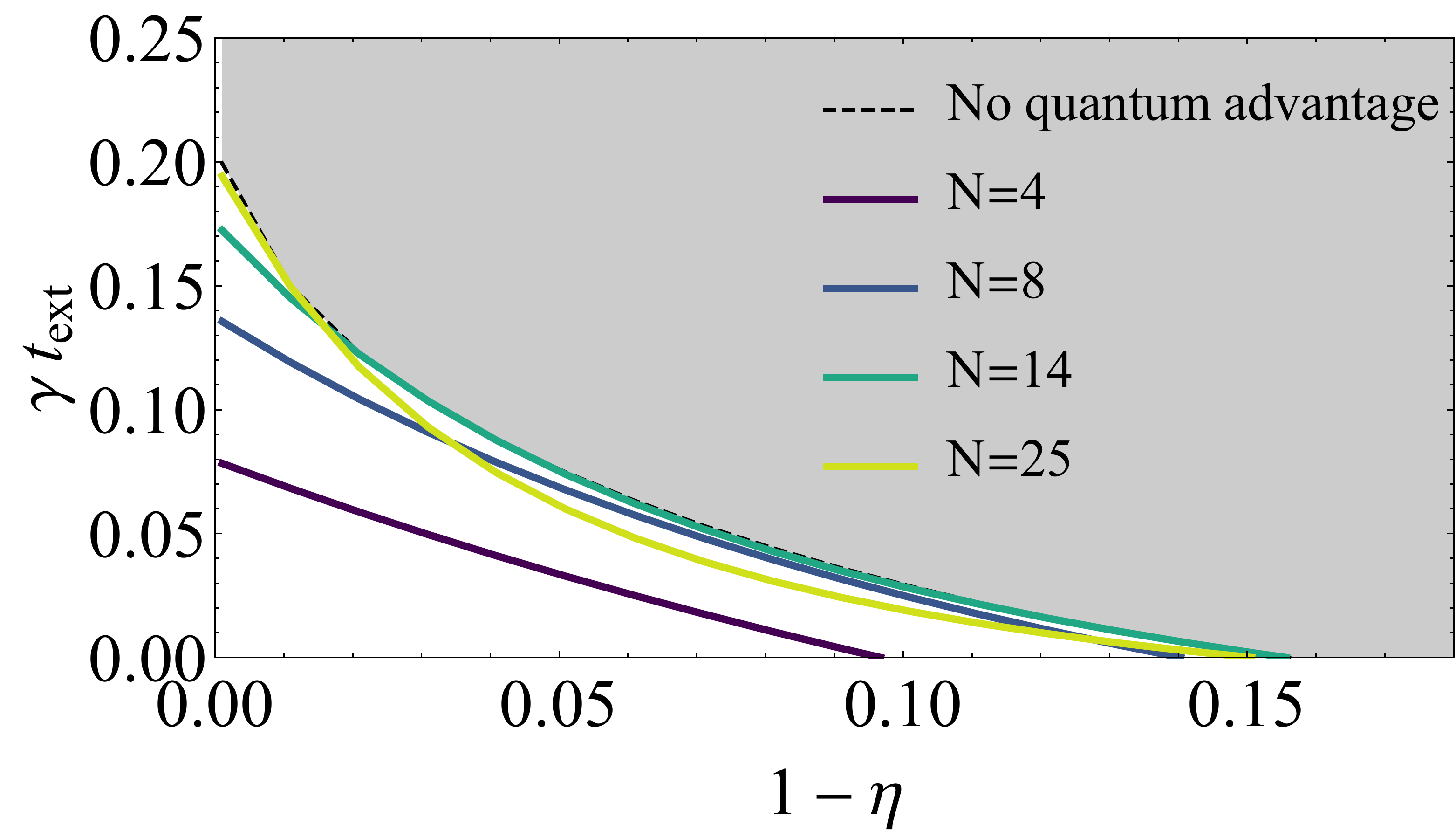}
\caption{We plot the values $\{\eta, t_{\rm ext} \}$ for which $(\Delta g)^{-2}_{\rm coh}/(\Delta g)^{-2}_{\rm AC}=1$  given different values  of  $n$, so that lower values of $\{\eta, t_{\rm ext} \}$ lead to quantum advantages. For NRM, we consider $g\approx 0$.
  }
  \label{figApp:tfscfiIIcas}
\end{figure}

\subsubsection{Error sources in the cavity QED and circuit QED implementation}

% Fock states are only generated probabilistically (see Fig.~\ref{fig:fig3}b), not all excitations decay into the cavity. 

In practice, there are several additional sources of error in this protocol that ought to be considered. 

First, we have assumed that all atoms/qubits are exactly on resonance with the cavity.
Realistically, they may have differing energies and be detuned from the cavity, which makes the photon exchange between cavity and atoms less efficient.
However, measuring the number of excited atoms after the interaction period still projects the cavity state onto a Fock state, such that disorder only affects the probability distribution of the produced Fock states, but not their fidelity.

Second, the fidelity of the generated Fock state is intrinsically limited by measurement error in the atomic readout.
High-fidelity atom readout has already been demonstrated for large numbers of atoms in cavity QED~\cite{Zhang2012,Hume2013}
and for intermediate numbers of qubits in circuit QED~\cite{Jeffrey2014,Walter2017,Dassonneville2020}.

Third, the atoms need to be reset to ground state in a time much faster than $1/g$. This is feasible if the cavity coupling is tunable (e.g., via an intermediate resonator in circuit QED) or if a fast decay channel is used (e.g., a fast optical transition of the atom).

Fourth, no photons may be lost throughout the experiment. This requires ultra-high finesse cavities or long qubit coherence times on the scale of $1/g$, i.e., this requires the strong coupling regime, which has been achieved in both cavity and circuit QED~\cite{Thompson1992,Brune1996,Yoshie2004,Wallraff2004,Reithmaier2004,Chiorescu2004}.
While the measurement is more challenging than the preparation, it relies on the same experimental figures of merit. A large advantage of our scheme is that one can post-select on successful runs of the experiment by requiring that at the end $N_a+N_b$ excitations are measured.

%\subsubsection{Fidelity of the produced intracavity state}

As an illustration, let us take the parameters from the experiment by Haas et al.~\cite{Haas2014}. These authors placed 40 atoms into a cavity of linewidth $\kappa=2\pi 52$MHz with an atom--cavity coupling of $J=2\pi 170$MHz. Given a time $\tau$, the probability that a photon leaks out of the cavity is hence $1-e^{-\tau \kappa }$, and the efficiency $\eta$ defined in \eqref{modelnoise} is given by $\eta=e^{-\tau \kappa }$.  
If we aim to prepare on average $n=8$ photons, Eq.~\eqref{eq:apptopt} gives a preparation time of $t_\mathrm{prep}=0.26$ns. The measurement time is given by \eqref{tmeas}, which for $n=8$ gives $t_{\rm meas}\approx  0.89 t_{\rm prep}$. Hence, we have $\eta = e^{- (t_{\rm prep}+t_{\rm meas}) \kappa } \approx 0.85 $ for the overall process of preparation and measurement of 8 photons. On top of that, there is a probability $p_{\rm error} \approx 0.04$ that one of the photons stays in the cavity (c.f. Fig. \ref{fig:fig3}), but from \eqref{perror} we see this has a negligible effect on the total efficiency $\eta$. Assuming that the atoms can be measured perfectly (current experiments show more than 99\% measurement fidelity~\cite{Zhang2012,Hume2013}), this leads to an overall measurement efficiency of $\eta  \approx 0.85 $. This can be contrasted with Fig.  \ref{figApp:tfscfiIIcas}, where we note that this experiment is on the verge of enabling quantum advantages in the proposed set-up. In order to increase the value of $\eta$, a natural strategy is to reduce the linewidth of the cavity (increasing the number of atoms would also result in a larger $\eta$, as discussed in the main text). 
In this sense, we should emphasize here that the experiment by Haas et al.\ does not include tuneable atom--cavity coupling and thus cannot directly be used to implement the scheme presented in the main text. To engineer a tuneable coupling, a Raman scheme has to be employed, and correspondingly much narrower cavities should be used~\cite{Reiserer2015}, which will significantly enhance the preparation fidelity.
%The probability that one  photon is lost  in this time is given by $1-\exp(-\kappa t_\mathrm{prep})$, and  is thus 7\% (the probability of not losing any of the 5 photons in this time, i.e., the fidelity of the prepared state after heralding is thus 68\%).  For the measurement,   the measurement time is given by \eqref{tmeas}, which for $n=5$ gives $t_{\rm meas}\approx  t_{\rm prep} $, so that the probabilty of not losing a photon is again 7\%.
%In this case, the probability of losing a single photon is thus 7\%, and hence .
%The probability of not losing a photon in this time, i.e., the fidelity of the prepared state after heralding is thus 68\% (corresponding to a per-photon loss probability of 7\%).
%Assuming a 99\% measurement fidelity on the atoms (surpassed by experiment~\cite{Zhang2012,Hume2013}), this corresponds to an efficiency of the measurement of %$1-\eta \approx 0.07$, which can be contrasted with Fig. \ref{figApp:tfscfiIIcas}.
%We should emphasize here that the experiment by Haas et al.\ does not include tuneable atom--cavity coupling and thus cannot directly be used to implement the scheme presented in the main text. To engineer a tuneable coupling, a Raman scheme has to be employed, and correspondingly much narrower cavities should be used~\cite{Reiserer2015}, which will significantly enhance the preparation fidelity.
%REWRITE THIS AND CLARIFY CONNECTION WITH $\eta$ and $p_{\rm error}$

\subsubsection{Non-linear regime and optimal time scale}

As a final comment, we note that it is in principle possible to use this protocol in the non-linear regime $n = \mathcal{O}(N_{\rm at})$. Indeed,  it might be useful in particular experimental implementations to take most use of the atoms at hand. We hence here discuss a few properties of $q(k)$ and $t_{\rm prep}$ in this regime.

We start by investigating the average photon number $n(t)$ in the cavity during the preparation stage. We see that $n(t)$ reaches its maximum value at $n(t_{\rm opt}) \approx 0.8 N_{\rm at}$. The distribution $q(k)$ for that optimal time is also illustrated in Fig.~\ref{fig:fig3appp}.

\begin{figure*}
		\centering
		\includegraphics[width=0.45\linewidth]{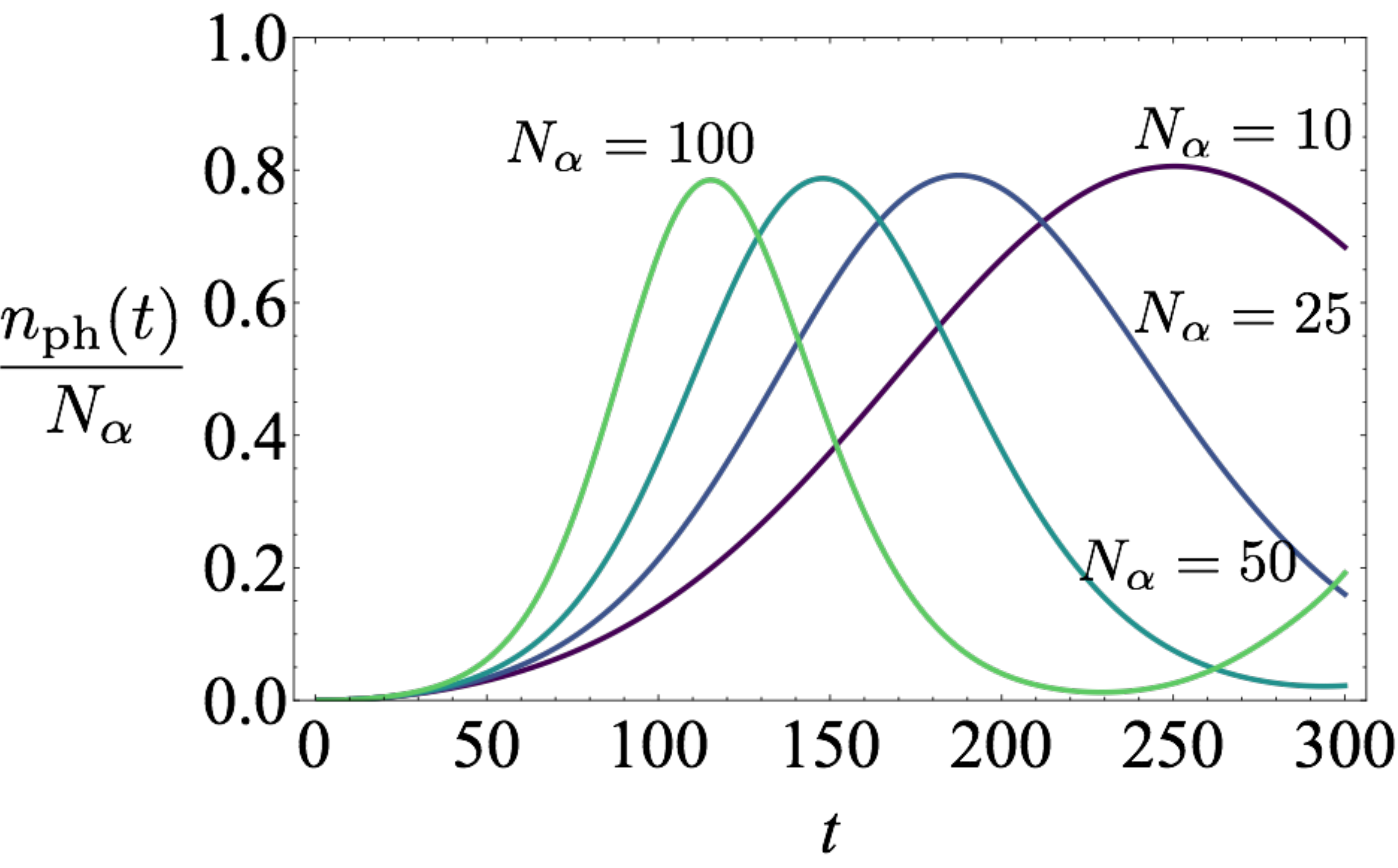} \hspace{1mm}
	    \includegraphics[width=0.4\linewidth]{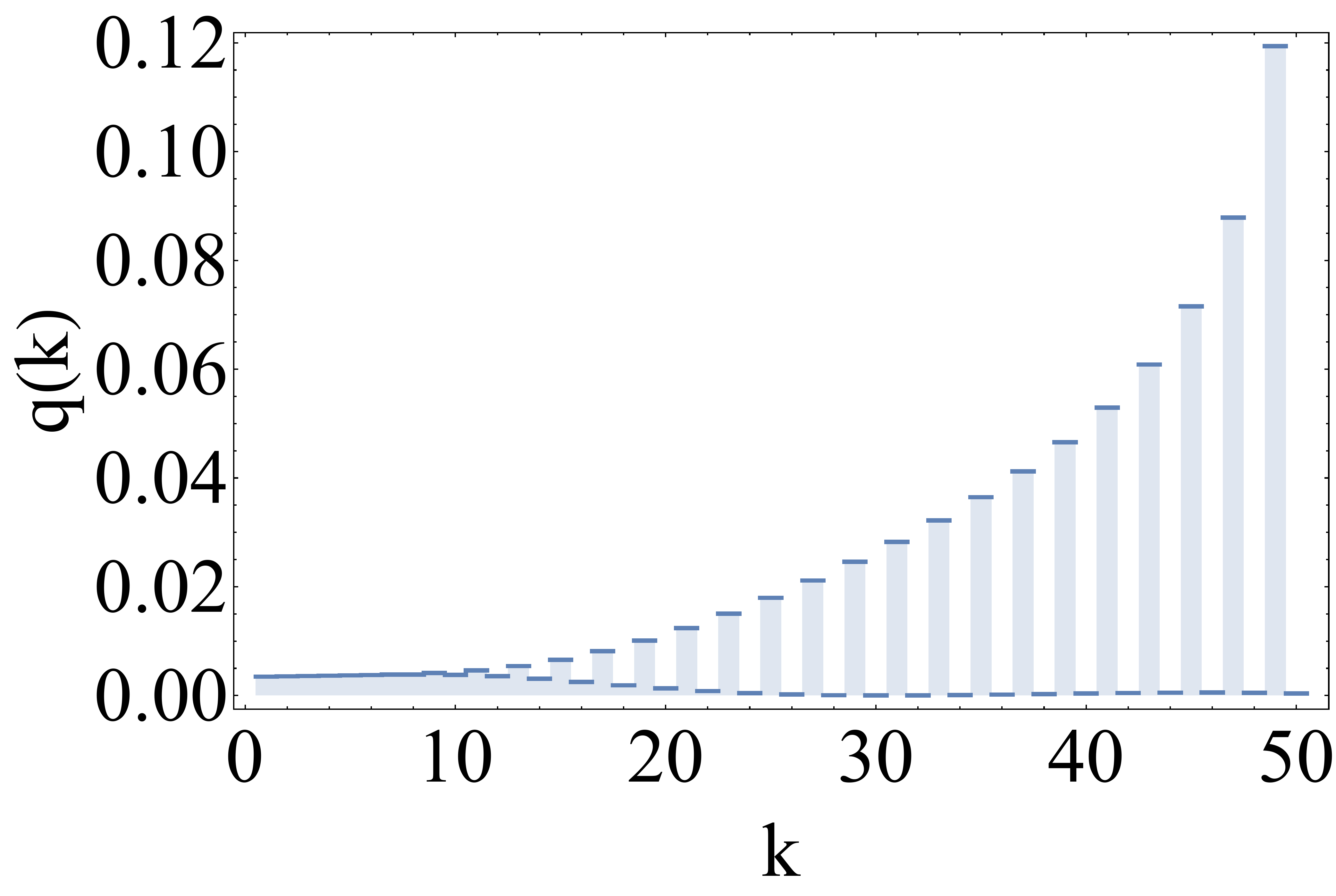} \hspace{2mm}
	    \caption{(a) Evolution of the average photon number $n(t)$  in the cavity  through the interaction (12). The initial state has $N_\alpha$ excited atoms and cavity in the ground state. The photon number is maximized at a time $t^*$ where $n(t^*)/N_\alpha \approx 0.8$. Parameters: $J=1$, $\omega_0=1$.  (b) Probability distribution of the photon number at $t^*$ for $N_\alpha=50$.  Parameters: $J=1$, $\omega_0=1$.  
 }
		\label{fig:fig3appp}
	\end{figure*}

Let us now show that $t_{\mathrm{opt}}$ is well approximated by $t_{\mathrm{opt}}\approx \log(4N_{\rm at})/(2J\sqrt{N_{\rm at}})$. Let us consider the Hamiltonian
\begin{equation}
  H_2=\frac12a b^2+\mathrm{H.c.},
  \label{eq:H_squeeze}
\end{equation}
where $a,b$ are the annihilation operators for two harmonic oscillators. $a$ corresponds to what formerly was a spin, and $b$ is just the same cavity mode as before. In the subspace spanned by the states $\ket{N-m}_a\otimes\ket{2m}_b$, where $\{\ket{m}\}$ are Fock states of an oscillator, the Hamiltonian is a bi-diagonal matrix
\begin{equation}
  	\mathcal H=\pmat{0 & a_0 & 0 &\cdots \\ a_0 & 0 & a_1 & \\ 0 & a_1 & 0 &\ddots \\
  	\vdots&& \ddots&\ddots}, 
  \label{eq:H_matrix}
\end{equation}
where
\begin{equation}
  a_m=\sqrt{(N-m)(m+1/2)(m+1)}\approx (m+1)\sqrt{N-m}.
  \label{eq:am}
\end{equation}
For large $N-m$ these are essentially the same matrix elements as the actual Hamiltonian (Eq. (12) in the main text) and one can check numerically that the dynamics is very similar for large enough $N$. In the classical limit, we replace the annihilation operators by the amplitudes of the corresponding coherent states. The classical (mean field) equations of motion read
\begin{equation}
  \frac{d}{dt}\begin{pmatrix}\alpha\\ \beta\end{pmatrix}=-i\begin{pmatrix}\beta^2/2\\ \beta^*\alpha\end{pmatrix},
  \label{eq:squeezing_eom}
\end{equation}
with initial conditions $\beta(0)=1$, $\alpha(0)=\sqrt{N}$.

\begin{figure}[tb]
  \centering
  \includegraphics[width=0.5\linewidth]{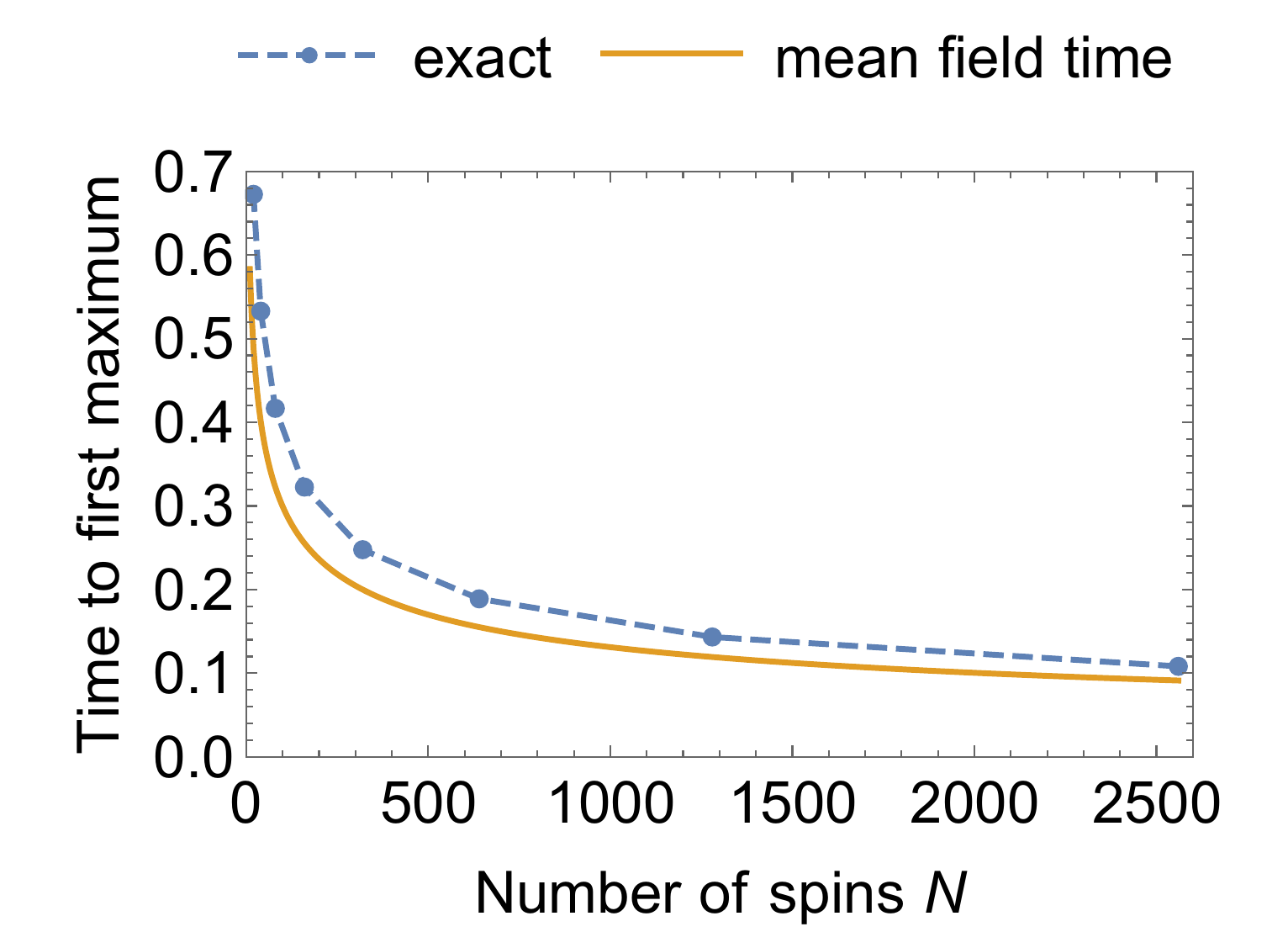}
  \caption{Comparison between the time scale obtained in mean field [Eq.~\eqref{eq:switch_timescale}] and the exact timescale calculated numerically.}
  \label{fig:time_scale}
\end{figure}

Now we do a change of variables, taking $\alpha=a\exp(ix)$, $\beta=b\exp(iy)$, such that
\begin{subequations}
  \begin{align}
  	\dot a &= -\frac{b^2}{2}\sin\phi,\qquad \dot x=-\frac{b^2}{2a}\cos\phi,\\   	\dot b &= +ab\sin\phi,\qquad      \dot y=-a\cos\phi,
 \end{align}
\end{subequations}
where $\phi=x-2y$.  Further defining the constant $K=a^2+b^2/2$, and the variable $r=a^2-b^2/2$, we obtain the equations of motion
\begin{subequations}
  \begin{align}
  	\dot\phi &= -\frac{K+3r}{\sqrt{2(K+r)}}\cos\phi ,\\
  	\dot r   &= \sqrt{2(K+r)}(K-r)\sin\phi.
\end{align}
\end{subequations}
This defines a vector field in a two-dimensional plane, whose closed orbits are the possible trajectories of the system. Since the phase $\phi$ is ill-defined in the initial state, we can choose it to be $\phi=\pi/2$.  In this case, the system is governed only by
\begin{equation}
  \dot r = \sqrt{2(K+r)}(K-r),
  \label{eq:switch_eom}
\end{equation}
which is readily integrated to yield
\begin{equation}
  \left[\tanh^{-1}\sqrt{\frac{1+r/K}{2}}\right]_{r(t_i)}^{r(t_f)}=\sqrt{2K}(t_f-t_i).
  \label{eq:swich_soln}
\end{equation}

Since initially, $r/K\simeq 1-2/N$ and $K\simeq N$ (taking $\alpha=\sqrt{iN}$ and $\beta=1$), we conclude that the natural timescale for the switch is
\begin{equation}
  T\simeq \frac{1}{2\sqrt N}\tanh^{-1}\left( 1-\frac{1}{4N} \right)\approx\frac{\log(4N)}{2\sqrt N}.
  \label{eq:switch_timescale}
\end{equation}
This converges to the actual optimal time scale in the limit of large $N$ as we show in Fig.~\ref{fig:time_scale}.

\newpage

\end{widetext}


\begin{thebibliography}{99}%
\makeatletter
\providecommand \@ifxundefined [1]{%
 \@ifx{#1\undefined}
}%
\providecommand \@ifnum [1]{%
 \ifnum #1\expandafter \@firstoftwo
 \else \expandafter \@secondoftwo
 \fi
}%
\providecommand \@ifx [1]{%
 \ifx #1\expandafter \@firstoftwo
 \else \expandafter \@secondoftwo
 \fi
}%
\providecommand \natexlab [1]{#1}%
\providecommand \enquote  [1]{``#1''}%
\providecommand \bibnamefont  [1]{#1}%
\providecommand \bibfnamefont [1]{#1}%
\providecommand \citenamefont [1]{#1}%
\providecommand \href@noop [0]{\@secondoftwo}%
\providecommand \href [0]{\begingroup \@sanitize@url \@href}%
\providecommand \@href[1]{\@@startlink{#1}\@@href}%
\providecommand \@@href[1]{\endgroup#1\@@endlink}%
\providecommand \@sanitize@url [0]{\catcode `\\12\catcode `\$12\catcode
  `\&12\catcode `\#12\catcode `\^12\catcode `\_12\catcode `\%12\relax}%
\providecommand \@@startlink[1]{}%
\providecommand \@@endlink[0]{}%
\providecommand \url  [0]{\begingroup\@sanitize@url \@url }%
\providecommand \@url [1]{\endgroup\@href {#1}{\urlprefix }}%
\providecommand \urlprefix  [0]{URL }%
\providecommand \Eprint [0]{\href }%
\providecommand \doibase [0]{http://dx.doi.org/}%
\providecommand \selectlanguage [0]{\@gobble}%
\providecommand \bibinfo  [0]{\@secondoftwo}%
\providecommand \bibfield  [0]{\@secondoftwo}%
\providecommand \translation [1]{[#1]}%
\providecommand \BibitemOpen [0]{}%
\providecommand \bibitemStop [0]{}%
\providecommand \bibitemNoStop [0]{.\EOS\space}%
\providecommand \EOS [0]{\spacefactor3000\relax}%
\providecommand \BibitemShut  [1]{\csname bibitem#1\endcsname}%
\let\auto@bib@innerbib\@empty
%</preamble>
\bibitem [{\citenamefont {Giovannetti}\ \emph {et~al.}(2011)\citenamefont
  {Giovannetti}, \citenamefont {Lloyd},\ and\ \citenamefont
  {Maccone}}]{Giovannetti2011}%
  \BibitemOpen
  \bibfield  {author} {\bibinfo {author} {\bibfnamefont {V.}~\bibnamefont
  {Giovannetti}}, \bibinfo {author} {\bibfnamefont {S.}~\bibnamefont {Lloyd}},
  \ and\ \bibinfo {author} {\bibfnamefont {L.}~\bibnamefont {Maccone}},\ }\href
  {\doibase 10.1038/nphoton.2011.35} {\bibfield  {journal} {\bibinfo  {journal}
  {Nature Photonics}\ }\textbf {\bibinfo {volume} {5}},\ \bibinfo {pages} {222}
  (\bibinfo {year} {2011})}\BibitemShut {NoStop}%
\bibitem [{\citenamefont {T{\'{o}}th}\ and\ \citenamefont
  {Apellaniz}(2014)}]{Tth2014}%
  \BibitemOpen
  \bibfield  {author} {\bibinfo {author} {\bibfnamefont {G.}~\bibnamefont
  {T{\'{o}}th}}\ and\ \bibinfo {author} {\bibfnamefont {I.}~\bibnamefont
  {Apellaniz}},\ }\href {\doibase 10.1088/1751-8113/47/42/424006} {\bibfield
  {journal} {\bibinfo  {journal} {Journal of Physics A: Mathematical and
  Theoretical}\ }\textbf {\bibinfo {volume} {47}},\ \bibinfo {pages} {424006}
  (\bibinfo {year} {2014})}\BibitemShut {NoStop}%
\bibitem [{\citenamefont {Demkowicz-Dobrza{\'{n}}ski}\ \emph
  {et~al.}(2015)\citenamefont {Demkowicz-Dobrza{\'{n}}ski}, \citenamefont
  {Jarzyna},\ and\ \citenamefont
  {Ko{\l}ody{\'{n}}ski}}]{demkowicz-dobrzanski15}%
  \BibitemOpen
  \bibfield  {author} {\bibinfo {author} {\bibfnamefont {R.}~\bibnamefont
  {Demkowicz-Dobrza{\'{n}}ski}}, \bibinfo {author} {\bibfnamefont
  {M.}~\bibnamefont {Jarzyna}}, \ and\ \bibinfo {author} {\bibfnamefont
  {J.}~\bibnamefont {Ko{\l}ody{\'{n}}ski}},\ }in\ \href {\doibase
  10.1016/bs.po.2015.02.003} {\emph {\bibinfo {booktitle} {Progress in
  Optics}}}\ (\bibinfo  {publisher} {Elsevier},\ \bibinfo {year} {2015})\ pp.\
  \bibinfo {pages} {345--435}\BibitemShut {NoStop}%
\bibitem [{\citenamefont {Dowling}\ and\ \citenamefont
  {Seshadreesan}(2015)}]{Dowling2015}%
  \BibitemOpen
  \bibfield  {author} {\bibinfo {author} {\bibfnamefont {J.~P.}\ \bibnamefont
  {Dowling}}\ and\ \bibinfo {author} {\bibfnamefont {K.~P.}\ \bibnamefont
  {Seshadreesan}},\ }\href {\doibase 10.1109/jlt.2014.2386795} {\bibfield
  {journal} {\bibinfo  {journal} {Journal of Lightwave Technology}\ }\textbf
  {\bibinfo {volume} {33}},\ \bibinfo {pages} {2359} (\bibinfo {year}
  {2015})}\BibitemShut {NoStop}%
\bibitem [{\citenamefont {Polino}\ \emph {et~al.}(2020)\citenamefont {Polino},
  \citenamefont {Valeri}, \citenamefont {Spagnolo},\ and\ \citenamefont
  {Sciarrino}}]{polino2020photonic}%
  \BibitemOpen
  \bibfield  {author} {\bibinfo {author} {\bibfnamefont {E.}~\bibnamefont
  {Polino}}, \bibinfo {author} {\bibfnamefont {M.}~\bibnamefont {Valeri}},
  \bibinfo {author} {\bibfnamefont {N.}~\bibnamefont {Spagnolo}}, \ and\
  \bibinfo {author} {\bibfnamefont {F.}~\bibnamefont {Sciarrino}},\ }\href
  {\doibase 10.1116/5.0007577} {\bibfield  {journal} {\bibinfo  {journal}
  {{AVS} Quantum Science}\ }\textbf {\bibinfo {volume} {2}},\ \bibinfo {pages}
  {024703} (\bibinfo {year} {2020})}\BibitemShut {NoStop}%
\bibitem [{\citenamefont {Caves}(1981)}]{Caves1981}%
  \BibitemOpen
  \bibfield  {author} {\bibinfo {author} {\bibfnamefont {C.~M.}\ \bibnamefont
  {Caves}},\ }\href {\doibase 10.1103/PhysRevD.23.1693} {\bibfield  {journal}
  {\bibinfo  {journal} {Phys. Rev. D}\ }\textbf {\bibinfo {volume} {23}},\
  \bibinfo {pages} {1693} (\bibinfo {year} {1981})}\BibitemShut {NoStop}%
\bibitem [{\citenamefont {Holland}\ and\ \citenamefont
  {Burnett}(1993)}]{holland93}%
  \BibitemOpen
  \bibfield  {author} {\bibinfo {author} {\bibfnamefont {M.~J.}\ \bibnamefont
  {Holland}}\ and\ \bibinfo {author} {\bibfnamefont {K.}~\bibnamefont
  {Burnett}},\ }\href {\doibase 10.1103/PhysRevLett.71.1355} {\bibfield
  {journal} {\bibinfo  {journal} {Phys. Rev. Lett.}\ }\textbf {\bibinfo
  {volume} {71}},\ \bibinfo {pages} {1355} (\bibinfo {year}
  {1993})}\BibitemShut {NoStop}%
\bibitem [{\citenamefont {Bollinger}\ \emph {et~al.}(1996)\citenamefont
  {Bollinger}, \citenamefont {Itano}, \citenamefont {Wineland},\ and\
  \citenamefont {Heinzen}}]{Bollinger1996}%
  \BibitemOpen
  \bibfield  {author} {\bibinfo {author} {\bibfnamefont {J.~J.~.}\ \bibnamefont
  {Bollinger}}, \bibinfo {author} {\bibfnamefont {W.~M.}\ \bibnamefont
  {Itano}}, \bibinfo {author} {\bibfnamefont {D.~J.}\ \bibnamefont {Wineland}},
  \ and\ \bibinfo {author} {\bibfnamefont {D.~J.}\ \bibnamefont {Heinzen}},\
  }\href {\doibase 10.1103/PhysRevA.54.R4649} {\bibfield  {journal} {\bibinfo
  {journal} {Phys. Rev. A}\ }\textbf {\bibinfo {volume} {54}},\ \bibinfo
  {pages} {R4649} (\bibinfo {year} {1996})}\BibitemShut {NoStop}%
\bibitem [{\citenamefont {Neuman}\ \emph {et~al.}(1999)\citenamefont {Neuman},
  \citenamefont {Chadd}, \citenamefont {Liou}, \citenamefont {Bergman},\ and\
  \citenamefont {Block}}]{Neuman1999}%
  \BibitemOpen
  \bibfield  {author} {\bibinfo {author} {\bibfnamefont {K.~C.}\ \bibnamefont
  {Neuman}}, \bibinfo {author} {\bibfnamefont {E.~H.}\ \bibnamefont {Chadd}},
  \bibinfo {author} {\bibfnamefont {G.~F.}\ \bibnamefont {Liou}}, \bibinfo
  {author} {\bibfnamefont {K.}~\bibnamefont {Bergman}}, \ and\ \bibinfo
  {author} {\bibfnamefont {S.~M.}\ \bibnamefont {Block}},\ }\href {\doibase
  10.1016/s0006-3495(99)77117-1} {\bibfield  {journal} {\bibinfo  {journal}
  {Biophysical Journal}\ }\textbf {\bibinfo {volume} {77}},\ \bibinfo {pages}
  {2856} (\bibinfo {year} {1999})}\BibitemShut {NoStop}%
\bibitem [{\citenamefont {Peterman}\ \emph {et~al.}(2003)\citenamefont
  {Peterman}, \citenamefont {Gittes},\ and\ \citenamefont
  {Schmidt}}]{Peterman2003}%
  \BibitemOpen
  \bibfield  {author} {\bibinfo {author} {\bibfnamefont {E.~J.}\ \bibnamefont
  {Peterman}}, \bibinfo {author} {\bibfnamefont {F.}~\bibnamefont {Gittes}}, \
  and\ \bibinfo {author} {\bibfnamefont {C.~F.}\ \bibnamefont {Schmidt}},\
  }\href {\doibase 10.1016/s0006-3495(03)74946-7} {\bibfield  {journal}
  {\bibinfo  {journal} {Biophysical Journal}\ }\textbf {\bibinfo {volume}
  {84}},\ \bibinfo {pages} {1308} (\bibinfo {year} {2003})}\BibitemShut
  {NoStop}%
\bibitem [{\citenamefont {Cole}(2014)}]{Cole2014}%
  \BibitemOpen
  \bibfield  {author} {\bibinfo {author} {\bibfnamefont {R.}~\bibnamefont
  {Cole}},\ }\href {\doibase 10.4161/cam.28348} {\bibfield  {journal} {\bibinfo
   {journal} {Cell Adhesion {\&} Migration}\ }\textbf {\bibinfo {volume} {8}},\
  \bibinfo {pages} {452} (\bibinfo {year} {2014})}\BibitemShut {NoStop}%
\bibitem [{\citenamefont {Taylor}\ \emph {et~al.}(2013)\citenamefont {Taylor},
  \citenamefont {Janousek}, \citenamefont {Daria}, \citenamefont {Knittel},
  \citenamefont {Hage}, \citenamefont {Bachor},\ and\ \citenamefont
  {Bowen}}]{Taylor2013}%
  \BibitemOpen
  \bibfield  {author} {\bibinfo {author} {\bibfnamefont {M.~A.}\ \bibnamefont
  {Taylor}}, \bibinfo {author} {\bibfnamefont {J.}~\bibnamefont {Janousek}},
  \bibinfo {author} {\bibfnamefont {V.}~\bibnamefont {Daria}}, \bibinfo
  {author} {\bibfnamefont {J.}~\bibnamefont {Knittel}}, \bibinfo {author}
  {\bibfnamefont {B.}~\bibnamefont {Hage}}, \bibinfo {author} {\bibfnamefont
  {H.-A.}\ \bibnamefont {Bachor}}, \ and\ \bibinfo {author} {\bibfnamefont
  {W.~P.}\ \bibnamefont {Bowen}},\ }\href {\doibase 10.1038/nphoton.2012.346}
  {\bibfield  {journal} {\bibinfo  {journal} {Nature Photonics}\ }\textbf
  {\bibinfo {volume} {7}},\ \bibinfo {pages} {229} (\bibinfo {year}
  {2013})}\BibitemShut {NoStop}%
\bibitem [{\citenamefont {Taylor}\ \emph {et~al.}(2014)\citenamefont {Taylor},
  \citenamefont {Janousek}, \citenamefont {Daria}, \citenamefont {Knittel},
  \citenamefont {Hage}, \citenamefont {Bachor},\ and\ \citenamefont
  {Bowen}}]{Taylor2014}%
  \BibitemOpen
  \bibfield  {author} {\bibinfo {author} {\bibfnamefont {M.~A.}\ \bibnamefont
  {Taylor}}, \bibinfo {author} {\bibfnamefont {J.}~\bibnamefont {Janousek}},
  \bibinfo {author} {\bibfnamefont {V.}~\bibnamefont {Daria}}, \bibinfo
  {author} {\bibfnamefont {J.}~\bibnamefont {Knittel}}, \bibinfo {author}
  {\bibfnamefont {B.}~\bibnamefont {Hage}}, \bibinfo {author} {\bibfnamefont
  {H.-A.}\ \bibnamefont {Bachor}}, \ and\ \bibinfo {author} {\bibfnamefont
  {W.~P.}\ \bibnamefont {Bowen}},\ }\href {\doibase 10.1103/PhysRevX.4.011017}
  {\bibfield  {journal} {\bibinfo  {journal} {Phys. Rev. X}\ }\textbf {\bibinfo
  {volume} {4}},\ \bibinfo {pages} {011017} (\bibinfo {year}
  {2014})}\BibitemShut {NoStop}%
\bibitem [{\citenamefont {Taylor}\ and\ \citenamefont
  {Bowen}(2016)}]{Taylor2016}%
  \BibitemOpen
  \bibfield  {author} {\bibinfo {author} {\bibfnamefont {M.~A.}\ \bibnamefont
  {Taylor}}\ and\ \bibinfo {author} {\bibfnamefont {W.~P.}\ \bibnamefont
  {Bowen}},\ }\href {\doibase 10.1016/j.physrep.2015.12.002} {\bibfield
  {journal} {\bibinfo  {journal} {Physics Reports}\ }\textbf {\bibinfo {volume}
  {615}},\ \bibinfo {pages} {1} (\bibinfo {year} {2016})}\BibitemShut {NoStop}%
\bibitem [{\citenamefont {Whittaker}\ \emph {et~al.}(2017)\citenamefont
  {Whittaker}, \citenamefont {Erven}, \citenamefont {Neville}, \citenamefont
  {Berry}, \citenamefont {O'Brien}, \citenamefont {Cable},\ and\ \citenamefont
  {Matthews}}]{Whittaker2017}%
  \BibitemOpen
  \bibfield  {author} {\bibinfo {author} {\bibfnamefont {R.}~\bibnamefont
  {Whittaker}}, \bibinfo {author} {\bibfnamefont {C.}~\bibnamefont {Erven}},
  \bibinfo {author} {\bibfnamefont {A.}~\bibnamefont {Neville}}, \bibinfo
  {author} {\bibfnamefont {M.}~\bibnamefont {Berry}}, \bibinfo {author}
  {\bibfnamefont {J.~L.}\ \bibnamefont {O'Brien}}, \bibinfo {author}
  {\bibfnamefont {H.}~\bibnamefont {Cable}}, \ and\ \bibinfo {author}
  {\bibfnamefont {J.~C.~F.}\ \bibnamefont {Matthews}},\ }\href {\doibase
  10.1088/1367-2630/aa5512} {\bibfield  {journal} {\bibinfo  {journal} {New
  Journal of Physics}\ }\textbf {\bibinfo {volume} {19}},\ \bibinfo {pages}
  {023013} (\bibinfo {year} {2017})}\BibitemShut {NoStop}%
\bibitem [{\citenamefont {Pototschnig}\ \emph {et~al.}(2011)\citenamefont
  {Pototschnig}, \citenamefont {Chassagneux}, \citenamefont {Hwang},
  \citenamefont {Zumofen}, \citenamefont {Renn},\ and\ \citenamefont
  {Sandoghdar}}]{Pototschnig2011}%
  \BibitemOpen
  \bibfield  {author} {\bibinfo {author} {\bibfnamefont {M.}~\bibnamefont
  {Pototschnig}}, \bibinfo {author} {\bibfnamefont {Y.}~\bibnamefont
  {Chassagneux}}, \bibinfo {author} {\bibfnamefont {J.}~\bibnamefont {Hwang}},
  \bibinfo {author} {\bibfnamefont {G.}~\bibnamefont {Zumofen}}, \bibinfo
  {author} {\bibfnamefont {A.}~\bibnamefont {Renn}}, \ and\ \bibinfo {author}
  {\bibfnamefont {V.}~\bibnamefont {Sandoghdar}},\ }\href {\doibase
  10.1103/PhysRevLett.107.063001} {\bibfield  {journal} {\bibinfo  {journal}
  {Phys. Rev. Lett.}\ }\textbf {\bibinfo {volume} {107}},\ \bibinfo {pages}
  {063001} (\bibinfo {year} {2011})}\BibitemShut {NoStop}%
\bibitem [{\citenamefont {Eckert}\ \emph {et~al.}(2007)\citenamefont {Eckert},
  \citenamefont {Romero-Isart}, \citenamefont {Rodriguez}, \citenamefont
  {Lewenstein}, \citenamefont {Polzik},\ and\ \citenamefont
  {Sanpera}}]{Eckert2007}%
  \BibitemOpen
  \bibfield  {author} {\bibinfo {author} {\bibfnamefont {K.}~\bibnamefont
  {Eckert}}, \bibinfo {author} {\bibfnamefont {O.}~\bibnamefont
  {Romero-Isart}}, \bibinfo {author} {\bibfnamefont {M.}~\bibnamefont
  {Rodriguez}}, \bibinfo {author} {\bibfnamefont {M.}~\bibnamefont
  {Lewenstein}}, \bibinfo {author} {\bibfnamefont {E.~S.}\ \bibnamefont
  {Polzik}}, \ and\ \bibinfo {author} {\bibfnamefont {A.}~\bibnamefont
  {Sanpera}},\ }\href {\doibase 10.1038/nphys776} {\bibfield  {journal}
  {\bibinfo  {journal} {Nature Physics}\ }\textbf {\bibinfo {volume} {4}},\
  \bibinfo {pages} {50} (\bibinfo {year} {2007})}\BibitemShut {NoStop}%
\bibitem [{\citenamefont {Wolfgramm}\ \emph {et~al.}(2012)\citenamefont
  {Wolfgramm}, \citenamefont {Vitelli}, \citenamefont {Beduini}, \citenamefont
  {Godbout},\ and\ \citenamefont {Mitchell}}]{Wolfgramm2012}%
  \BibitemOpen
  \bibfield  {author} {\bibinfo {author} {\bibfnamefont {F.}~\bibnamefont
  {Wolfgramm}}, \bibinfo {author} {\bibfnamefont {C.}~\bibnamefont {Vitelli}},
  \bibinfo {author} {\bibfnamefont {F.~A.}\ \bibnamefont {Beduini}}, \bibinfo
  {author} {\bibfnamefont {N.}~\bibnamefont {Godbout}}, \ and\ \bibinfo
  {author} {\bibfnamefont {M.~W.}\ \bibnamefont {Mitchell}},\ }\href {\doibase
  10.1038/nphoton.2012.300} {\bibfield  {journal} {\bibinfo  {journal} {Nature
  Photonics}\ }\textbf {\bibinfo {volume} {7}},\ \bibinfo {pages} {28}
  (\bibinfo {year} {2012})}\BibitemShut {NoStop}%
\bibitem [{\citenamefont {Ono}\ \emph {et~al.}(2013)\citenamefont {Ono},
  \citenamefont {Okamoto},\ and\ \citenamefont {Takeuchi}}]{Ono2013}%
  \BibitemOpen
  \bibfield  {author} {\bibinfo {author} {\bibfnamefont {T.}~\bibnamefont
  {Ono}}, \bibinfo {author} {\bibfnamefont {R.}~\bibnamefont {Okamoto}}, \ and\
  \bibinfo {author} {\bibfnamefont {S.}~\bibnamefont {Takeuchi}},\ }\href
  {\doibase 10.1038/ncomms3426} {\bibfield  {journal} {\bibinfo  {journal}
  {Nature Communications}\ }\textbf {\bibinfo {volume} {4}} (\bibinfo {year}
  {2013}),\ 10.1038/ncomms3426}\BibitemShut {NoStop}%
\bibitem [{\citenamefont {Huelga}\ \emph {et~al.}(1997)\citenamefont {Huelga},
  \citenamefont {Macchiavello}, \citenamefont {Pellizzari}, \citenamefont
  {Ekert}, \citenamefont {Plenio},\ and\ \citenamefont {Cirac}}]{Huelga1997}%
  \BibitemOpen
  \bibfield  {author} {\bibinfo {author} {\bibfnamefont {S.~F.}\ \bibnamefont
  {Huelga}}, \bibinfo {author} {\bibfnamefont {C.}~\bibnamefont
  {Macchiavello}}, \bibinfo {author} {\bibfnamefont {T.}~\bibnamefont
  {Pellizzari}}, \bibinfo {author} {\bibfnamefont {A.~K.}\ \bibnamefont
  {Ekert}}, \bibinfo {author} {\bibfnamefont {M.~B.}\ \bibnamefont {Plenio}}, \
  and\ \bibinfo {author} {\bibfnamefont {J.~I.}\ \bibnamefont {Cirac}},\ }\href
  {\doibase 10.1103/PhysRevLett.79.3865} {\bibfield  {journal} {\bibinfo
  {journal} {Phys. Rev. Lett.}\ }\textbf {\bibinfo {volume} {79}},\ \bibinfo
  {pages} {3865} (\bibinfo {year} {1997})}\BibitemShut {NoStop}%
\bibitem [{\citenamefont {Shaji}\ and\ \citenamefont
  {Caves}(2007)}]{Shaji2007}%
  \BibitemOpen
  \bibfield  {author} {\bibinfo {author} {\bibfnamefont {A.}~\bibnamefont
  {Shaji}}\ and\ \bibinfo {author} {\bibfnamefont {C.~M.}\ \bibnamefont
  {Caves}},\ }\href {\doibase 10.1103/PhysRevA.76.032111} {\bibfield  {journal}
  {\bibinfo  {journal} {Phys. Rev. A}\ }\textbf {\bibinfo {volume} {76}},\
  \bibinfo {pages} {032111} (\bibinfo {year} {2007})}\BibitemShut {NoStop}%
\bibitem [{\citenamefont {Hayes}\ \emph {et~al.}(2018)\citenamefont {Hayes},
  \citenamefont {Dooley}, \citenamefont {Munro}, \citenamefont {Nemoto},\ and\
  \citenamefont {Dunningham}}]{Hayes2018}%
  \BibitemOpen
  \bibfield  {author} {\bibinfo {author} {\bibfnamefont {A.~J.}\ \bibnamefont
  {Hayes}}, \bibinfo {author} {\bibfnamefont {S.}~\bibnamefont {Dooley}},
  \bibinfo {author} {\bibfnamefont {W.~J.}\ \bibnamefont {Munro}}, \bibinfo
  {author} {\bibfnamefont {K.}~\bibnamefont {Nemoto}}, \ and\ \bibinfo {author}
  {\bibfnamefont {J.}~\bibnamefont {Dunningham}},\ }\href {\doibase
  10.1088/2058-9565/aac30b} {\bibfield  {journal} {\bibinfo  {journal} {Quantum
  Science and Technology}\ }\textbf {\bibinfo {volume} {3}},\ \bibinfo {pages}
  {035007} (\bibinfo {year} {2018})}\BibitemShut {NoStop}%
\bibitem [{\citenamefont {Matsuzaki}\ \emph {et~al.}(2011)\citenamefont
  {Matsuzaki}, \citenamefont {Benjamin},\ and\ \citenamefont
  {Fitzsimons}}]{Matsuzaki2011}%
  \BibitemOpen
  \bibfield  {author} {\bibinfo {author} {\bibfnamefont {Y.}~\bibnamefont
  {Matsuzaki}}, \bibinfo {author} {\bibfnamefont {S.~C.}\ \bibnamefont
  {Benjamin}}, \ and\ \bibinfo {author} {\bibfnamefont {J.}~\bibnamefont
  {Fitzsimons}},\ }\href {\doibase 10.1103/PhysRevA.84.012103} {\bibfield
  {journal} {\bibinfo  {journal} {Phys. Rev. A}\ }\textbf {\bibinfo {volume}
  {84}},\ \bibinfo {pages} {012103} (\bibinfo {year} {2011})}\BibitemShut
  {NoStop}%
\bibitem [{\citenamefont {Chin}\ \emph {et~al.}(2012)\citenamefont {Chin},
  \citenamefont {Huelga},\ and\ \citenamefont {Plenio}}]{Chin2012}%
  \BibitemOpen
  \bibfield  {author} {\bibinfo {author} {\bibfnamefont {A.~W.}\ \bibnamefont
  {Chin}}, \bibinfo {author} {\bibfnamefont {S.~F.}\ \bibnamefont {Huelga}}, \
  and\ \bibinfo {author} {\bibfnamefont {M.~B.}\ \bibnamefont {Plenio}},\
  }\href {\doibase 10.1103/PhysRevLett.109.233601} {\bibfield  {journal}
  {\bibinfo  {journal} {Phys. Rev. Lett.}\ }\textbf {\bibinfo {volume} {109}},\
  \bibinfo {pages} {233601} (\bibinfo {year} {2012})}\BibitemShut {NoStop}%
\bibitem [{\citenamefont {Chaves}\ \emph {et~al.}(2013)\citenamefont {Chaves},
  \citenamefont {Brask}, \citenamefont {Markiewicz}, \citenamefont
  {Ko\l{}ody\ifmmode~\acute{n}\else \'{n}\fi{}ski},\ and\ \citenamefont
  {Ac\'{\i}n}}]{Chaves2013}%
  \BibitemOpen
  \bibfield  {author} {\bibinfo {author} {\bibfnamefont {R.}~\bibnamefont
  {Chaves}}, \bibinfo {author} {\bibfnamefont {J.~B.}\ \bibnamefont {Brask}},
  \bibinfo {author} {\bibfnamefont {M.}~\bibnamefont {Markiewicz}}, \bibinfo
  {author} {\bibfnamefont {J.}~\bibnamefont {Ko\l{}ody\ifmmode~\acute{n}\else
  \'{n}\fi{}ski}}, \ and\ \bibinfo {author} {\bibfnamefont {A.}~\bibnamefont
  {Ac\'{\i}n}},\ }\href {\doibase 10.1103/PhysRevLett.111.120401} {\bibfield
  {journal} {\bibinfo  {journal} {Phys. Rev. Lett.}\ }\textbf {\bibinfo
  {volume} {111}},\ \bibinfo {pages} {120401} (\bibinfo {year}
  {2013})}\BibitemShut {NoStop}%
\bibitem [{\citenamefont {Brask}\ \emph {et~al.}(2015)\citenamefont {Brask},
  \citenamefont {Chaves},\ and\ \citenamefont {Ko\l{}ody\ifmmode~\acute{n}\else
  \'{n}\fi{}ski}}]{Brask2015}%
  \BibitemOpen
  \bibfield  {author} {\bibinfo {author} {\bibfnamefont {J.~B.}\ \bibnamefont
  {Brask}}, \bibinfo {author} {\bibfnamefont {R.}~\bibnamefont {Chaves}}, \
  and\ \bibinfo {author} {\bibfnamefont {J.}~\bibnamefont
  {Ko\l{}ody\ifmmode~\acute{n}\else \'{n}\fi{}ski}},\ }\href {\doibase
  10.1103/PhysRevX.5.031010} {\bibfield  {journal} {\bibinfo  {journal} {Phys.
  Rev. X}\ }\textbf {\bibinfo {volume} {5}},\ \bibinfo {pages} {031010}
  (\bibinfo {year} {2015})}\BibitemShut {NoStop}%
\bibitem [{\citenamefont {Smirne}\ \emph {et~al.}(2016)\citenamefont {Smirne},
  \citenamefont {Ko\l{}ody\ifmmode~\acute{n}\else \'{n}\fi{}ski}, \citenamefont
  {Huelga},\ and\ \citenamefont {Demkowicz-Dobrza\ifmmode~\acute{n}\else
  \'{n}\fi{}ski}}]{Smirne2016}%
  \BibitemOpen
  \bibfield  {author} {\bibinfo {author} {\bibfnamefont {A.}~\bibnamefont
  {Smirne}}, \bibinfo {author} {\bibfnamefont {J.}~\bibnamefont
  {Ko\l{}ody\ifmmode~\acute{n}\else \'{n}\fi{}ski}}, \bibinfo {author}
  {\bibfnamefont {S.~F.}\ \bibnamefont {Huelga}}, \ and\ \bibinfo {author}
  {\bibfnamefont {R.}~\bibnamefont {Demkowicz-Dobrza\ifmmode~\acute{n}\else
  \'{n}\fi{}ski}},\ }\href {\doibase 10.1103/PhysRevLett.116.120801} {\bibfield
   {journal} {\bibinfo  {journal} {Phys. Rev. Lett.}\ }\textbf {\bibinfo
  {volume} {116}},\ \bibinfo {pages} {120801} (\bibinfo {year}
  {2016})}\BibitemShut {NoStop}%
\bibitem [{\citenamefont {Woolley}\ \emph {et~al.}(2008)\citenamefont
  {Woolley}, \citenamefont {Milburn},\ and\ \citenamefont
  {Caves}}]{Woolley2008}%
  \BibitemOpen
  \bibfield  {author} {\bibinfo {author} {\bibfnamefont {M.~J.}\ \bibnamefont
  {Woolley}}, \bibinfo {author} {\bibfnamefont {G.~J.}\ \bibnamefont
  {Milburn}}, \ and\ \bibinfo {author} {\bibfnamefont {C.~M.}\ \bibnamefont
  {Caves}},\ }\href {\doibase 10.1088/1367-2630/10/12/125018} {\bibfield
  {journal} {\bibinfo  {journal} {New Journal of Physics}\ }\textbf {\bibinfo
  {volume} {10}},\ \bibinfo {pages} {125018} (\bibinfo {year}
  {2008})}\BibitemShut {NoStop}%
\bibitem [{\citenamefont {Beau}\ and\ \citenamefont {del
  Campo}(2017)}]{Beau2017}%
  \BibitemOpen
  \bibfield  {author} {\bibinfo {author} {\bibfnamefont {M.}~\bibnamefont
  {Beau}}\ and\ \bibinfo {author} {\bibfnamefont {A.}~\bibnamefont {del
  Campo}},\ }\href {\doibase 10.1103/PhysRevLett.119.010403} {\bibfield
  {journal} {\bibinfo  {journal} {Phys. Rev. Lett.}\ }\textbf {\bibinfo
  {volume} {119}},\ \bibinfo {pages} {010403} (\bibinfo {year}
  {2017})}\BibitemShut {NoStop}%
\bibitem [{\citenamefont {Naghiloo}\ \emph {et~al.}(2017)\citenamefont
  {Naghiloo}, \citenamefont {Jordan},\ and\ \citenamefont
  {Murch}}]{Naghiloo2017}%
  \BibitemOpen
  \bibfield  {author} {\bibinfo {author} {\bibfnamefont {M.}~\bibnamefont
  {Naghiloo}}, \bibinfo {author} {\bibfnamefont {A.~N.}\ \bibnamefont
  {Jordan}}, \ and\ \bibinfo {author} {\bibfnamefont {K.~W.}\ \bibnamefont
  {Murch}},\ }\href {\doibase 10.1103/PhysRevLett.119.180801} {\bibfield
  {journal} {\bibinfo  {journal} {Phys. Rev. Lett.}\ }\textbf {\bibinfo
  {volume} {119}},\ \bibinfo {pages} {180801} (\bibinfo {year}
  {2017})}\BibitemShut {NoStop}%
\bibitem [{\citenamefont {Pang}\ and\ \citenamefont {Jordan}(2017)}]{Pang2017}%
  \BibitemOpen
  \bibfield  {author} {\bibinfo {author} {\bibfnamefont {S.}~\bibnamefont
  {Pang}}\ and\ \bibinfo {author} {\bibfnamefont {A.~N.}\ \bibnamefont
  {Jordan}},\ }\href {\doibase 10.1038/ncomms14695} {\bibfield  {journal}
  {\bibinfo  {journal} {Nature Communications}\ }\textbf {\bibinfo {volume}
  {8}} (\bibinfo {year} {2017}),\ 10.1038/ncomms14695}\BibitemShut {NoStop}%
\bibitem [{\citenamefont {Sun}\ \emph {et~al.}(2020)\citenamefont {Sun},
  \citenamefont {He}, \citenamefont {You}, \citenamefont {Lv}, \citenamefont
  {Li}, \citenamefont {Lloyd},\ and\ \citenamefont
  {Wang}}]{sun2020exponentially}%
  \BibitemOpen
  \bibfield  {author} {\bibinfo {author} {\bibfnamefont {L.}~\bibnamefont
  {Sun}}, \bibinfo {author} {\bibfnamefont {X.}~\bibnamefont {He}}, \bibinfo
  {author} {\bibfnamefont {C.}~\bibnamefont {You}}, \bibinfo {author}
  {\bibfnamefont {C.}~\bibnamefont {Lv}}, \bibinfo {author} {\bibfnamefont
  {B.}~\bibnamefont {Li}}, \bibinfo {author} {\bibfnamefont {S.}~\bibnamefont
  {Lloyd}}, \ and\ \bibinfo {author} {\bibfnamefont {X.}~\bibnamefont {Wang}},\
  }\href@noop {} {\bibfield  {journal} {\bibinfo  {journal} {arXiv preprint
  arXiv:2004.01216}\ } (\bibinfo {year} {2020})}\BibitemShut {NoStop}%
\bibitem [{\citenamefont {Dorner}(2012)}]{Dorner2012}%
  \BibitemOpen
  \bibfield  {author} {\bibinfo {author} {\bibfnamefont {U.}~\bibnamefont
  {Dorner}},\ }\href {\doibase 10.1088/1367-2630/14/4/043011} {\bibfield
  {journal} {\bibinfo  {journal} {New Journal of Physics}\ }\textbf {\bibinfo
  {volume} {14}},\ \bibinfo {pages} {043011} (\bibinfo {year}
  {2012})}\BibitemShut {NoStop}%
\bibitem [{\citenamefont {D\"ur}\ \emph {et~al.}(2014)\citenamefont {D\"ur},
  \citenamefont {Skotiniotis}, \citenamefont {Fr\"owis},\ and\ \citenamefont
  {Kraus}}]{Dur2014}%
  \BibitemOpen
  \bibfield  {author} {\bibinfo {author} {\bibfnamefont {W.}~\bibnamefont
  {D\"ur}}, \bibinfo {author} {\bibfnamefont {M.}~\bibnamefont {Skotiniotis}},
  \bibinfo {author} {\bibfnamefont {F.}~\bibnamefont {Fr\"owis}}, \ and\
  \bibinfo {author} {\bibfnamefont {B.}~\bibnamefont {Kraus}},\ }\href
  {\doibase 10.1103/PhysRevLett.112.080801} {\bibfield  {journal} {\bibinfo
  {journal} {Phys. Rev. Lett.}\ }\textbf {\bibinfo {volume} {112}},\ \bibinfo
  {pages} {080801} (\bibinfo {year} {2014})}\BibitemShut {NoStop}%
\bibitem [{\citenamefont {Kessler}\ \emph {et~al.}(2014)\citenamefont
  {Kessler}, \citenamefont {Lovchinsky}, \citenamefont {Sushkov},\ and\
  \citenamefont {Lukin}}]{Kessler2014}%
  \BibitemOpen
  \bibfield  {author} {\bibinfo {author} {\bibfnamefont {E.~M.}\ \bibnamefont
  {Kessler}}, \bibinfo {author} {\bibfnamefont {I.}~\bibnamefont {Lovchinsky}},
  \bibinfo {author} {\bibfnamefont {A.~O.}\ \bibnamefont {Sushkov}}, \ and\
  \bibinfo {author} {\bibfnamefont {M.~D.}\ \bibnamefont {Lukin}},\ }\href
  {\doibase 10.1103/PhysRevLett.112.150802} {\bibfield  {journal} {\bibinfo
  {journal} {Phys. Rev. Lett.}\ }\textbf {\bibinfo {volume} {112}},\ \bibinfo
  {pages} {150802} (\bibinfo {year} {2014})}\BibitemShut {NoStop}%
\bibitem [{\citenamefont {Arrad}\ \emph {et~al.}(2014)\citenamefont {Arrad},
  \citenamefont {Vinkler}, \citenamefont {Aharonov},\ and\ \citenamefont
  {Retzker}}]{Arrad2014}%
  \BibitemOpen
  \bibfield  {author} {\bibinfo {author} {\bibfnamefont {G.}~\bibnamefont
  {Arrad}}, \bibinfo {author} {\bibfnamefont {Y.}~\bibnamefont {Vinkler}},
  \bibinfo {author} {\bibfnamefont {D.}~\bibnamefont {Aharonov}}, \ and\
  \bibinfo {author} {\bibfnamefont {A.}~\bibnamefont {Retzker}},\ }\href
  {\doibase 10.1103/PhysRevLett.112.150801} {\bibfield  {journal} {\bibinfo
  {journal} {Phys. Rev. Lett.}\ }\textbf {\bibinfo {volume} {112}},\ \bibinfo
  {pages} {150801} (\bibinfo {year} {2014})}\BibitemShut {NoStop}%
\bibitem [{\citenamefont {Escher}\ \emph {et~al.}(2011)\citenamefont {Escher},
  \citenamefont {de~Matos~Filho},\ and\ \citenamefont
  {Davidovich}}]{escher2011general}%
  \BibitemOpen
  \bibfield  {author} {\bibinfo {author} {\bibfnamefont {B.}~\bibnamefont
  {Escher}}, \bibinfo {author} {\bibfnamefont {R.}~\bibnamefont
  {de~Matos~Filho}}, \ and\ \bibinfo {author} {\bibfnamefont {L.}~\bibnamefont
  {Davidovich}},\ }\href {https://www.nature.com/articles/nphys1958} {\bibfield
   {journal} {\bibinfo  {journal} {Nature Physics}\ }\textbf {\bibinfo {volume}
  {7}},\ \bibinfo {pages} {406} (\bibinfo {year} {2011})}\BibitemShut {NoStop}%
\bibitem [{\citenamefont {Demkowicz-Dobrza{\'{n}}ski}\ \emph
  {et~al.}(2012)\citenamefont {Demkowicz-Dobrza{\'{n}}ski}, \citenamefont
  {Ko{\l}ody{\'{n}}ski},\ and\ \citenamefont
  {Gu{\c{t}}{\u{a}}}}]{DemkowiczDobrzaski2012}%
  \BibitemOpen
  \bibfield  {author} {\bibinfo {author} {\bibfnamefont {R.}~\bibnamefont
  {Demkowicz-Dobrza{\'{n}}ski}}, \bibinfo {author} {\bibfnamefont
  {J.}~\bibnamefont {Ko{\l}ody{\'{n}}ski}}, \ and\ \bibinfo {author}
  {\bibfnamefont {M.}~\bibnamefont {Gu{\c{t}}{\u{a}}}},\ }\href {\doibase
  10.1038/ncomms2067} {\bibfield  {journal} {\bibinfo  {journal} {Nature
  Communications}\ }\textbf {\bibinfo {volume} {3}} (\bibinfo {year} {2012}),\
  10.1038/ncomms2067}\BibitemShut {NoStop}%
\bibitem [{\citenamefont {Ko{\l}ody{\'{n}}ski}\ and\ \citenamefont
  {Demkowicz-Dobrza{\'{n}}ski}(2013)}]{Koodyski2013}%
  \BibitemOpen
  \bibfield  {author} {\bibinfo {author} {\bibfnamefont {J.}~\bibnamefont
  {Ko{\l}ody{\'{n}}ski}}\ and\ \bibinfo {author} {\bibfnamefont
  {R.}~\bibnamefont {Demkowicz-Dobrza{\'{n}}ski}},\ }\href {\doibase
  10.1088/1367-2630/15/7/073043} {\bibfield  {journal} {\bibinfo  {journal}
  {New Journal of Physics}\ }\textbf {\bibinfo {volume} {15}},\ \bibinfo
  {pages} {073043} (\bibinfo {year} {2013})}\BibitemShut {NoStop}%
\bibitem [{\citenamefont {Knysh}\ \emph {et~al.}(2014)\citenamefont {Knysh},
  \citenamefont {Chen},\ and\ \citenamefont {Durkin}}]{knysh2014true}%
  \BibitemOpen
  \bibfield  {author} {\bibinfo {author} {\bibfnamefont {S.~I.}\ \bibnamefont
  {Knysh}}, \bibinfo {author} {\bibfnamefont {E.~H.}\ \bibnamefont {Chen}}, \
  and\ \bibinfo {author} {\bibfnamefont {G.~A.}\ \bibnamefont {Durkin}},\
  }\href {https://arxiv.org/abs/1402.0495} {\bibfield  {journal} {\bibinfo
  {journal} {arXiv preprint arXiv:1402.0495}\ } (\bibinfo {year}
  {2014})}\BibitemShut {NoStop}%
\bibitem [{\citenamefont {Haroche}\ and\ \citenamefont
  {Raimond}(2006)}]{Haroche2006}%
  \BibitemOpen
  \bibfield  {author} {\bibinfo {author} {\bibfnamefont {S.}~\bibnamefont
  {Haroche}}\ and\ \bibinfo {author} {\bibfnamefont {J.-M.}\ \bibnamefont
  {Raimond}},\ }\href {\doibase 10.1093/acprof:oso/9780198509141.001.0001}
  {\emph {\bibinfo {title} {{Exploring the Quantum}}}},\ Vol.~\bibinfo {volume}
  {5}\ (\bibinfo  {publisher} {Oxford University Press},\ \bibinfo {year}
  {2006})\ p.~\bibinfo {pages} {11}\BibitemShut {NoStop}%
\bibitem [{\citenamefont {Blais}\ \emph {et~al.}(2020)\citenamefont {Blais},
  \citenamefont {Girvin},\ and\ \citenamefont {Oliver}}]{Blais2020}%
  \BibitemOpen
  \bibfield  {author} {\bibinfo {author} {\bibfnamefont {A.}~\bibnamefont
  {Blais}}, \bibinfo {author} {\bibfnamefont {S.~M.}\ \bibnamefont {Girvin}}, \
  and\ \bibinfo {author} {\bibfnamefont {W.~D.}\ \bibnamefont {Oliver}},\
  }\href {\doibase 10.1038/s41567-020-0806-z} {\bibfield  {journal} {\bibinfo
  {journal} {Nature Physics}\ }\textbf {\bibinfo {volume} {16}},\ \bibinfo
  {pages} {247} (\bibinfo {year} {2020})}\BibitemShut {NoStop}%
\bibitem [{\citenamefont {Dicke}(1954)}]{dicke54}%
  \BibitemOpen
  \bibfield  {author} {\bibinfo {author} {\bibfnamefont {R.~H.}\ \bibnamefont
  {Dicke}},\ }\href {\doibase 10.1103/physrev.93.99} {\bibfield  {journal}
  {\bibinfo  {journal} {Phys. Rev.}\ }\textbf {\bibinfo {volume} {93}},\
  \bibinfo {pages} {99} (\bibinfo {year} {1954})}\BibitemShut {NoStop}%
\bibitem [{\citenamefont {Haas}\ \emph {et~al.}(2014)\citenamefont {Haas},
  \citenamefont {Volz}, \citenamefont {Gehr}, \citenamefont {Reichel},\ and\
  \citenamefont {Esteve}}]{Haas2014}%
  \BibitemOpen
  \bibfield  {author} {\bibinfo {author} {\bibfnamefont {F.}~\bibnamefont
  {Haas}}, \bibinfo {author} {\bibfnamefont {J.}~\bibnamefont {Volz}}, \bibinfo
  {author} {\bibfnamefont {R.}~\bibnamefont {Gehr}}, \bibinfo {author}
  {\bibfnamefont {J.}~\bibnamefont {Reichel}}, \ and\ \bibinfo {author}
  {\bibfnamefont {J.}~\bibnamefont {Esteve}},\ }\href {\doibase
  10.1126/science.1248905} {\bibfield  {journal} {\bibinfo  {journal}
  {Science}\ }\textbf {\bibinfo {volume} {344}},\ \bibinfo {pages} {180}
  (\bibinfo {year} {2014})}\BibitemShut {NoStop}%
\bibitem [{\citenamefont {Norcia}\ \emph {et~al.}(2016)\citenamefont {Norcia},
  \citenamefont {Winchester}, \citenamefont {Cline},\ and\ \citenamefont
  {Thompson}}]{Norcia2016}%
  \BibitemOpen
  \bibfield  {author} {\bibinfo {author} {\bibfnamefont {M.~A.}\ \bibnamefont
  {Norcia}}, \bibinfo {author} {\bibfnamefont {M.~N.}\ \bibnamefont
  {Winchester}}, \bibinfo {author} {\bibfnamefont {J.~R.~K.}\ \bibnamefont
  {Cline}}, \ and\ \bibinfo {author} {\bibfnamefont {J.~K.}\ \bibnamefont
  {Thompson}},\ }\href {\doibase 10.1126/sciadv.1601231} {\bibfield  {journal}
  {\bibinfo  {journal} {Science Advances}\ }\textbf {\bibinfo {volume} {2}},\
  \bibinfo {pages} {e1601231} (\bibinfo {year} {2016})}\BibitemShut {NoStop}%
\bibitem [{\citenamefont {Hosseini}\ \emph {et~al.}(2017)\citenamefont
  {Hosseini}, \citenamefont {Duan}, \citenamefont {Beck}, \citenamefont
  {Chen},\ and\ \citenamefont {Vuleti\ifmmode~\acute{c}\else
  \'{c}\fi{}}}]{Hosseini2017}%
  \BibitemOpen
  \bibfield  {author} {\bibinfo {author} {\bibfnamefont {M.}~\bibnamefont
  {Hosseini}}, \bibinfo {author} {\bibfnamefont {Y.}~\bibnamefont {Duan}},
  \bibinfo {author} {\bibfnamefont {K.~M.}\ \bibnamefont {Beck}}, \bibinfo
  {author} {\bibfnamefont {Y.-T.}\ \bibnamefont {Chen}}, \ and\ \bibinfo
  {author} {\bibfnamefont {V.}~\bibnamefont {Vuleti\ifmmode~\acute{c}\else
  \'{c}\fi{}}},\ }\href {\doibase 10.1103/PhysRevLett.118.183601} {\bibfield
  {journal} {\bibinfo  {journal} {Phys. Rev. Lett.}\ }\textbf {\bibinfo
  {volume} {118}},\ \bibinfo {pages} {183601} (\bibinfo {year}
  {2017})}\BibitemShut {NoStop}%
\bibitem [{\citenamefont {Kim}\ \emph {et~al.}(2018)\citenamefont {Kim},
  \citenamefont {Yang}, \citenamefont {Oh},\ and\ \citenamefont
  {An}}]{Kim2018}%
  \BibitemOpen
  \bibfield  {author} {\bibinfo {author} {\bibfnamefont {J.}~\bibnamefont
  {Kim}}, \bibinfo {author} {\bibfnamefont {D.}~\bibnamefont {Yang}}, \bibinfo
  {author} {\bibfnamefont {S.-h.}\ \bibnamefont {Oh}}, \ and\ \bibinfo {author}
  {\bibfnamefont {K.}~\bibnamefont {An}},\ }\href {\doibase
  10.1126/science.aar2179} {\bibfield  {journal} {\bibinfo  {journal}
  {Science}\ }\textbf {\bibinfo {volume} {359}},\ \bibinfo {pages} {662}
  (\bibinfo {year} {2018})}\BibitemShut {NoStop}%
\bibitem [{\citenamefont {{Frisk Kockum}}\ \emph {et~al.}(2019)\citenamefont
  {{Frisk Kockum}}, \citenamefont {Miranowicz}, \citenamefont {{De Liberato}},
  \citenamefont {Savasta},\ and\ \citenamefont {Nori}}]{Kockum2019a}%
  \BibitemOpen
  \bibfield  {author} {\bibinfo {author} {\bibfnamefont {A.}~\bibnamefont
  {{Frisk Kockum}}}, \bibinfo {author} {\bibfnamefont {A.}~\bibnamefont
  {Miranowicz}}, \bibinfo {author} {\bibfnamefont {S.}~\bibnamefont {{De
  Liberato}}}, \bibinfo {author} {\bibfnamefont {S.}~\bibnamefont {Savasta}}, \
  and\ \bibinfo {author} {\bibfnamefont {F.}~\bibnamefont {Nori}},\ }\href
  {\doibase 10.1038/s42254-018-0006-2} {\bibfield  {journal} {\bibinfo
  {journal} {Nature Reviews Physics}\ }\textbf {\bibinfo {volume} {1}},\
  \bibinfo {pages} {19} (\bibinfo {year} {2019})},\ \Eprint
  {http://arxiv.org/abs/1807.11636} {arXiv:1807.11636} \BibitemShut {NoStop}%
\bibitem [{\citenamefont {Mlynek}\ \emph {et~al.}(2014)\citenamefont {Mlynek},
  \citenamefont {Abdumalikov}, \citenamefont {Eichler},\ and\ \citenamefont
  {Wallraff}}]{mlynek14}%
  \BibitemOpen
  \bibfield  {author} {\bibinfo {author} {\bibfnamefont {J.~A.}\ \bibnamefont
  {Mlynek}}, \bibinfo {author} {\bibfnamefont {A.~A.}\ \bibnamefont
  {Abdumalikov}}, \bibinfo {author} {\bibfnamefont {C.}~\bibnamefont
  {Eichler}}, \ and\ \bibinfo {author} {\bibfnamefont {A.}~\bibnamefont
  {Wallraff}},\ }\href {\doibase 10.1038/ncomms6186} {\bibfield  {journal}
  {\bibinfo  {journal} {Nat. Commun.}\ }\textbf {\bibinfo {volume} {5}},\
  \bibinfo {pages} {5186} (\bibinfo {year} {2014})}\BibitemShut {NoStop}%
\bibitem [{\citenamefont {Goban}\ \emph {et~al.}(2015)\citenamefont {Goban},
  \citenamefont {Hung}, \citenamefont {Hood}, \citenamefont {Yu}, \citenamefont
  {Muniz}, \citenamefont {Painter},\ and\ \citenamefont {Kimble}}]{goban15}%
  \BibitemOpen
  \bibfield  {author} {\bibinfo {author} {\bibfnamefont {A.}~\bibnamefont
  {Goban}}, \bibinfo {author} {\bibfnamefont {C.-L.}\ \bibnamefont {Hung}},
  \bibinfo {author} {\bibfnamefont {J.}~\bibnamefont {Hood}}, \bibinfo {author}
  {\bibfnamefont {S.-P.}\ \bibnamefont {Yu}}, \bibinfo {author} {\bibfnamefont
  {J.}~\bibnamefont {Muniz}}, \bibinfo {author} {\bibfnamefont
  {O.}~\bibnamefont {Painter}}, \ and\ \bibinfo {author} {\bibfnamefont
  {H.}~\bibnamefont {Kimble}},\ }\href {\doibase
  10.1103/physrevlett.115.063601} {\bibfield  {journal} {\bibinfo  {journal}
  {Phys. Rev. Lett.}\ }\textbf {\bibinfo {volume} {115}},\ \bibinfo {pages}
  {063601} (\bibinfo {year} {2015})}\BibitemShut {NoStop}%
\bibitem [{\citenamefont {Solano}\ \emph {et~al.}(2017)\citenamefont {Solano},
  \citenamefont {Barberis-Blostein}, \citenamefont {Fatemi}, \citenamefont
  {Orozco},\ and\ \citenamefont {Rolston}}]{Solano2017}%
  \BibitemOpen
  \bibfield  {author} {\bibinfo {author} {\bibfnamefont {P.}~\bibnamefont
  {Solano}}, \bibinfo {author} {\bibfnamefont {P.}~\bibnamefont
  {Barberis-Blostein}}, \bibinfo {author} {\bibfnamefont {F.~K.}\ \bibnamefont
  {Fatemi}}, \bibinfo {author} {\bibfnamefont {L.~A.}\ \bibnamefont {Orozco}},
  \ and\ \bibinfo {author} {\bibfnamefont {S.~L.}\ \bibnamefont {Rolston}},\
  }\href {\doibase 10.1038/s41467-017-01994-3} {\bibfield  {journal} {\bibinfo
  {journal} {Nature Communications}\ }\textbf {\bibinfo {volume} {8}},\
  \bibinfo {pages} {1857} (\bibinfo {year} {2017})}\BibitemShut {NoStop}%
\bibitem [{\citenamefont {O{\textquotesingle}Brien}\ \emph
  {et~al.}(2009)\citenamefont {O{\textquotesingle}Brien}, \citenamefont
  {Furusawa},\ and\ \citenamefont {Vu{\v{c}}kovi{\'{c}}}}]{OBrien2009}%
  \BibitemOpen
  \bibfield  {author} {\bibinfo {author} {\bibfnamefont {J.~L.}\ \bibnamefont
  {O{\textquotesingle}Brien}}, \bibinfo {author} {\bibfnamefont
  {A.}~\bibnamefont {Furusawa}}, \ and\ \bibinfo {author} {\bibfnamefont
  {J.}~\bibnamefont {Vu{\v{c}}kovi{\'{c}}}},\ }\href {\doibase
  10.1038/nphoton.2009.229} {\bibfield  {journal} {\bibinfo  {journal} {Nature
  Photonics}\ }\textbf {\bibinfo {volume} {3}},\ \bibinfo {pages} {687}
  (\bibinfo {year} {2009})}\BibitemShut {NoStop}%
\bibitem [{\citenamefont {Gonz\'alez-Tudela}\ \emph {et~al.}(2017)\citenamefont
  {Gonz\'alez-Tudela}, \citenamefont {Paulisch}, \citenamefont {Kimble},\ and\
  \citenamefont {Cirac}}]{gonzalez-tudela17}%
  \BibitemOpen
  \bibfield  {author} {\bibinfo {author} {\bibfnamefont {A.}~\bibnamefont
  {Gonz\'alez-Tudela}}, \bibinfo {author} {\bibfnamefont {V.}~\bibnamefont
  {Paulisch}}, \bibinfo {author} {\bibfnamefont {H.~J.}\ \bibnamefont
  {Kimble}}, \ and\ \bibinfo {author} {\bibfnamefont {J.~I.}\ \bibnamefont
  {Cirac}},\ }\href {\doibase 10.1103/PhysRevLett.118.213601} {\bibfield
  {journal} {\bibinfo  {journal} {Phys. Rev. Lett.}\ }\textbf {\bibinfo
  {volume} {118}},\ \bibinfo {pages} {213601} (\bibinfo {year}
  {2017})}\BibitemShut {NoStop}%
\bibitem [{\citenamefont {Paulisch}\ \emph {et~al.}(2018)\citenamefont
  {Paulisch}, \citenamefont {Kimble}, \citenamefont {Cirac},\ and\
  \citenamefont {Gonz\'alez-Tudela}}]{Paulisch2018}%
  \BibitemOpen
  \bibfield  {author} {\bibinfo {author} {\bibfnamefont {V.}~\bibnamefont
  {Paulisch}}, \bibinfo {author} {\bibfnamefont {H.~J.}\ \bibnamefont
  {Kimble}}, \bibinfo {author} {\bibfnamefont {J.~I.}\ \bibnamefont {Cirac}}, \
  and\ \bibinfo {author} {\bibfnamefont {A.}~\bibnamefont
  {Gonz\'alez-Tudela}},\ }\href {\doibase 10.1103/PhysRevA.97.053831}
  {\bibfield  {journal} {\bibinfo  {journal} {Phys. Rev. A}\ }\textbf {\bibinfo
  {volume} {97}},\ \bibinfo {pages} {053831} (\bibinfo {year}
  {2018})}\BibitemShut {NoStop}%
\bibitem [{\citenamefont {Uria}\ \emph {et~al.}(2020)\citenamefont {Uria},
  \citenamefont {Solano},\ and\ \citenamefont {Hermann-Avigliano}}]{Uria2020}%
  \BibitemOpen
  \bibfield  {author} {\bibinfo {author} {\bibfnamefont {M.}~\bibnamefont
  {Uria}}, \bibinfo {author} {\bibfnamefont {P.}~\bibnamefont {Solano}}, \ and\
  \bibinfo {author} {\bibfnamefont {C.}~\bibnamefont {Hermann-Avigliano}},\
  }\href {\doibase 10.1103/PhysRevLett.125.093603} {\bibfield  {journal}
  {\bibinfo  {journal} {Phys. Rev. Lett.}\ }\textbf {\bibinfo {volume} {125}},\
  \bibinfo {pages} {093603} (\bibinfo {year} {2020})}\BibitemShut {NoStop}%
\bibitem [{\citenamefont {Groiseau}\ \emph {et~al.}(2020)\citenamefont
  {Groiseau}, \citenamefont {Elliott}, \citenamefont {Masson},\ and\
  \citenamefont {Parkins}}]{groiseau2020deterministic}%
  \BibitemOpen
  \bibfield  {author} {\bibinfo {author} {\bibfnamefont {C.}~\bibnamefont
  {Groiseau}}, \bibinfo {author} {\bibfnamefont {A.~E.}\ \bibnamefont
  {Elliott}}, \bibinfo {author} {\bibfnamefont {S.~J.}\ \bibnamefont {Masson}},
  \ and\ \bibinfo {author} {\bibfnamefont {S.}~\bibnamefont {Parkins}},\
  }\href@noop {} {\bibfield  {journal} {\bibinfo  {journal} {arXiv preprint
  arXiv:2012.00246}\ } (\bibinfo {year} {2020})}\BibitemShut {NoStop}%
\bibitem [{\citenamefont {Romero}\ \emph {et~al.}(2009)\citenamefont {Romero},
  \citenamefont {Garc{\'{i}}a-Ripoll},\ and\ \citenamefont
  {Solano}}]{Romero2009}%
  \BibitemOpen
  \bibfield  {author} {\bibinfo {author} {\bibfnamefont {G.}~\bibnamefont
  {Romero}}, \bibinfo {author} {\bibfnamefont {J.~J.}\ \bibnamefont
  {Garc{\'{i}}a-Ripoll}}, \ and\ \bibinfo {author} {\bibfnamefont
  {E.}~\bibnamefont {Solano}},\ }\href {\doibase
  10.1103/PhysRevLett.102.173602} {\bibfield  {journal} {\bibinfo  {journal}
  {Physical Review Letters}\ }\textbf {\bibinfo {volume} {102}},\ \bibinfo
  {pages} {173602} (\bibinfo {year} {2009})}\BibitemShut {NoStop}%
\bibitem [{\citenamefont {Peropadre}\ \emph {et~al.}(2011)\citenamefont
  {Peropadre}, \citenamefont {Romero}, \citenamefont {Johansson}, \citenamefont
  {Wilson}, \citenamefont {Solano},\ and\ \citenamefont
  {Garc{\'{i}}a-Ripoll}}]{Peropadre2011}%
  \BibitemOpen
  \bibfield  {author} {\bibinfo {author} {\bibfnamefont {B.}~\bibnamefont
  {Peropadre}}, \bibinfo {author} {\bibfnamefont {G.}~\bibnamefont {Romero}},
  \bibinfo {author} {\bibfnamefont {G.}~\bibnamefont {Johansson}}, \bibinfo
  {author} {\bibfnamefont {C.~M.}\ \bibnamefont {Wilson}}, \bibinfo {author}
  {\bibfnamefont {E.}~\bibnamefont {Solano}}, \ and\ \bibinfo {author}
  {\bibfnamefont {J.~J.}\ \bibnamefont {Garc{\'{i}}a-Ripoll}},\ }\href
  {\doibase 10.1103/PhysRevA.84.063834} {\bibfield  {journal} {\bibinfo
  {journal} {Physical Review A}\ }\textbf {\bibinfo {volume} {84}},\ \bibinfo
  {pages} {063834} (\bibinfo {year} {2011})}\BibitemShut {NoStop}%
\bibitem [{\citenamefont {Malz}\ and\ \citenamefont
  {Cirac}(2020)}]{malz2019number}%
  \BibitemOpen
  \bibfield  {author} {\bibinfo {author} {\bibfnamefont {D.}~\bibnamefont
  {Malz}}\ and\ \bibinfo {author} {\bibfnamefont {J.~I.}\ \bibnamefont
  {Cirac}},\ }\href {\doibase 10.1103/PhysRevResearch.2.033091} {\bibfield
  {journal} {\bibinfo  {journal} {Phys. Rev. Research}\ }\textbf {\bibinfo
  {volume} {2}},\ \bibinfo {pages} {033091} (\bibinfo {year}
  {2020})}\BibitemShut {NoStop}%
\bibitem [{\citenamefont {Paulisch}\ \emph
  {et~al.}(2019{\natexlab{a}})\citenamefont {Paulisch}, \citenamefont
  {Perarnau-Llobet}, \citenamefont {Gonz{\'{a}}lez-Tudela},\ and\ \citenamefont
  {Cirac}}]{Paulisch2019}%
  \BibitemOpen
  \bibfield  {author} {\bibinfo {author} {\bibfnamefont {V.}~\bibnamefont
  {Paulisch}}, \bibinfo {author} {\bibfnamefont {M.}~\bibnamefont
  {Perarnau-Llobet}}, \bibinfo {author} {\bibfnamefont {A.}~\bibnamefont
  {Gonz{\'{a}}lez-Tudela}}, \ and\ \bibinfo {author} {\bibfnamefont {J.~I.}\
  \bibnamefont {Cirac}},\ }\href {\doibase 10.1103/PhysRevA.99.043807}
  {\bibfield  {journal} {\bibinfo  {journal} {Physical Review A}\ }\textbf
  {\bibinfo {volume} {99}},\ \bibinfo {pages} {043807} (\bibinfo {year}
  {2019}{\natexlab{a}})}\BibitemShut {NoStop}%
\bibitem [{\citenamefont {Perarnau-Llobet}\ \emph {et~al.}(2020)\citenamefont
  {Perarnau-Llobet}, \citenamefont {Gonz{\'{a}}lez-Tudela},\ and\ \citenamefont
  {Cirac}}]{PerarnauLlobet2020}%
  \BibitemOpen
  \bibfield  {author} {\bibinfo {author} {\bibfnamefont {M.}~\bibnamefont
  {Perarnau-Llobet}}, \bibinfo {author} {\bibfnamefont {A.}~\bibnamefont
  {Gonz{\'{a}}lez-Tudela}}, \ and\ \bibinfo {author} {\bibfnamefont {J.~I.}\
  \bibnamefont {Cirac}},\ }\href {\doibase 10.1088/2058-9565/ab6ce5} {\bibfield
   {journal} {\bibinfo  {journal} {Quantum Science and Technology}\ }\textbf
  {\bibinfo {volume} {5}},\ \bibinfo {pages} {025003} (\bibinfo {year}
  {2020})}\BibitemShut {NoStop}%
\bibitem [{\citenamefont {Helstrom}(1976)}]{helstrom76}%
  \BibitemOpen
  \bibfield  {author} {\bibinfo {author} {\bibfnamefont {C.~W.}\ \bibnamefont
  {Helstrom}},\ }\href
  {http://www.ebook.de/de/product/15210504/quantum_detection_and_estimation_theory.html}
  {\emph {\bibinfo {title} {Quantum Detection and Estimation Theory}}}\
  (\bibinfo  {publisher} {Elsevier Science},\ \bibinfo {year}
  {1976})\BibitemShut {NoStop}%
\bibitem [{\citenamefont {Holevo}(1982)}]{holevo82}%
  \BibitemOpen
  \bibfield  {author} {\bibinfo {author} {\bibfnamefont {A.~S.}\ \bibnamefont
  {Holevo}},\ }\href
  {https://www.amazon.com/Probabilistic-Statistical-Aspects-Statistics-Probability/dp/0444863338?SubscriptionId=0JYN1NVW651KCA56C102&tag=techkie-20&linkCode=xm2&camp=2025&creative=165953&creativeASIN=0444863338}
  {\emph {\bibinfo {title} {Probabilistic and Statistical Aspects of Quantum
  Theory (Statistics \& Probability) (English and Russian Edition)}}}\
  (\bibinfo  {publisher} {Elsevier Science},\ \bibinfo {year}
  {1982})\BibitemShut {NoStop}%
\bibitem [{\citenamefont {Braunstein}\ and\ \citenamefont
  {Caves}(1994)}]{Braunstein1994}%
  \BibitemOpen
  \bibfield  {author} {\bibinfo {author} {\bibfnamefont {S.~L.}\ \bibnamefont
  {Braunstein}}\ and\ \bibinfo {author} {\bibfnamefont {C.~M.}\ \bibnamefont
  {Caves}},\ }\href {\doibase 10.1103/PhysRevLett.72.3439} {\bibfield
  {journal} {\bibinfo  {journal} {Phys. Rev. Lett.}\ }\textbf {\bibinfo
  {volume} {72}},\ \bibinfo {pages} {3439} (\bibinfo {year}
  {1994})}\BibitemShut {NoStop}%
\bibitem [{\citenamefont {Fujiwara}\ and\ \citenamefont
  {Imai}(2008)}]{Fujiwara2008}%
  \BibitemOpen
  \bibfield  {author} {\bibinfo {author} {\bibfnamefont {A.}~\bibnamefont
  {Fujiwara}}\ and\ \bibinfo {author} {\bibfnamefont {H.}~\bibnamefont
  {Imai}},\ }\href {\doibase 10.1088/1751-8113/41/25/255304} {\bibfield
  {journal} {\bibinfo  {journal} {Journal of Physics A: Mathematical and
  Theoretical}\ }\textbf {\bibinfo {volume} {41}},\ \bibinfo {pages} {255304}
  (\bibinfo {year} {2008})}\BibitemShut {NoStop}%
\bibitem [{\citenamefont {Knysh}\ \emph {et~al.}(2011)\citenamefont {Knysh},
  \citenamefont {Smelyanskiy},\ and\ \citenamefont {Durkin}}]{Knysh2011}%
  \BibitemOpen
  \bibfield  {author} {\bibinfo {author} {\bibfnamefont {S.}~\bibnamefont
  {Knysh}}, \bibinfo {author} {\bibfnamefont {V.~N.}\ \bibnamefont
  {Smelyanskiy}}, \ and\ \bibinfo {author} {\bibfnamefont {G.~A.}\ \bibnamefont
  {Durkin}},\ }\href {\doibase 10.1103/PhysRevA.83.021804} {\bibfield
  {journal} {\bibinfo  {journal} {Phys. Rev. A}\ }\textbf {\bibinfo {volume}
  {83}},\ \bibinfo {pages} {021804} (\bibinfo {year} {2011})}\BibitemShut
  {NoStop}%
\bibitem [{\citenamefont {Campos}\ \emph {et~al.}(2003)\citenamefont {Campos},
  \citenamefont {Gerry},\ and\ \citenamefont {Benmoussa}}]{campos03}%
  \BibitemOpen
  \bibfield  {author} {\bibinfo {author} {\bibfnamefont {R.~A.}\ \bibnamefont
  {Campos}}, \bibinfo {author} {\bibfnamefont {C.~C.}\ \bibnamefont {Gerry}}, \
  and\ \bibinfo {author} {\bibfnamefont {A.}~\bibnamefont {Benmoussa}},\ }\href
  {\doibase 10.1103/physreva.68.023810} {\bibfield  {journal} {\bibinfo
  {journal} {Phys. Rev. A}\ }\textbf {\bibinfo {volume} {68}},\ \bibinfo
  {pages} {023810} (\bibinfo {year} {2003})}\BibitemShut {NoStop}%
\bibitem [{\citenamefont {Datta}\ \emph {et~al.}(2011)\citenamefont {Datta},
  \citenamefont {Zhang}, \citenamefont {Thomas-Peter}, \citenamefont {Dorner},
  \citenamefont {Smith},\ and\ \citenamefont {Walmsley}}]{Datta2011}%
  \BibitemOpen
  \bibfield  {author} {\bibinfo {author} {\bibfnamefont {A.}~\bibnamefont
  {Datta}}, \bibinfo {author} {\bibfnamefont {L.}~\bibnamefont {Zhang}},
  \bibinfo {author} {\bibfnamefont {N.}~\bibnamefont {Thomas-Peter}}, \bibinfo
  {author} {\bibfnamefont {U.}~\bibnamefont {Dorner}}, \bibinfo {author}
  {\bibfnamefont {B.~J.}\ \bibnamefont {Smith}}, \ and\ \bibinfo {author}
  {\bibfnamefont {I.~A.}\ \bibnamefont {Walmsley}},\ }\href {\doibase
  10.1103/PhysRevA.83.063836} {\bibfield  {journal} {\bibinfo  {journal} {Phys.
  Rev. A}\ }\textbf {\bibinfo {volume} {83}},\ \bibinfo {pages} {063836}
  (\bibinfo {year} {2011})}\BibitemShut {NoStop}%
\bibitem [{\citenamefont {Paris}(2009)}]{PARIS2009}%
  \BibitemOpen
  \bibfield  {author} {\bibinfo {author} {\bibfnamefont {M.~G.~A.}\
  \bibnamefont {Paris}},\ }\href {\doibase 10.1142/s0219749909004839}
  {\bibfield  {journal} {\bibinfo  {journal} {International Journal of Quantum
  Information}\ }\textbf {\bibinfo {volume} {07}},\ \bibinfo {pages} {125}
  (\bibinfo {year} {2009})}\BibitemShut {NoStop}%
\bibitem [{\citenamefont {Kim}\ \emph {et~al.}(1998)\citenamefont {Kim},
  \citenamefont {Pfister}, \citenamefont {Holland}, \citenamefont {Noh},\ and\
  \citenamefont {Hall}}]{Taesoo1998}%
  \BibitemOpen
  \bibfield  {author} {\bibinfo {author} {\bibfnamefont {T.}~\bibnamefont
  {Kim}}, \bibinfo {author} {\bibfnamefont {O.}~\bibnamefont {Pfister}},
  \bibinfo {author} {\bibfnamefont {M.~J.}\ \bibnamefont {Holland}}, \bibinfo
  {author} {\bibfnamefont {J.}~\bibnamefont {Noh}}, \ and\ \bibinfo {author}
  {\bibfnamefont {J.~L.}\ \bibnamefont {Hall}},\ }\href {\doibase
  10.1103/PhysRevA.57.4004} {\bibfield  {journal} {\bibinfo  {journal} {Phys.
  Rev. A}\ }\textbf {\bibinfo {volume} {57}},\ \bibinfo {pages} {4004}
  (\bibinfo {year} {1998})}\BibitemShut {NoStop}%
\bibitem [{\citenamefont {Pezz\'e}\ and\ \citenamefont
  {Smerzi}(2013)}]{Pezze2013}%
  \BibitemOpen
  \bibfield  {author} {\bibinfo {author} {\bibfnamefont {L.}~\bibnamefont
  {Pezz\'e}}\ and\ \bibinfo {author} {\bibfnamefont {A.}~\bibnamefont
  {Smerzi}},\ }\href {\doibase 10.1103/PhysRevLett.110.163604} {\bibfield
  {journal} {\bibinfo  {journal} {Phys. Rev. Lett.}\ }\textbf {\bibinfo
  {volume} {110}},\ \bibinfo {pages} {163604} (\bibinfo {year}
  {2013})}\BibitemShut {NoStop}%
\bibitem [{\citenamefont {Zhong}\ \emph {et~al.}(2017)\citenamefont {Zhong},
  \citenamefont {Huang}, \citenamefont {Wang},\ and\ \citenamefont
  {Zhu}}]{Zhong2017}%
  \BibitemOpen
  \bibfield  {author} {\bibinfo {author} {\bibfnamefont {W.}~\bibnamefont
  {Zhong}}, \bibinfo {author} {\bibfnamefont {Y.}~\bibnamefont {Huang}},
  \bibinfo {author} {\bibfnamefont {X.}~\bibnamefont {Wang}}, \ and\ \bibinfo
  {author} {\bibfnamefont {S.-L.}\ \bibnamefont {Zhu}},\ }\href {\doibase
  10.1103/PhysRevA.95.052304} {\bibfield  {journal} {\bibinfo  {journal} {Phys.
  Rev. A}\ }\textbf {\bibinfo {volume} {95}},\ \bibinfo {pages} {052304}
  (\bibinfo {year} {2017})}\BibitemShut {NoStop}%
\bibitem [{\citenamefont {Yurke}\ \emph {et~al.}(1986)\citenamefont {Yurke},
  \citenamefont {McCall},\ and\ \citenamefont {Klauder}}]{Yurke1986}%
  \BibitemOpen
  \bibfield  {author} {\bibinfo {author} {\bibfnamefont {B.}~\bibnamefont
  {Yurke}}, \bibinfo {author} {\bibfnamefont {S.~L.}\ \bibnamefont {McCall}}, \
  and\ \bibinfo {author} {\bibfnamefont {J.~R.}\ \bibnamefont {Klauder}},\
  }\href {\doibase 10.1103/PhysRevA.33.4033} {\bibfield  {journal} {\bibinfo
  {journal} {Phys. Rev. A}\ }\textbf {\bibinfo {volume} {33}},\ \bibinfo
  {pages} {4033} (\bibinfo {year} {1986})}\BibitemShut {NoStop}%
\bibitem [{\citenamefont {Olivares}\ and\ \citenamefont
  {Paris}(2007)}]{Olivares2007}%
  \BibitemOpen
  \bibfield  {author} {\bibinfo {author} {\bibfnamefont {S.}~\bibnamefont
  {Olivares}}\ and\ \bibinfo {author} {\bibfnamefont {M.~G.~A.}\ \bibnamefont
  {Paris}},\ }\href {\doibase 10.1134/S0030400X07080103} {\bibfield  {journal}
  {\bibinfo  {journal} {Optics and Spectroscopy}\ }\textbf {\bibinfo {volume}
  {103}},\ \bibinfo {pages} {231} (\bibinfo {year} {2007})}\BibitemShut
  {NoStop}%
\bibitem [{\citenamefont {\ifmmode~\check{S}\else \v{S}\fi{}afr\'anek}\ and\
  \citenamefont {Fuentes}(2016)}]{Fuentes2016}%
  \BibitemOpen
  \bibfield  {author} {\bibinfo {author} {\bibfnamefont {D.}~\bibnamefont
  {\ifmmode~\check{S}\else \v{S}\fi{}afr\'anek}}\ and\ \bibinfo {author}
  {\bibfnamefont {I.}~\bibnamefont {Fuentes}},\ }\href {\doibase
  10.1103/PhysRevA.94.062313} {\bibfield  {journal} {\bibinfo  {journal} {Phys.
  Rev. A}\ }\textbf {\bibinfo {volume} {94}},\ \bibinfo {pages} {062313}
  (\bibinfo {year} {2016})}\BibitemShut {NoStop}%
\bibitem [{\citenamefont {Huver}\ \emph {et~al.}(2008)\citenamefont {Huver},
  \citenamefont {Wildfeuer},\ and\ \citenamefont {Dowling}}]{Huver2008}%
  \BibitemOpen
  \bibfield  {author} {\bibinfo {author} {\bibfnamefont {S.~D.}\ \bibnamefont
  {Huver}}, \bibinfo {author} {\bibfnamefont {C.~F.}\ \bibnamefont
  {Wildfeuer}}, \ and\ \bibinfo {author} {\bibfnamefont {J.~P.}\ \bibnamefont
  {Dowling}},\ }\href {\doibase 10.1103/PhysRevA.78.063828} {\bibfield
  {journal} {\bibinfo  {journal} {Phys. Rev. A}\ }\textbf {\bibinfo {volume}
  {78}},\ \bibinfo {pages} {063828} (\bibinfo {year} {2008})}\BibitemShut
  {NoStop}%
\bibitem [{\citenamefont {Dorner}\ \emph {et~al.}(2009)\citenamefont {Dorner},
  \citenamefont {Demkowicz-Dobrzanski}, \citenamefont {Smith}, \citenamefont
  {Lundeen}, \citenamefont {Wasilewski}, \citenamefont {Banaszek},\ and\
  \citenamefont {Walmsley}}]{Dorner2009}%
  \BibitemOpen
  \bibfield  {author} {\bibinfo {author} {\bibfnamefont {U.}~\bibnamefont
  {Dorner}}, \bibinfo {author} {\bibfnamefont {R.}~\bibnamefont
  {Demkowicz-Dobrzanski}}, \bibinfo {author} {\bibfnamefont {B.~J.}\
  \bibnamefont {Smith}}, \bibinfo {author} {\bibfnamefont {J.~S.}\ \bibnamefont
  {Lundeen}}, \bibinfo {author} {\bibfnamefont {W.}~\bibnamefont {Wasilewski}},
  \bibinfo {author} {\bibfnamefont {K.}~\bibnamefont {Banaszek}}, \ and\
  \bibinfo {author} {\bibfnamefont {I.~A.}\ \bibnamefont {Walmsley}},\ }\href
  {\doibase 10.1103/PhysRevLett.102.040403} {\bibfield  {journal} {\bibinfo
  {journal} {Phys. Rev. Lett.}\ }\textbf {\bibinfo {volume} {102}},\ \bibinfo
  {pages} {040403} (\bibinfo {year} {2009})}\BibitemShut {NoStop}%
\bibitem [{\citenamefont {Demkowicz-Dobrzanski}\ \emph
  {et~al.}(2009)\citenamefont {Demkowicz-Dobrzanski}, \citenamefont {Dorner},
  \citenamefont {Smith}, \citenamefont {Lundeen}, \citenamefont {Wasilewski},
  \citenamefont {Banaszek},\ and\ \citenamefont {Walmsley}}]{Demkowicz2009}%
  \BibitemOpen
  \bibfield  {author} {\bibinfo {author} {\bibfnamefont {R.}~\bibnamefont
  {Demkowicz-Dobrzanski}}, \bibinfo {author} {\bibfnamefont {U.}~\bibnamefont
  {Dorner}}, \bibinfo {author} {\bibfnamefont {B.~J.}\ \bibnamefont {Smith}},
  \bibinfo {author} {\bibfnamefont {J.~S.}\ \bibnamefont {Lundeen}}, \bibinfo
  {author} {\bibfnamefont {W.}~\bibnamefont {Wasilewski}}, \bibinfo {author}
  {\bibfnamefont {K.}~\bibnamefont {Banaszek}}, \ and\ \bibinfo {author}
  {\bibfnamefont {I.~A.}\ \bibnamefont {Walmsley}},\ }\href {\doibase
  10.1103/PhysRevA.80.013825} {\bibfield  {journal} {\bibinfo  {journal} {Phys.
  Rev. A}\ }\textbf {\bibinfo {volume} {80}},\ \bibinfo {pages} {013825}
  (\bibinfo {year} {2009})}\BibitemShut {NoStop}%
\bibitem [{\citenamefont {Cable}\ and\ \citenamefont
  {Durkin}(2010)}]{Cable2010}%
  \BibitemOpen
  \bibfield  {author} {\bibinfo {author} {\bibfnamefont {H.}~\bibnamefont
  {Cable}}\ and\ \bibinfo {author} {\bibfnamefont {G.~A.}\ \bibnamefont
  {Durkin}},\ }\href {\doibase 10.1103/PhysRevLett.105.013603} {\bibfield
  {journal} {\bibinfo  {journal} {Phys. Rev. Lett.}\ }\textbf {\bibinfo
  {volume} {105}},\ \bibinfo {pages} {013603} (\bibinfo {year}
  {2010})}\BibitemShut {NoStop}%
\bibitem [{\citenamefont {Calsamiglia}\ \emph {et~al.}(2016)\citenamefont
  {Calsamiglia}, \citenamefont {Gendra}, \citenamefont {Mu{\~{n}}oz-Tapia},\
  and\ \citenamefont {Bagan}}]{Calsamiglia2016}%
  \BibitemOpen
  \bibfield  {author} {\bibinfo {author} {\bibfnamefont {J.}~\bibnamefont
  {Calsamiglia}}, \bibinfo {author} {\bibfnamefont {B.}~\bibnamefont {Gendra}},
  \bibinfo {author} {\bibfnamefont {R.}~\bibnamefont {Mu{\~{n}}oz-Tapia}}, \
  and\ \bibinfo {author} {\bibfnamefont {E.}~\bibnamefont {Bagan}},\ }\href
  {\doibase 10.1088/1367-2630/18/10/103049} {\bibfield  {journal} {\bibinfo
  {journal} {New Journal of Physics}\ }\textbf {\bibinfo {volume} {18}},\
  \bibinfo {pages} {103049} (\bibinfo {year} {2016})}\BibitemShut {NoStop}%
\bibitem [{\citenamefont {Matsubara}\ \emph {et~al.}(2019)\citenamefont
  {Matsubara}, \citenamefont {Facchi}, \citenamefont {Giovannetti},\ and\
  \citenamefont {Yuasa}}]{Matsubara2019}%
  \BibitemOpen
  \bibfield  {author} {\bibinfo {author} {\bibfnamefont {T.}~\bibnamefont
  {Matsubara}}, \bibinfo {author} {\bibfnamefont {P.}~\bibnamefont {Facchi}},
  \bibinfo {author} {\bibfnamefont {V.}~\bibnamefont {Giovannetti}}, \ and\
  \bibinfo {author} {\bibfnamefont {K.}~\bibnamefont {Yuasa}},\ }\href
  {\doibase 10.1088/1367-2630/ab0604} {\bibfield  {journal} {\bibinfo
  {journal} {New Journal of Physics}\ }\textbf {\bibinfo {volume} {21}},\
  \bibinfo {pages} {033014} (\bibinfo {year} {2019})}\BibitemShut {NoStop}%
\bibitem [{\citenamefont {Wolf}\ \emph {et~al.}(2019)\citenamefont {Wolf},
  \citenamefont {Shi}, \citenamefont {Heip}, \citenamefont {Gessner},
  \citenamefont {Pezz{\`{e}}}, \citenamefont {Smerzi}, \citenamefont {Schulte},
  \citenamefont {Hammerer},\ and\ \citenamefont {Schmidt}}]{Wolf2019}%
  \BibitemOpen
  \bibfield  {author} {\bibinfo {author} {\bibfnamefont {F.}~\bibnamefont
  {Wolf}}, \bibinfo {author} {\bibfnamefont {C.}~\bibnamefont {Shi}}, \bibinfo
  {author} {\bibfnamefont {J.~C.}\ \bibnamefont {Heip}}, \bibinfo {author}
  {\bibfnamefont {M.}~\bibnamefont {Gessner}}, \bibinfo {author} {\bibfnamefont
  {L.}~\bibnamefont {Pezz{\`{e}}}}, \bibinfo {author} {\bibfnamefont
  {A.}~\bibnamefont {Smerzi}}, \bibinfo {author} {\bibfnamefont
  {M.}~\bibnamefont {Schulte}}, \bibinfo {author} {\bibfnamefont
  {K.}~\bibnamefont {Hammerer}}, \ and\ \bibinfo {author} {\bibfnamefont
  {P.~O.}\ \bibnamefont {Schmidt}},\ }\href {\doibase
  10.1038/s41467-019-10576-4} {\bibfield  {journal} {\bibinfo  {journal}
  {Nature Communications}\ }\textbf {\bibinfo {volume} {10}} (\bibinfo {year}
  {2019}),\ 10.1038/s41467-019-10576-4}\BibitemShut {NoStop}%
\bibitem [{\citenamefont {Paulisch}\ \emph
  {et~al.}(2019{\natexlab{b}})\citenamefont {Paulisch}, \citenamefont
  {Perarnau-Llobet}, \citenamefont {Gonz\'alez-Tudela},\ and\ \citenamefont
  {Cirac}}]{paulisch18b}%
  \BibitemOpen
  \bibfield  {author} {\bibinfo {author} {\bibfnamefont {V.}~\bibnamefont
  {Paulisch}}, \bibinfo {author} {\bibfnamefont {M.}~\bibnamefont
  {Perarnau-Llobet}}, \bibinfo {author} {\bibfnamefont {A.}~\bibnamefont
  {Gonz\'alez-Tudela}}, \ and\ \bibinfo {author} {\bibfnamefont {J.~I.}\
  \bibnamefont {Cirac}},\ }\href {\doibase 10.1103/PhysRevA.99.043807}
  {\bibfield  {journal} {\bibinfo  {journal} {Phys. Rev. A}\ }\textbf {\bibinfo
  {volume} {99}},\ \bibinfo {pages} {043807} (\bibinfo {year}
  {2019}{\natexlab{b}})}\BibitemShut {NoStop}%
\bibitem [{\citenamefont {Walther}\ \emph {et~al.}(2006)\citenamefont
  {Walther}, \citenamefont {Varcoe}, \citenamefont {Englert},\ and\
  \citenamefont {Becker}}]{Walther2006}%
  \BibitemOpen
  \bibfield  {author} {\bibinfo {author} {\bibfnamefont {H.}~\bibnamefont
  {Walther}}, \bibinfo {author} {\bibfnamefont {B.~T.~H.}\ \bibnamefont
  {Varcoe}}, \bibinfo {author} {\bibfnamefont {B.-G.}\ \bibnamefont {Englert}},
  \ and\ \bibinfo {author} {\bibfnamefont {T.}~\bibnamefont {Becker}},\ }\href
  {\doibase 10.1088/0034-4885/69/5/R02} {\bibfield  {journal} {\bibinfo
  {journal} {Reports on Progress in Physics}\ }\textbf {\bibinfo {volume}
  {69}},\ \bibinfo {pages} {1325} (\bibinfo {year} {2006})}\BibitemShut
  {NoStop}%
\bibitem [{\citenamefont {Hennrich}\ \emph {et~al.}(2000)\citenamefont
  {Hennrich}, \citenamefont {Legero}, \citenamefont {Kuhn},\ and\ \citenamefont
  {Rempe}}]{Hennrich2000}%
  \BibitemOpen
  \bibfield  {author} {\bibinfo {author} {\bibfnamefont {M.}~\bibnamefont
  {Hennrich}}, \bibinfo {author} {\bibfnamefont {T.}~\bibnamefont {Legero}},
  \bibinfo {author} {\bibfnamefont {A.}~\bibnamefont {Kuhn}}, \ and\ \bibinfo
  {author} {\bibfnamefont {G.}~\bibnamefont {Rempe}},\ }\href {\doibase
  10.1103/PhysRevLett.85.4872} {\bibfield  {journal} {\bibinfo  {journal}
  {Physical Review Letters}\ }\textbf {\bibinfo {volume} {85}},\ \bibinfo
  {pages} {4872} (\bibinfo {year} {2000})}\BibitemShut {NoStop}%
\bibitem [{\citenamefont {Mivehvar}\ \emph {et~al.}(2021)\citenamefont
  {Mivehvar}, \citenamefont {Piazza}, \citenamefont {Donner},\ and\
  \citenamefont {Ritsch}}]{Mivehvar2021}%
  \BibitemOpen
  \bibfield  {author} {\bibinfo {author} {\bibfnamefont {F.}~\bibnamefont
  {Mivehvar}}, \bibinfo {author} {\bibfnamefont {F.}~\bibnamefont {Piazza}},
  \bibinfo {author} {\bibfnamefont {T.}~\bibnamefont {Donner}}, \ and\ \bibinfo
  {author} {\bibfnamefont {H.}~\bibnamefont {Ritsch}},\ }\href
  {http://arxiv.org/abs/2102.04473} {\  (\bibinfo {year} {2021})},\ \Eprint
  {http://arxiv.org/abs/2102.04473} {arXiv:2102.04473} \BibitemShut {NoStop}%
\bibitem [{\citenamefont {Zhang}\ \emph {et~al.}(2012)\citenamefont {Zhang},
  \citenamefont {McConnell}, \citenamefont {{\'{C}}uk}, \citenamefont {Lin},
  \citenamefont {Schleier-Smith}, \citenamefont {Leroux},\ and\ \citenamefont
  {Vuleti{\'{c}}}}]{Zhang2012}%
  \BibitemOpen
  \bibfield  {author} {\bibinfo {author} {\bibfnamefont {H.}~\bibnamefont
  {Zhang}}, \bibinfo {author} {\bibfnamefont {R.}~\bibnamefont {McConnell}},
  \bibinfo {author} {\bibfnamefont {S.}~\bibnamefont {{\'{C}}uk}}, \bibinfo
  {author} {\bibfnamefont {Q.}~\bibnamefont {Lin}}, \bibinfo {author}
  {\bibfnamefont {M.~H.}\ \bibnamefont {Schleier-Smith}}, \bibinfo {author}
  {\bibfnamefont {I.~D.}\ \bibnamefont {Leroux}}, \ and\ \bibinfo {author}
  {\bibfnamefont {V.}~\bibnamefont {Vuleti{\'{c}}}},\ }\href {\doibase
  10.1103/PhysRevLett.109.133603} {\bibfield  {journal} {\bibinfo  {journal}
  {Physical Review Letters}\ }\textbf {\bibinfo {volume} {109}},\ \bibinfo
  {pages} {133603} (\bibinfo {year} {2012})}\BibitemShut {NoStop}%
\bibitem [{\citenamefont {Hume}\ \emph {et~al.}(2013)\citenamefont {Hume},
  \citenamefont {Stroescu}, \citenamefont {Joos}, \citenamefont {Muessel},
  \citenamefont {Strobel},\ and\ \citenamefont {Oberthaler}}]{Hume2013}%
  \BibitemOpen
  \bibfield  {author} {\bibinfo {author} {\bibfnamefont {D.~B.}\ \bibnamefont
  {Hume}}, \bibinfo {author} {\bibfnamefont {I.}~\bibnamefont {Stroescu}},
  \bibinfo {author} {\bibfnamefont {M.}~\bibnamefont {Joos}}, \bibinfo {author}
  {\bibfnamefont {W.}~\bibnamefont {Muessel}}, \bibinfo {author} {\bibfnamefont
  {H.}~\bibnamefont {Strobel}}, \ and\ \bibinfo {author} {\bibfnamefont
  {M.~K.}\ \bibnamefont {Oberthaler}},\ }\href {\doibase
  10.1103/PhysRevLett.111.253001} {\bibfield  {journal} {\bibinfo  {journal}
  {Physical Review Letters}\ }\textbf {\bibinfo {volume} {111}},\ \bibinfo
  {pages} {253001} (\bibinfo {year} {2013})},\ \Eprint
  {http://arxiv.org/abs/1307.7598} {arXiv:1307.7598} \BibitemShut {NoStop}%
\bibitem [{\citenamefont {Jeffrey}\ \emph {et~al.}(2014)\citenamefont
  {Jeffrey}, \citenamefont {Sank}, \citenamefont {Mutus}, \citenamefont
  {White}, \citenamefont {Kelly}, \citenamefont {Barends}, \citenamefont
  {Chen}, \citenamefont {Chen}, \citenamefont {Chiaro}, \citenamefont
  {Dunsworth}, \citenamefont {Megrant}, \citenamefont {O'Malley}, \citenamefont
  {Neill}, \citenamefont {Roushan}, \citenamefont {Vainsencher}, \citenamefont
  {Wenner}, \citenamefont {Cleland},\ and\ \citenamefont
  {Martinis}}]{Jeffrey2014}%
  \BibitemOpen
  \bibfield  {author} {\bibinfo {author} {\bibfnamefont {E.}~\bibnamefont
  {Jeffrey}}, \bibinfo {author} {\bibfnamefont {D.}~\bibnamefont {Sank}},
  \bibinfo {author} {\bibfnamefont {J.~Y.}\ \bibnamefont {Mutus}}, \bibinfo
  {author} {\bibfnamefont {T.~C.}\ \bibnamefont {White}}, \bibinfo {author}
  {\bibfnamefont {J.}~\bibnamefont {Kelly}}, \bibinfo {author} {\bibfnamefont
  {R.}~\bibnamefont {Barends}}, \bibinfo {author} {\bibfnamefont
  {Y.}~\bibnamefont {Chen}}, \bibinfo {author} {\bibfnamefont {Z.}~\bibnamefont
  {Chen}}, \bibinfo {author} {\bibfnamefont {B.}~\bibnamefont {Chiaro}},
  \bibinfo {author} {\bibfnamefont {A.}~\bibnamefont {Dunsworth}}, \bibinfo
  {author} {\bibfnamefont {A.}~\bibnamefont {Megrant}}, \bibinfo {author}
  {\bibfnamefont {P.~J.~J.}\ \bibnamefont {O'Malley}}, \bibinfo {author}
  {\bibfnamefont {C.}~\bibnamefont {Neill}}, \bibinfo {author} {\bibfnamefont
  {P.}~\bibnamefont {Roushan}}, \bibinfo {author} {\bibfnamefont
  {A.}~\bibnamefont {Vainsencher}}, \bibinfo {author} {\bibfnamefont
  {J.}~\bibnamefont {Wenner}}, \bibinfo {author} {\bibfnamefont {A.~N.}\
  \bibnamefont {Cleland}}, \ and\ \bibinfo {author} {\bibfnamefont {J.~M.}\
  \bibnamefont {Martinis}},\ }\href {\doibase 10.1103/PhysRevLett.112.190504}
  {\bibfield  {journal} {\bibinfo  {journal} {Physical Review Letters}\
  }\textbf {\bibinfo {volume} {112}},\ \bibinfo {pages} {190504} (\bibinfo
  {year} {2014})}\BibitemShut {NoStop}%
\bibitem [{\citenamefont {Walter}\ \emph {et~al.}(2017)\citenamefont {Walter},
  \citenamefont {Kurpiers}, \citenamefont {Gasparinetti}, \citenamefont
  {Magnard}, \citenamefont {Poto{\v{c}}nik}, \citenamefont {Salath{\'{e}}},
  \citenamefont {Pechal}, \citenamefont {Mondal}, \citenamefont {Oppliger},
  \citenamefont {Eichler},\ and\ \citenamefont {Wallraff}}]{Walter2017}%
  \BibitemOpen
  \bibfield  {author} {\bibinfo {author} {\bibfnamefont {T.}~\bibnamefont
  {Walter}}, \bibinfo {author} {\bibfnamefont {P.}~\bibnamefont {Kurpiers}},
  \bibinfo {author} {\bibfnamefont {S.}~\bibnamefont {Gasparinetti}}, \bibinfo
  {author} {\bibfnamefont {P.}~\bibnamefont {Magnard}}, \bibinfo {author}
  {\bibfnamefont {A.}~\bibnamefont {Poto{\v{c}}nik}}, \bibinfo {author}
  {\bibfnamefont {Y.}~\bibnamefont {Salath{\'{e}}}}, \bibinfo {author}
  {\bibfnamefont {M.}~\bibnamefont {Pechal}}, \bibinfo {author} {\bibfnamefont
  {M.}~\bibnamefont {Mondal}}, \bibinfo {author} {\bibfnamefont
  {M.}~\bibnamefont {Oppliger}}, \bibinfo {author} {\bibfnamefont
  {C.}~\bibnamefont {Eichler}}, \ and\ \bibinfo {author} {\bibfnamefont
  {A.}~\bibnamefont {Wallraff}},\ }\href {\doibase
  10.1103/PhysRevApplied.7.054020} {\bibfield  {journal} {\bibinfo  {journal}
  {Physical Review Applied}\ }\textbf {\bibinfo {volume} {7}},\ \bibinfo
  {pages} {054020} (\bibinfo {year} {2017})}\BibitemShut {NoStop}%
\bibitem [{\citenamefont {Dassonneville}\ \emph {et~al.}(2020)\citenamefont
  {Dassonneville}, \citenamefont {Ramos}, \citenamefont {Milchakov},
  \citenamefont {Planat}, \citenamefont {Dumur}, \citenamefont {Foroughi},
  \citenamefont {Puertas}, \citenamefont {Leger}, \citenamefont {Bharadwaj},
  \citenamefont {Delaforce}, \citenamefont {Naud}, \citenamefont
  {Hasch-Guichard}, \citenamefont {Garc{\'{i}}a-Ripoll}, \citenamefont {Roch},\
  and\ \citenamefont {Buisson}}]{Dassonneville2020}%
  \BibitemOpen
  \bibfield  {author} {\bibinfo {author} {\bibfnamefont {R.}~\bibnamefont
  {Dassonneville}}, \bibinfo {author} {\bibfnamefont {T.}~\bibnamefont
  {Ramos}}, \bibinfo {author} {\bibfnamefont {V.}~\bibnamefont {Milchakov}},
  \bibinfo {author} {\bibfnamefont {L.}~\bibnamefont {Planat}}, \bibinfo
  {author} {\bibnamefont {Dumur}}, \bibinfo {author} {\bibfnamefont
  {F.}~\bibnamefont {Foroughi}}, \bibinfo {author} {\bibfnamefont
  {J.}~\bibnamefont {Puertas}}, \bibinfo {author} {\bibfnamefont
  {S.}~\bibnamefont {Leger}}, \bibinfo {author} {\bibfnamefont
  {K.}~\bibnamefont {Bharadwaj}}, \bibinfo {author} {\bibfnamefont
  {J.}~\bibnamefont {Delaforce}}, \bibinfo {author} {\bibfnamefont
  {C.}~\bibnamefont {Naud}}, \bibinfo {author} {\bibfnamefont {W.}~\bibnamefont
  {Hasch-Guichard}}, \bibinfo {author} {\bibfnamefont {J.~J.}\ \bibnamefont
  {Garc{\'{i}}a-Ripoll}}, \bibinfo {author} {\bibfnamefont {N.}~\bibnamefont
  {Roch}}, \ and\ \bibinfo {author} {\bibfnamefont {O.}~\bibnamefont
  {Buisson}},\ }\href {\doibase 10.1103/PhysRevX.10.011045} {\bibfield
  {journal} {\bibinfo  {journal} {Physical Review X}\ }\textbf {\bibinfo
  {volume} {10}},\ \bibinfo {pages} {11045} (\bibinfo {year}
  {2020})}\BibitemShut {NoStop}%
\bibitem [{\citenamefont {Mitchell}(2017)}]{Mitchell2017}%
  \BibitemOpen
  \bibfield  {author} {\bibinfo {author} {\bibfnamefont {M.~W.}\ \bibnamefont
  {Mitchell}},\ }\href {\doibase 10.1088/2058-9565/aa80c0} {\bibfield
  {journal} {\bibinfo  {journal} {Quantum Science and Technology}\ }\textbf
  {\bibinfo {volume} {2}},\ \bibinfo {pages} {044005} (\bibinfo {year}
  {2017})}\BibitemShut {NoStop}%
\bibitem [{\citenamefont {Thompson}\ \emph {et~al.}(1992)\citenamefont
  {Thompson}, \citenamefont {Rempe},\ and\ \citenamefont
  {Kimble}}]{Thompson1992}%
  \BibitemOpen
  \bibfield  {author} {\bibinfo {author} {\bibfnamefont {R.~J.}\ \bibnamefont
  {Thompson}}, \bibinfo {author} {\bibfnamefont {G.}~\bibnamefont {Rempe}}, \
  and\ \bibinfo {author} {\bibfnamefont {H.~J.}\ \bibnamefont {Kimble}},\
  }\href {\doibase 10.1103/PhysRevLett.68.1132} {\bibfield  {journal} {\bibinfo
   {journal} {Physical Review Letters}\ }\textbf {\bibinfo {volume} {68}},\
  \bibinfo {pages} {1132} (\bibinfo {year} {1992})}\BibitemShut {NoStop}%
\bibitem [{\citenamefont {Brune}\ \emph {et~al.}(1996)\citenamefont {Brune},
  \citenamefont {Schmidt-Kaler}, \citenamefont {Maali}, \citenamefont {Dreyer},
  \citenamefont {Hagley}, \citenamefont {Raimond},\ and\ \citenamefont
  {Haroche}}]{Brune1996}%
  \BibitemOpen
  \bibfield  {author} {\bibinfo {author} {\bibfnamefont {M.}~\bibnamefont
  {Brune}}, \bibinfo {author} {\bibfnamefont {F.}~\bibnamefont
  {Schmidt-Kaler}}, \bibinfo {author} {\bibfnamefont {A.}~\bibnamefont
  {Maali}}, \bibinfo {author} {\bibfnamefont {J.}~\bibnamefont {Dreyer}},
  \bibinfo {author} {\bibfnamefont {E.}~\bibnamefont {Hagley}}, \bibinfo
  {author} {\bibfnamefont {J.~M.}\ \bibnamefont {Raimond}}, \ and\ \bibinfo
  {author} {\bibfnamefont {S.}~\bibnamefont {Haroche}},\ }\href {\doibase
  10.1103/PhysRevLett.76.1800} {\bibfield  {journal} {\bibinfo  {journal}
  {Physical Review Letters}\ }\textbf {\bibinfo {volume} {76}},\ \bibinfo
  {pages} {1800} (\bibinfo {year} {1996})}\BibitemShut {NoStop}%
\bibitem [{\citenamefont {Yoshie}\ \emph {et~al.}(2004)\citenamefont {Yoshie},
  \citenamefont {Scherer}, \citenamefont {Hendrickson}, \citenamefont
  {Khitrova}, \citenamefont {Gibbs}, \citenamefont {Rupper}, \citenamefont
  {Ell}, \citenamefont {Shchekin},\ and\ \citenamefont {Deppe}}]{Yoshie2004}%
  \BibitemOpen
  \bibfield  {author} {\bibinfo {author} {\bibfnamefont {T.}~\bibnamefont
  {Yoshie}}, \bibinfo {author} {\bibfnamefont {A.}~\bibnamefont {Scherer}},
  \bibinfo {author} {\bibfnamefont {J.}~\bibnamefont {Hendrickson}}, \bibinfo
  {author} {\bibfnamefont {G.}~\bibnamefont {Khitrova}}, \bibinfo {author}
  {\bibfnamefont {H.~M.}\ \bibnamefont {Gibbs}}, \bibinfo {author}
  {\bibfnamefont {G.}~\bibnamefont {Rupper}}, \bibinfo {author} {\bibfnamefont
  {C.}~\bibnamefont {Ell}}, \bibinfo {author} {\bibfnamefont {O.~B.}\
  \bibnamefont {Shchekin}}, \ and\ \bibinfo {author} {\bibfnamefont {D.~G.}\
  \bibnamefont {Deppe}},\ }\href {\doibase 10.1038/nature03119} {\bibfield
  {journal} {\bibinfo  {journal} {Nature}\ }\textbf {\bibinfo {volume} {432}},\
  \bibinfo {pages} {200} (\bibinfo {year} {2004})}\BibitemShut {NoStop}%
\bibitem [{\citenamefont {Wallraff}\ \emph {et~al.}(2004)\citenamefont
  {Wallraff}, \citenamefont {Schuster}, \citenamefont {Blais}, \citenamefont
  {Frunzio}, \citenamefont {Huang}, \citenamefont {Majer}, \citenamefont
  {Kumar}, \citenamefont {Girvin},\ and\ \citenamefont
  {Schoelkopf}}]{Wallraff2004}%
  \BibitemOpen
  \bibfield  {author} {\bibinfo {author} {\bibfnamefont {A.}~\bibnamefont
  {Wallraff}}, \bibinfo {author} {\bibfnamefont {D.~I.}\ \bibnamefont
  {Schuster}}, \bibinfo {author} {\bibfnamefont {A.}~\bibnamefont {Blais}},
  \bibinfo {author} {\bibfnamefont {L.}~\bibnamefont {Frunzio}}, \bibinfo
  {author} {\bibfnamefont {R.-S.}\ \bibnamefont {Huang}}, \bibinfo {author}
  {\bibfnamefont {J.}~\bibnamefont {Majer}}, \bibinfo {author} {\bibfnamefont
  {S.}~\bibnamefont {Kumar}}, \bibinfo {author} {\bibfnamefont {S.~M.}\
  \bibnamefont {Girvin}}, \ and\ \bibinfo {author} {\bibfnamefont {R.~J.}\
  \bibnamefont {Schoelkopf}},\ }\href {\doibase 10.1038/nature02851} {\bibfield
   {journal} {\bibinfo  {journal} {Nature}\ }\textbf {\bibinfo {volume}
  {431}},\ \bibinfo {pages} {162} (\bibinfo {year} {2004})}\BibitemShut
  {NoStop}%
\bibitem [{\citenamefont {Reithmaier}\ \emph {et~al.}(2004)\citenamefont
  {Reithmaier}, \citenamefont {S{\k{e}}k}, \citenamefont
  {L{\"{o}}ffler}, \citenamefont {Hofmann}, \citenamefont {Kuhn}, \citenamefont
  {Reitzenstein}, \citenamefont {Keldysh}, \citenamefont {Kulakovskii},
  \citenamefont {Reinecke},\ and\ \citenamefont {Forchel}}]{Reithmaier2004}%
  \BibitemOpen
  \bibfield  {author} {\bibinfo {author} {\bibfnamefont {J.~P.}\ \bibnamefont
  {Reithmaier}}, \bibinfo {author} {\bibfnamefont {G.}~\bibnamefont
  {S{\k{e}}k}}, \bibinfo {author} {\bibfnamefont
  {A.}~\bibnamefont {L{\"{o}}ffler}}, \bibinfo {author} {\bibfnamefont
  {C.}~\bibnamefont {Hofmann}}, \bibinfo {author} {\bibfnamefont
  {S.}~\bibnamefont {Kuhn}}, \bibinfo {author} {\bibfnamefont {S.}~\bibnamefont
  {Reitzenstein}}, \bibinfo {author} {\bibfnamefont {L.~V.}\ \bibnamefont
  {Keldysh}}, \bibinfo {author} {\bibfnamefont {V.~D.}\ \bibnamefont
  {Kulakovskii}}, \bibinfo {author} {\bibfnamefont {T.~L.}\ \bibnamefont
  {Reinecke}}, \ and\ \bibinfo {author} {\bibfnamefont {A.}~\bibnamefont
  {Forchel}},\ }\href {\doibase 10.1038/nature02969} {\bibfield  {journal}
  {\bibinfo  {journal} {Nature}\ }\textbf {\bibinfo {volume} {432}},\ \bibinfo
  {pages} {197} (\bibinfo {year} {2004})}\BibitemShut {NoStop}%
\bibitem [{\citenamefont {Chiorescu}\ \emph {et~al.}(2004)\citenamefont
  {Chiorescu}, \citenamefont {Bertet}, \citenamefont {Semba}, \citenamefont
  {Nakamura}, \citenamefont {Harmans},\ and\ \citenamefont
  {Mooij}}]{Chiorescu2004}%
  \BibitemOpen
  \bibfield  {author} {\bibinfo {author} {\bibfnamefont {I.}~\bibnamefont
  {Chiorescu}}, \bibinfo {author} {\bibfnamefont {P.}~\bibnamefont {Bertet}},
  \bibinfo {author} {\bibfnamefont {K.}~\bibnamefont {Semba}}, \bibinfo
  {author} {\bibfnamefont {Y.}~\bibnamefont {Nakamura}}, \bibinfo {author}
  {\bibfnamefont {C.~J. P.~M.}\ \bibnamefont {Harmans}}, \ and\ \bibinfo
  {author} {\bibfnamefont {J.~E.}\ \bibnamefont {Mooij}},\ }\href {\doibase
  10.1038/nature02831} {\bibfield  {journal} {\bibinfo  {journal} {Nature}\
  }\textbf {\bibinfo {volume} {431}},\ \bibinfo {pages} {159} (\bibinfo {year}
  {2004})}\BibitemShut {NoStop}%
\bibitem [{\citenamefont {Reiserer}\ and\ \citenamefont
  {Rempe}(2015)}]{Reiserer2015}%
  \BibitemOpen
  \bibfield  {author} {\bibinfo {author} {\bibfnamefont {A.}~\bibnamefont
  {Reiserer}}\ and\ \bibinfo {author} {\bibfnamefont {G.}~\bibnamefont
  {Rempe}},\ }\href {\doibase 10.1103/RevModPhys.87.1379} {\bibfield  {journal}
  {\bibinfo  {journal} {Reviews of Modern Physics}\ }\textbf {\bibinfo {volume}
  {87}},\ \bibinfo {pages} {1379} (\bibinfo {year} {2015})}\BibitemShut
  {NoStop}%
\end{thebibliography}
\end{document}